\documentclass[sigconf, acm, screen]{acmart}
\newtheorem{myDef}{\bf Definition}

\newtheorem{myEmp}{\bf Example}

\newtheorem{myProblem}{\bf Problem}

\definecolor{deepgreen}{rgb}{0.0, 0.5, 0.0}

\usepackage[tikz]{bclogo}

\usepackage[utf8]{inputenc}  
\usepackage[T1]{fontenc}     
\usepackage{lipsum}
\usepackage{xcolor}
\usepackage{tabularx} 
\usepackage{enumitem}
\usepackage{subfigure}
\usepackage{multirow}
\usepackage{algorithm}      
\usepackage{algorithmic}    
\usepackage{amsmath}        
\newcommand{\spara}[1]{\smallskip\noindent{\bf #1}}
\newcommand{\stitle}[1]{\vspace{1.5ex}\noindent{\bf #1}}

\setcopyright{acmcopyright}
\copyrightyear{2026}
\acmYear{2026}
\acmDOI{XXXXXXX.XXXXXXX}

\acmConference[SIGMOD '26]{ACM International Conference on Management of Data}{May 31--June 5, 2026}{Bengaluru, India}

\acmPrice{15.00}
\acmISBN{978-1-4503-XXXX-X/18/06}

\setcopyright{acmcopyright}

\begin{document}

\title{
In-context Clustering-based Entity Resolution
with Large Language Models: A Design Space Exploration
}

\author{Jiajie Fu}
\affiliation{%
  \institution{Zhejiang University}
}
\email{jiajiefu@zju.edu.cn}

\author{Haitong Tang}
\affiliation{%
  \institution{Zhejiang University}
}
\email{tht@zju.edu.cn}

\author{Arijit Khan}
\affiliation{%
  \institution{Aalborg University}
}
\email{arijitk@cs.aau.dk}

\author{Sharad Mehrotra}
\affiliation{%
  \institution{University of California, Irvine}
}
\email{sharad@ics.uci.edu}

\author{Xiangyu Ke}
\affiliation{%
  \institution{Zhejiang University}
}
\email{xiangyu.ke@zju.edu.cn}

\author{Yunjun Gao}
\affiliation{%
  \institution{Zhejiang University}
}
\email{gaoyj@zju.edu.cn}

\begin{abstract}
Entity Resolution (ER) is a fundamental data quality improvement task that identifies and links records referring to the same real-world entity. Traditional ER approaches often rely on pairwise comparisons, which can be costly regarding both time and monetary resources, especially when large datasets are involved. Recently, Large Language Models (LLMs) have demonstrated promising results in ER tasks. Still, existing methods typically focus on pairwise matching, missing the potential of LLMs to directly perform clustering in a more cost-effective and scalable manner. In this paper, we propose a novel {\em in-context clustering} approach for ER, where LLMs are used to cluster records directly, reducing both time complexity and monetary costs. We systematically investigate the design space for in-context clustering, analyzing the impact of factors such as set size, diversity, variation, and ordering of records on clustering performance. Based on these insights, we develop
{\sf LLM-CER} (\underline{LLM}-powered \underline{C}lustering-based
\underline{ER}) that obtains high-quality ER results while minimizing LLM API calls. Our approach addresses key challenges, including efficient cluster merging and LLM's hallucination, providing a scalable and effective solution for ER. 
Extensive experiments on nine real-world datasets demonstrate that our method significantly improves result quality, achieving up to 150\% higher accuracy, 10\% increase in the FP-measure, and reducing API calls by up to 5$\times$, while maintaining a comparable monetary cost to the most cost-effective baseline.
\end{abstract}

\maketitle

\section{Introduction}
\label{sec:intro}

Entity Resolution (ER) is a fundamental data quality task that identifies and disambiguates real-world entities represented by different identifiers or descriptions across multiple records. 
In ER, a {\em record} denotes a specific data instance; an {\em entity} corresponds to a unique object represented by one or more potentially duplicate records; and {\em attributes} are the properties of each record, which may be numerical (e.g., price), categorical (e.g., brand), or textual (e.g., descriptions).
ER aims to link and cluster these records, ensuring a clean and unified dataset~\cite{Christen12,ChristophidesEP21,ElmagarmidIV07,2010Naumann}. 
Traditional ER is computationally expensive, particularly for large datasets, as it often requires evaluating {\em all possible pairs} of records to detect duplicates. 
To reduce this cost, ER typically includes a pre-processing step, {\em blocking} or {\em filtering}~\cite{papadakis2020blocking}, which partitions the dataset into (possibly overlapping) blocks, where records in different blocks are unlikely to refer to the same entity.
The subsequent {\em resolution} phase, which involves matching and combining records, is the more resource-intensive step, ensuring that duplicates are grouped within the same cluster.

\begin{figure}[t]
    \hspace*{-7mm}
    \centering    \hspace{5mm}\includegraphics[width=0.467\textwidth]{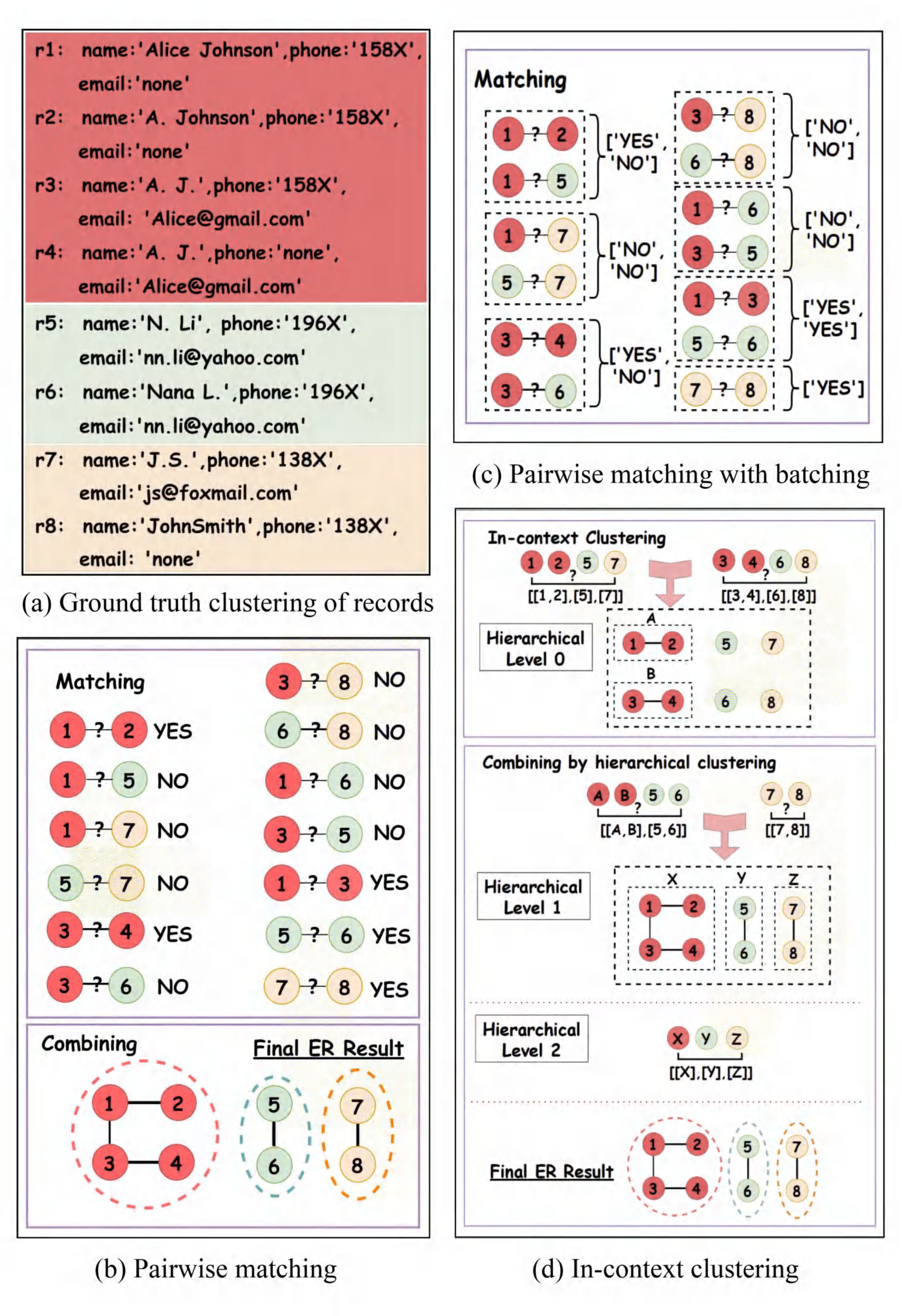}
    \caption{(a) Ground truth ER result. (b) Pairwise matching and (c) Pairwise matching with batching (state-of-the-art) vs. (d) In-context clustering (this work) for ER. The combining phase is omitted for pairwise matching with batching \cite{fan2024cost}, as it remains the same with pairwise matching \cite{narayan2022can}.}
\label{fig:intro example}
\end{figure}

Existing ER solutions can be broadly divided into three categories: crowdsourcing-based frameworks~\cite{VerroiosG15}, machine learning (ML) and deep learning-oriented methods \cite{mests2018distributed}, and those leveraging large language models (LLMs) \cite{li2024booster}, such as {\sf GPT} or {\sf LLAMA}. 
Crowdsourcing frameworks rely on human intelligence to achieve high-quality results but incur significant costs. For example, platforms like Amazon Mechanical Turk (AMT) typically charge around USD 0.02 per Human Intelligence Task (HIT) involving pairwise comparisons~\cite{wang2012crowder}. For a dataset containing 1,000 records, naive pairwise ER would require approximately $O(10^6)$ comparisons—amounting to around USD 20,000 costs. To mitigate such expenses, hybrid human-machine frameworks like {\sf CrowdER}~\cite{wang2012crowder} employ automated techniques to eliminate obvious non-matches, reserving human effort for only the most uncertain pairs.
Machine-based approaches, such as rule-based \cite{BenjellounGMSWW09}, ML-based \cite{BilenkoM03}, and deep learning-based \cite{mests2018distributed} works, formulate pairwise matching as a classification problem. 
These models can be trained on labeled data to provide faster inference. Still, they require {\em task-specific supervision} and {\em large amounts of labeled data}, both of which are expensive and time-consuming to acquire. 
For instance, AMT charges USD 0.08 per labeling task~\cite{fan2024cost}, making the data annotation process costly and difficult. 
Recently emerged large language models, e.g., {\sf GPT-4},  are pre-trained on massive corpora in a self-supervised manner \cite{liu2021self} and have achieved remarkable results in a series of natural language processing and data science tasks like text generation \cite{qu2023layoutllm}, machine translation \cite{enis2024llm}, data research and education \cite{AmerYahiaBCLSXY23}. 
General-purpose LLMs circumvent the need for expensive task-specific supervision and labeled data, thanks to simpler prompt-based interactions, making them easier to operate and iterate in downstream applications. 
Moreover, their lower costs—e.g., {\sf GPT-4o-mini} charges only USD 0.15 per million tokens~\cite{apicost}—and superior performance on zero-shot tasks~\cite{kojima2022large} make them an attractive alternative for entity resolution tasks.


Recently, several works have pursued LLMs for ER. 
Narayan et al. \cite{narayan2022can} pioneered the exploration of {\sf GPT-3}'s potential for ER in a few-shot manner using task demonstrations. Their approach has exhibited notable performance improvements compared to methods based on pre-trained language models (PLMs), e.g., {\sf Ditto} \cite{li2023effective}. 
{\sf ZeroMatch}~\cite{ZhangGCS24} takes a more advanced approach by using parameter-efficient fine-tuning, maintaining many pre-trained LLM variants from different domains, and selecting the most appropriate variant at inference time for zero-shot ER. 
Further advancements in prompt engineering for ER have been investigated in works such as~\cite{NananukulSK24, WLCHWZS25, PeetersB23, A24}. 
Li et al. \cite{li2024booster} develop {\sf BoostER}, a framework that applies LLMs to select the best answer from multiple different partitionings generated by existing ER tools. Their focus is on the next record pair selection problem, reducing the uncertainty of the current ER result. 
Additionally, batch prompting, which involves packing multiple pairwise questions into a single prompt to reduce the monetary cost of calling LLM APIs, has been studied in~\cite{fan2024cost, zhang2023large}. 

However, all state-of-the-art LLM-based ER methods rely on {\em comparing record pairs} and do not fully exploit the emergent capability of LLMs to {\em directly cluster a set of records}. 
Recent studies have shown that LLMs are surprisingly effective at clustering text data in a zero-shot or few-shot manner, preserving semantic meaning and context \cite{0002GGLWN24,0001WS23,abs-2405-00988}. 
Given that entity resolution is essentially a clustering problem, LLMs' clustering capabilities can be directly leveraged to reduce both the time and monetary costs associated with end-to-end ER. 
To bridge this gap, we -- for the first time -- formulate and investigate the problem of in-context clustering-based entity resolution, that is, instructing an LLM to cluster different record sets to achieve the end-to-end ER task at a significantly lower cost and time. Consider the following example.

\begin{myEmp}
Figure 1(b) compares existing pairwise matching \cite{narayan2022can} with our proposed in-context clustering approach for an ER task. Figure 1(a) shows the records with ground-truth entity assignments, where each color represents an entity.

Pairwise questioning-based ER consists of two phases: matching and combining (excluding blocking for simplicity).
Matching aims to determine if a pair of records refers to the same entity. 
Transitivity can reduce comparisons; for example, if records $a$ and $b$ are the same entity, and $b$ and $c$ are the same (or different), we can infer that $a$ and $c$ are also the same (or different).
As shown in Figure 1(b), the matching phase concludes when all record pairs are compared explicitly, or their status is inferred through transitivity or anti-transitivity.  
In the combining phase, the final ER result is obtained by clustering records belonging to the same entity.

In contrast, our in-context clustering approach directly creates input record sets for clustering by the LLM, as depicted in Figure 1(d) (with a maximum size of 4). 
Records within each LLM cluster are considered the same entity (i.e., transitivity holds). 
For example, in the LLM-based clustering of the set $(1,2,5,7)$, records $1$ and $2$ are clustered together, indicating they are the same entity, while $2$ and $5$ belong to different entities (i.e., anti-transitivity), as reflected in their assignment to clusters $A = [1,2]$ and $ [5]$.
In cases where transitivity or anti-transitivity relationships are unknown, clusters are packed to form the next level of input sets, such as $(A, B, 5, 6)$. 
This hierarchical process continues, merging clusters until all record pairs from different clusters satisfy anti-transitivity, yielding the final ER output.
\end{myEmp}

Unlike pairwise matching, which classifies two records at a time and heavily relies on transitivity, in-context clustering {\em partitions multiple records simultaneously} and {\em hierarchically merges them}. 
Each pairwise comparison requires an LLM API call, driving up both cost and time. In the worst case, the ER process requires comparing all pairs of records when they belong to different entities. 
As illustrated in Figure 1(b), pairwise matching requires 13 comparisons (i.e., 13 LLM API calls) for 8 records, even with the help of transitivity. 
Our method, however, requires only 5 record sets for clustering, leading to fewer LLM API calls, lower costs, and reduced running time. 
This efficiency becomes particularly significant for large-scale datasets, where processing costs and ER time are crucial.

\begin{myEmp}
    An improved baseline, pairwise matching with batching \cite{fan2024cost}, reduces costs by processing multiple pairwise questions in batches.
    With the same constraint of 4 records per LLM API call for a fair comparison (i.e., 2 pairwise questions in a batch), Figure 1(c) shows that batching reduces the number of API calls to 7, still higher than that of 5 by our in-context clustering approach.
    This inefficiency arises because the pairwise questioning format inherently introduces redundancy when conveying information to the LLM. 
    For instance, exploring all possible relationships among four records require 6 pairwise questions, which necessitates 3 batches of 2 pairwise questions each. 
    Even with transitivity and anti-transitivity among 4 records, the number of batches can only be reduced to 2.
    In contrast, our approach can capture all relationships among 4 records in a single API call, 
    thus reducing the number of API calls. 
    A key follow-up question is the highest possible information density (i.e., maximum set size constraint in our clustering solution) that the LLM can effectively process in a single API call\footnote{The number of relationships to explore grows quadratically with the set size.}. This motivates our in-depth investigation in \S~\ref{sec:experiment on key factors}.
\end{myEmp}

Although LLMs have demonstrated strong performance in ER via pairwise matching~\cite{fan2024cost, narayan2022can}, the application of in-context clustering for LLM-based ER remains largely unexplored. 
We observe that directly applying in-context clustering to ER, without a carefully designed mechanism, introduces several key challenges:
{\bf (1)} {\em Clustering Performance Variability:} The clustering performance of an LLM can vary significantly based on how the record sets are configured. 
Determining the optimal design for these sets—specifically in terms of size, diversity, and the ordering of records—is crucial. 
For instance, if the record set is too large, the LLM's performance may degrade due to long context lengths~\cite{li2024long}, whereas a smaller set size will result in excessive LLM API calls, thereby increasing the monetary cost. 
{\sf (2)} {\em Hallucination Risks:} LLMs are prone to hallucinations, where the model generates outputs that may not be grounded in the input data~\cite{gu2023hallucinations}. 
As a result, the outputs of in-context clustering may not always be accurate. Addressing this challenge requires the development of effective result verification strategies, such as implementing LLM ``guardrails'' and record set regeneration.
{\bf (3)} {\em Efficient Result Combination:} Combining the outputs of the in-context clustering efficiently presents another challenge. The process involves balancing the trade-off between the quality of the end-to-end ER result and the number of LLM API calls, which directly affects the monetary cost.
To address these challenges, we extensively explore the design space and introduce several novel algorithms that optimize the creation of input record sets, the validation of LLM clustering outputs, and the merging of clustering results. 

Our empirical evaluations reveal that in-context clustering generally performs best when each record set contains 9 records drawn from 4 distinct entities, with roughly equal representation per entity.
For domain-specific datasets, however, the optimal configuration tends to involve fewer records and entities, as factors such as attribute type and noise can significantly impact clustering quality. 
Performance further improves when records from the same entity appear consecutively within the set.
Building on these insights, we propose an effective record set creation strategy (\S~\ref{sec:in_context_cluster}) that balances set size, diversity, and intra-set variation to maximize clustering performance while minimizing monetary cost. 
Across nine real-world datasets, our method {\sf LLM-CER} (\underline{LLM}-powered \underline{C}lustering-based
\underline{ER}) yields substantial improvements—up to 150\% in Accuracy (ACC) and 10\% in FP-measure—while reducing API calls by up to 5$\times$ and maintaining comparable monetary cost to the most cost-effective baseline (\S~\ref{sec:exp_end2end}).
Additionally, scalability tests confirm that our method remains both effective and efficient as dataset size increases (\S~\ref{sec:scalability}).

\spara{Contribution.}
Our main contributions are summarized below. 
\begin{itemize}[left=4pt]
\item We are the first to apply a clustering-based approach using LLMs for entity resolution, exploring the design space of clustering-based LLM ER (\S~\ref{sub:batch scheme}).
\item We identify four key factors—set size, set diversity, set variation, and record order—and analyze their impact on the performance of in-context clustering (\S~\ref{sec:experiment on key factors}).
\item We propose the Next Record Set Creation algorithm, which effectively addresses the four identified factors. To mitigate potential LLM misclassifications, we design a Misclustering Detection Guardrail and a Record Set Regeneration strategy for further refinement of results (\S~\ref{sec:in_context_cluster}). Our experimental findings demonstrate that {\sf MDG} incurs 10\% time overhead, while improving the FP-measure by up to 75\% (\S~\ref{sec:exp_block_filter}). 
\item We propose a Hierarchical Cluster Merging approach that generates record sets hierarchically and performs efficient merging, improving the quality of cluster merging and overall entity resolution performance (\S~\ref{sec:cluster_merge}). 
\item We empirically evaluate our clustering-based end-to-end ER algorithm on nine real-world datasets. The results show that our method significantly improves ER quality, achieving up to 150\% higher ACC, reduces API calls by up to 5$\times$, while maintaining competitive cost efficiency compared to the most cost-effective baseline and demonstrating strong scalability (\S~\ref{sec:exp}).
\end{itemize} 

\vspace{-2.5mm}
\section{Related Work}
\label{sec:related works}

ER problems can be broadly classified into 
deduplication (dirty ER) and record linkage (clean-clean ER). The former partitions a single record collection into multiple entity clusters. 
The latter involves matching records from two separate, typically overlapping but duplicate-free, collections (e.g., two tables) and identifying pairs of records that refer to the same entity \cite{KondaDCDABLPZNP16}.
Dirty ER is generally more challenging than clean-clean ER due to inherent data quality issues, such as spelling errors, missing values, and noise. In contrast, clean-clean ER typically involves tables that are already cleaned and standardized~\cite{christophides2020overview}. 
This work focuses on the dirty ER problem, while our solution is also applicable to the clean-clean ER problem by considering the union of records across tables as a single collection. We categorize related work as follows. 

\vspace{-0.8mm}
\spara{PLM and LLM-based Entity Resolution.} 
Before the advent of large language models, pre-trained transformer language models (PLMs) were commonly used for ER tasks \cite{LiMWSW21}. 
{\sf DeepBlocker} \cite{Thirumuruganathan21} evaluates various deep learning methods, including {\sf SBERT} (Sentence BERT) \cite{Reimers2019SentenceBERTSE} for blocking.
{\sf Ditto} \cite{li2023effective} exploits PLMs such as {\sf BERT} \cite{devlin2018bert}, {\sf DistilBERT} \cite{SDCW19}, or {\sf RoBERTa} \cite{abs-1907-11692} for clean-clean ER. 
{\sf JointBERT} develops a dual-objective training method for {\sf BERT}, combining binary matching and multi-class classification, to predict both match/non-match decisions and entity identifiers in training pairs \cite{PeetersB21}.
{\sf RobEM} aims to enhance the robustness of PLM-based entity matching models \cite{RastaghiKR22}. 

PLM-based ER approaches typically require task-specific labeled data and fine-tuning, which are resource-intensive. 
More recently, LLMs, a.k.a. foundation models, have demonstrated strong performance in data cleaning and integration tasks, including ER, at a lower cost. This is primarily due to the more straightforward prompt-based interactions without requiring task-specific model re-training or labeled data. 
State-of-the-art LLM-based ER approaches are discussed in \S~\ref{sec:intro}. Among them, \cite{fan2024cost,zhang2023large,WLCHWZS25} are the closest to ours. 
Both \cite{fan2024cost,zhang2023large} introduce batch prompting, where multiple pairwise questions are packaged into a single batch for the LLM.
However, our approach differs by packing multiple records into a set for direct, in-context clustering via the LLM, allowing for more efficient and scalable ER. 
Furthermore, {\sf ComEM} \cite{WLCHWZS25} employs advanced prompts such as ``match'', ``compare'', or ``select'' to explore interactions across multiple records beyond pairwise questions. 
However, unlike ours, it does not leverage the direct clustering capacity of LLMs for a record set. As pairwise questioning is a special case of our in-context clustering (with set size = 2), our framework generalizes the state-of-the-art ER approaches, offering improved cost-effectiveness and efficiency. 

Besides, our work {\sf LLM-CER} differs from existing LLM-based ER methods in its comprehensive end-to-end pipeline, which includes blocking/filtering (except for \cite{fan2024cost}), and addresses critical issues like LLM hallucinations, which are not explicitly handled before. 

\vspace{-0.8mm}
\spara{Crowdsourcing-based Entity Resolution.} 
A key concern in crowdsourced entity resolution (ER) is minimizing the number of questions posed to workers, which directly impacts the overall cost.
Transitivity is commonly used to reduce the number of questions in \cite{WangLKFF13,VesdapuntBD14}. Closer to our approach, {\sf CrowdER} \cite{wang2012crowder} develops a clustering-based method where crowd workers are directly tasked with clustering a set of records. 
Various models for selecting high-quality questions have been developed in \cite{ChaiLLDF16,WhangLG13}. {\sf ZenCrowd} combines algorithmic and crowdsourcing-based matching techniques within a probabilistic framework \cite{DemartiniDC12}, while {\sf Roomba} uses a decision-theoretic approach to select matches to confirm based on their utility \cite{JefferyFH08}.
More recent works \cite{VerroiosG15,WangXL15,GruenheidKNG13,YKK17} consider crowd errors in ER tasks. 
{\sf Corleone} crowdsources the entire ER workflow, including blocking rules learning \cite{GDDNRSZ14}, and {\sf Falcon} scales Corleone using RDBMS-style query execution over a Hadoop cluster \cite{DasCDNKDARP17}.

Although {\sf CrowdER} \cite{wang2012crowder} provides record sets to human taskers for direct clustering, there are fundamental differences between their approach and our in-context clustering using LLMs. 
(1) {\sf CrowdER} aims to minimize the number of record sets required (equivalently the number of HITs), given a predefined set size, to cover and resolve all uncertain record pairs in a block. 
In contrast, we adopt a hierarchical approach for record set creation. Unlike {\sf CrowdER}, which generates all record sets {\em at once} and covers all uncertain record pairs, our method incrementally generates record sets across multiple layers. 
In the initial layers, we focus on generating non-overlapping sets that cover all records, {\em without immediately resolving all uncertain pairs}. As we progress to higher layers, we leverage transitivity and anti-transitivity to {\em identify and reduce unnecessary record sets}. 
This approach minimizes the total number of sets, thus reducing the number of LLM API calls and ultimately lowering monetary costs. 
According to our experimental results over real-world datasets and considering the same set size as well as the same blocking approach (e.g., {\sf LSH}~\cite{mests2018distributed}), the number of record sets (equivalently the number of HITs) needed for CrowdER can be 2-5$\times$ higher than the number of record sets (equivalently the number of LLM API calls) required by ours. 
(2) Our combining approach differs from CrowdER's in several key ways. 
The initial record sets generated by our algorithm are non-overlapping. Based on clustering results via the LLM, we conduct hierarchical record sets generation and further in-context clustering to merge those clusters. 
Our hierarchical record sets generation process maintains the optimal record set configuration, it also leverages transitivity and anti-transitivity to reduce the number of LLM API calls. 
In summary, direct and robust LLM-based clustering is employed both in our matching and combining phases. In contrast, {\sf CrowdER} allows overlaps of records in different HITs, and their cluster merging process relies on these overlaps to indirectly facilitate merging through transitivity. This could suffer due to human errors as two clusters could be incorrectly merged. 
(3) Finally, our approach includes {\em a guardrail mechanism to assess the quality of the LLM’s clustering results}. This allows us to identify and discard incorrect clustering results and regenerate better record sets. In contrast, {\sf CrowdER} assumes that crowdsourcing results are always accurate, it lacks a built-in process to validate or modify the clustering outcomes. 
These differences ensure that our method not only minimizes the number of record sets (equivalently the number of LLM API calls) but also maintains higher quality of the final ER results.

\vspace{-1mm}
\spara{Filtering and Blocking.} 
Blocking is a critical preprocessing step in ER, aimed at reducing computational overhead by partitioning records into groups and ensuring that only potentially matching records are compared.  
Blocking methods could be classified as rule-based, sorting-based, and hash-based. 
Rule-based methods organize tuples by applying fixed keys or decision rules created by experts or derived from heuristics~\cite{qian2017active}. 
Sorting-based methods cluster records by rapidly sorting them according to their textual similarities, which are determined using different similarity functions~\cite{levandowsky1971distance}. 
Hash-based approaches utilize hashing techniques, such as Min-Hashing and Locality-Sensitive Hashing (LSH), to place records into different buckets~\cite{mests2018distributed}. 
Filtering complements blocking by eliminating record pairs that are guaranteed not to match, thus focusing comparisons only on the remaining pairs to improve computational efficiency. 
Prominent filtering techniques are categorized into prefix-based, partition-based, and tree-based. We refer to \cite{papadakis2020blocking} for details.

\vspace{-1mm}
\section{Preliminaries}
\label{sec:preliminaries}
\vspace{-1mm}

\begin{figure}[tb!]
\centering
\vspace{-2mm}
\includegraphics[width=0.49\textwidth]{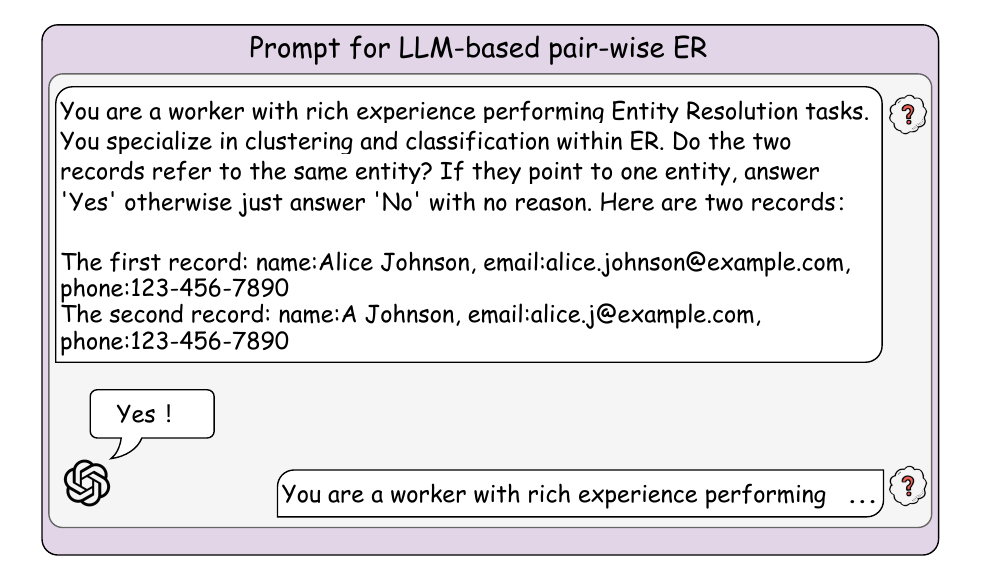}
\vspace{-9.5mm}
\caption{\centering{Example zero-shot prompt for pairwise ER}}
\label{fig:pair-wise prompt}
\vspace{-4mm}
\end{figure}

We discuss the background on the entity resolution (ER) problem (\S \ref{sec:ER}), followed by how pre-trained LLMs can be adapted for ER in a zero-shot prompting paradigm (\S \ref{sec:llm_adapt}).  

\vspace{-1.5mm}
\subsection{Entity Resolution}
\label{sec:ER}
\vspace{-1mm}

An ER algorithm receives an input set of records $\mathcal{R}$ and returns a partition of them: $\mathbb{C}=\{\mathcal{C}_1,\mathcal{C}_2,…,\mathcal{C}_n\}$, such that $\mathcal{C}_i \cap \mathcal{C}_j = \emptyset\,$ for all $i,j$, and $\cup_i \mathcal{C}_i = \mathcal{R}$. Each $\mathcal{C}_i$ is called a \textit{cluster} or a \textit{group} of $\mathcal{R}$ and represents a distinct real-world entity. 
If two records $r_i,r_j$ refer to the same (different) entity, they are denoted as $r_i=r_j$ ($r_i \neq r_j$). 

There are three key steps in entity resolution, {\em blocking/ filtering},  {\em matching}, and {\em combining} \cite{ChristophidesEP21, 2021Papadakis, get12theory}. 
Blocking and filtering aim to separate almost dissimilar records and keep similar records in the same block to reduce the number of comparisons in the subsequent matching step and make the algorithm more efficient \cite{PSTP20}. 
Matching determines whether records in the same block refer to the same entity. 
Finally, combining creates the final clusters of entities by inferring indirect matching relations following the matching step.  

In pairwise ER, a pairwise similarity function is used over each pair of records to find the matching pairs. In this case, each pair is referred to as a {\em question} posed to the similarity function. An ER algorithm generally obeys {\em transitivity} and {\em anti-transitivity} rules, which can be employed in the matching and combining phases to further reduce the number of questions \cite{BenjellounGMSWW09}.

\spara{Transitivity.} Given three records $r_1,r_2$, and $r_3$, if $r_1 = r_2$ and $r_2 = r_3$, then we have $r_1 = r_3$.

\spara{Anti-transitivity.} Given three records $r_1,r_2$, and $r_3$, if $r_1 = r_2$ and $r_2 \neq r_3$, then we have $r_1 \neq r_3$.

A clustering $\mathbb{C}$ of the input set of records $\mathcal{R}$ is transitively closed. For instance, if a pairwise similarity function decides $r_1 = r_2$ and $r_2 = r_3$, then we can infer $r_1 = r_3$ without employing the similarity function on the pair $(r_1, r_3)$. This reduces the number of questions.

\begin{figure}[tb!]
\centering
\vspace{-1mm}
\includegraphics[width=0.48\textwidth]{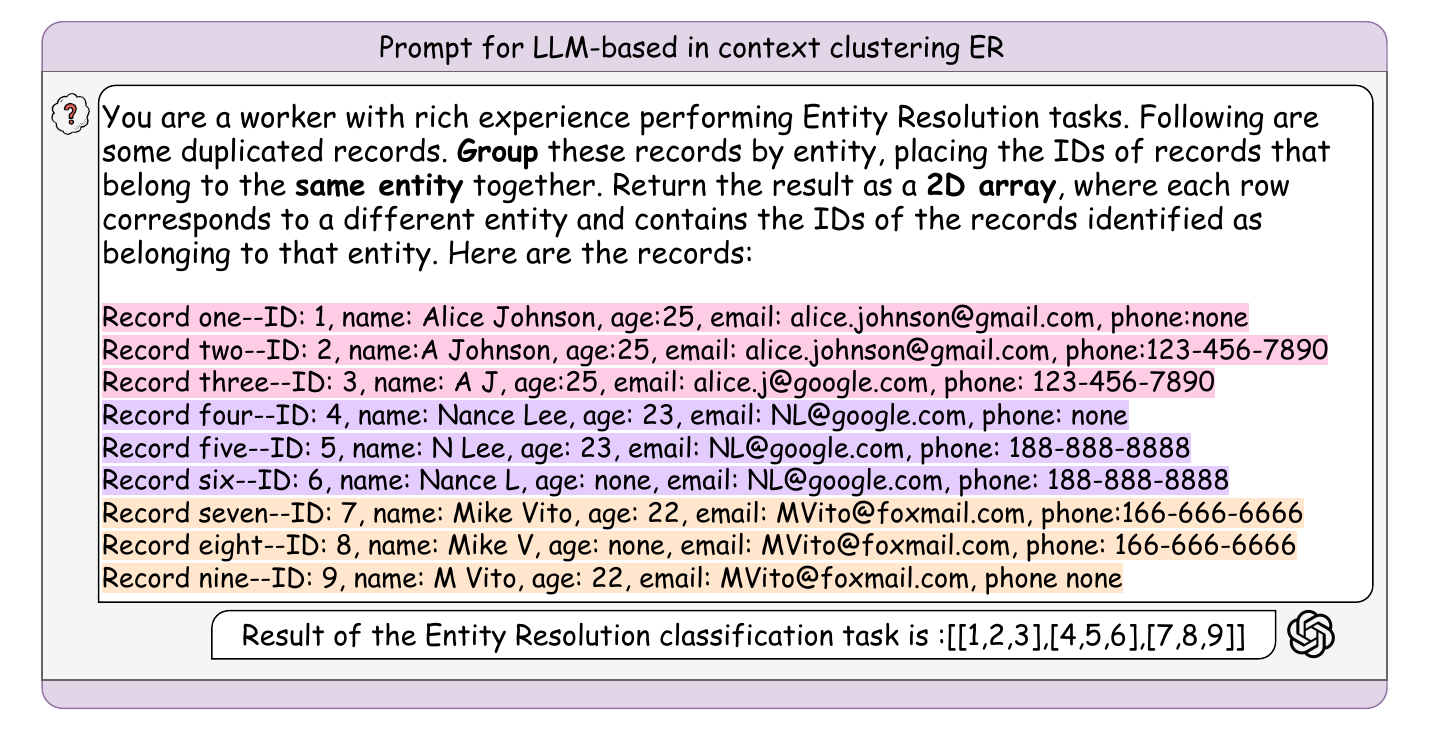}
\vspace{-8mm}
\caption{Example zero-shot prompt for in-context clustering} 
\label{fig:cluster ER}
\vspace{-6mm}
\end{figure}

\vspace{-1.5mm}
\subsection{LLM: Adaptation Techniques}
\label{sec:llm_adapt}
\vspace{-1mm}

Deep neural network–based models have long been effective as similarity functions for record matching, typically taking a pair of records as input and predicting whether they refer to the same real-world entity \cite{li2021improving,nie2019deep,hou2019gradual}. However, these methods demand large quantities of high-quality, task-specific labeled data, which can be costly to obtain. In contrast, large language models (LLMs)—pre-trained on massive text corpora via self-supervision—have recently demonstrated strong zero-shot performance on ER matching tasks \cite{narayan2022can,WLCHWZS25,fan2024cost,li2024booster,PeetersSB25,A24}. Their extensive prior knowledge and superior classification capabilities~\cite{brown2020language,chen2023label,sun2023text} make them well suited to entity disambiguation. Figure~\ref{fig:pair-wise prompt} illustrates a typical LLM prompt for pairwise ER. 
Our approach adopts a purely {\em zero-shot} strategy: rather than fine-tuning or providing in-context examples, it relies entirely on the LLM's internal representations to assess record similarity. 
Consequently, the underlying similarity function remains latent, rather than implemented as an explicit, hand-crafted metric.

\section{Problem: In-context Clustering}
\label{sub:batch scheme}

We first introduce our novel record set for in-context clustering and its key factors (\S \ref{sec:record set}), followed by experiments on how these factors impact clustering on individual record sets (\S \ref{sec:experiment on key factors}). Based on these results, we formally define the end-to-end ER problem (\S \ref{sec:problem for er}).

\vspace{-2mm}
\subsection{Record Set and Key Factors}
\label{sec:record set}
\vspace{-1mm}

Different from pairwise input schemes used by recent LLM-based ER solutions ~\cite{li2024booster,fan2024cost}, we combine the advantages of LLMs and innovatively propose to input a record set. In particular, we pack a number of records in a set as the input prompt and ask the LLM to output a clustering of the records in this set. Such an in-context clustering scheme can make good use of the LLM's ability to process batch data~\cite{zhang2023large}, as well as cluster text data preserving semantic similarity \cite{0001WS23}. 
Direct clustering of larger record sets reduces the number of questions to be prepared, subsequently minimizing the number of LLM API calls, the number of LLM tokens, and thereby obtaining high-quality answers at a lower cost and time.

Given a set of records as input to the LLM, the goal is to partition the records into clusters. 
We refer to this paradigm as "{\em in-context clustering}".
Three key factors may influence the performance of the LLM within this clustering-based scheme: set size, set diversity, and set variation. 
The definitions are given below. 

{\em{i) Set size}}\,: the number of records in an input set.

{\em{ii) Set diversity}}\,: the number of distinct entities (i.e., clusters) within an input set.

{\em{iii) Set variation}}\,: the variability in cluster sizes within an input set, quantified by the {\bf coefficient of variation}: 
\begin{equation}
    \label{set deviation}
    variation(S) = \frac{\sigma(S)}{\mu(S)}
\end{equation}
where $\sigma(S)$ and $\mu(S)$ denote the standard deviation and mean of the cluster sizes, respectively, in the input set $S$. 
This metric measures the degree of variability in cluster sizes relative to the mean.  

Additionally, our follow-up experimental results demonstrate the importance of {\em ordering similar records sequentially inside a record set}, this enhances an LLM’s in-context clustering performance by facilitating better context differentiation \cite{brown2020language}. Similar findings were also demonstrated by prior RAG work~\cite{rag24}, which shows that ordering semantically similar sentences together improves LLM output.

Figure~\ref{fig:cluster ER} presents an example of an LLM prompt for our novel in-context clustering-based entity resolution. Instead of pairwise ER, the prompt explicitly asks to cluster the records in an input set. 

\begin{myEmp}
As shown in Figure~\ref{fig:cluster ER}, we use an LLM to group a record set containing 9 records, thus the \textbf {set size}  is 9. The \textbf {set diversity} of this record set is 3, since it contains three distinct entities (clusters), which are represented by three different colors. The mean of cluster sizes in this record set is $(3+3+3)/3=3$, while the standard deviation of cluster sizes in the record set is \(\sqrt{\frac{(3-3)^2 + (3-3)^2 + (3-3)^2}{3}} = 0\), indicating a \textbf {set variation} of 0, i.e., no variation in relation to three cluster sizes.
\end{myEmp}

\subsection{Empirical Evaluations for the Key Factors}
\label{sec:experiment on key factors}

We examine the impact of varying \textbf{set size ($S_s$)}, \textbf{set diversity ($S_d$)}, and \textbf{set variation ($S_v$)} on the LLM's performance for in-context clustering of individual record sets. 
We further explore how the order of records in a set affects the LLM's performance by implementing: {\textbf{sequential record order}}, where records belonging to the same entity are placed consecutively in the record set; and {\textbf{random record order}}, where the records are randomly placed. 
These analyses help us identify the {\em optimal information density and structure} that LLMs can handle within a single API call, hence {\em guiding our design} for creating effective record sets and fully utilizing the knowledge provided by blocking techniques (\S~\ref{sec:methodology}).

We define three levels of set variations (\( S_v \)) to represent different degrees of imbalance within the record sets: {\bf balanced} (\( S_v < 0.3 \)), {\bf relatively balanced} (\( 0.3 \leq S_v \leq 0.7 \)), and {\bf unbalanced} (\( S_v > 0.7 \)). 
For each fixed $S_s$, $S_d$, $S_v$, and record ordering, we uniformly at random select 200 questions (i.e., record sets) to query the LLM, provided that the total number of questions available under the given parameter settings exceeds 200.
The ground truth for the questions corresponds to the correct clustering of the records within each record set. 
More detailed experimental settings, evaluation metrics, and dataset descriptions can be found in \S~\ref{sec:exp}.

\begin{figure}[tb!]
\centering
\vspace{-1.5mm}
\includegraphics[width=0.35\textwidth]{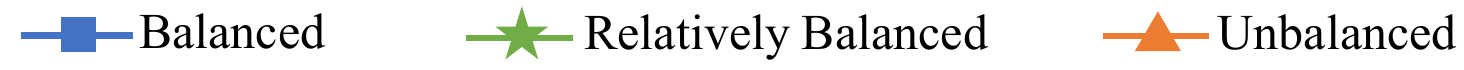}\\
  \vspace{-1.5mm}
  \hspace{-7.5mm}
  \subfigcapskip=-2mm
  \subfigure[Music 20K]{
   \includegraphics[width=1.17in]{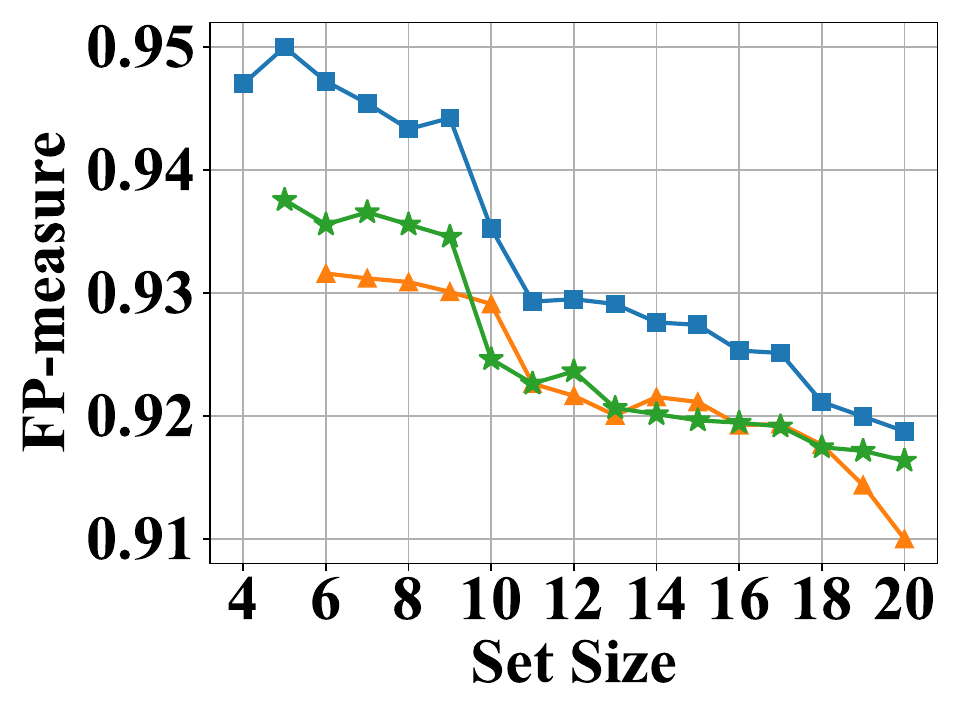}
   \label{fig:fp-music-diver4}
  }
 \hspace{ -2.7mm}
  \subfigure[Cora]{
   \includegraphics[width=1.17in]{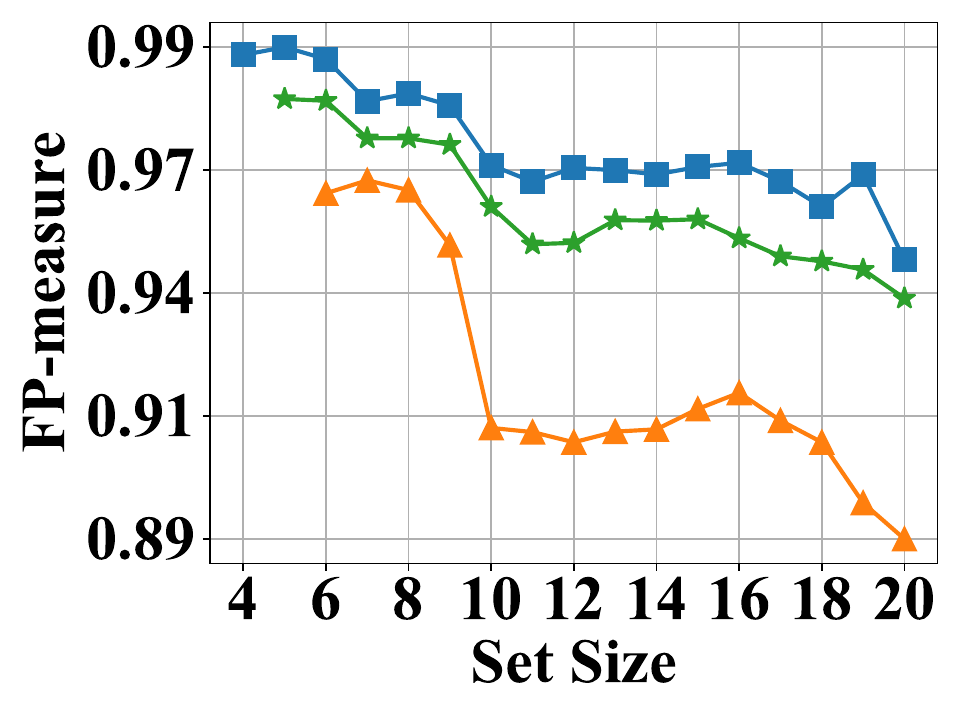} 
   \label{fig:fp-cora-diver4}
  }
  \hspace{-2.7mm}
  \subfigure[Alaska]{
   \centering
   \includegraphics[width=1.17in]{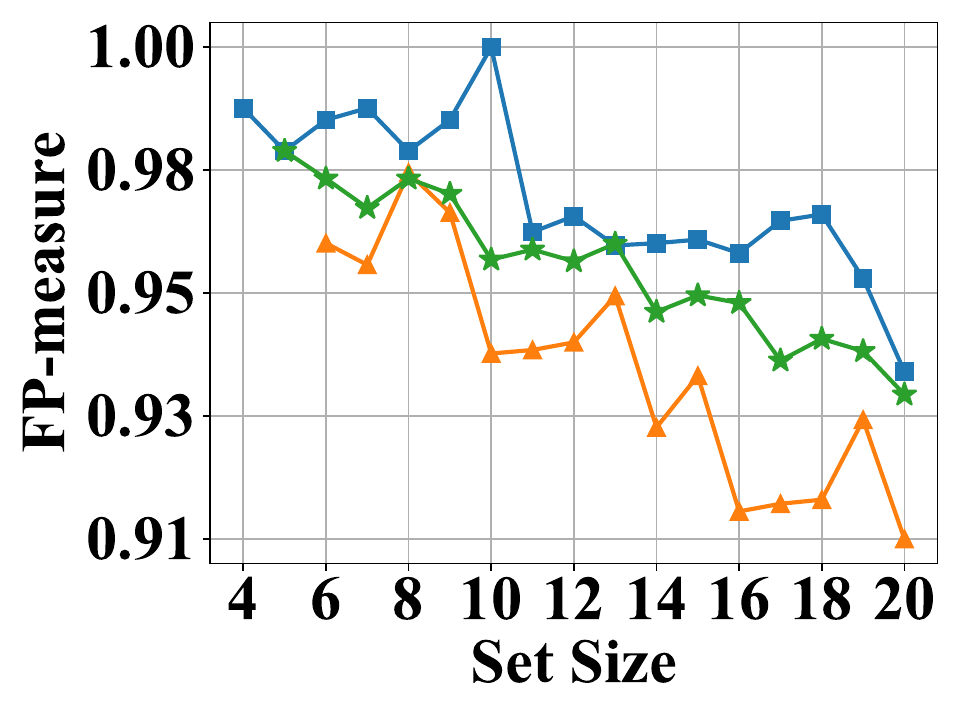}
   \label{fig:fp-sigmod-diver4}
  }
\vspace{-5mm}
\caption{ \centering {Clustering performance vs. set size and variation} } 
\vspace{-4mm}
\label{fig:cluster-Ss-Sv-diver4}
\end{figure}

Figure~\ref{fig:cluster-Ss-Sv-diver4} illustrates the in-context clustering performance of the LLM on individual record sets under varying $S_s$ and $S_v$. 
The results indicate that LLM performance {\em consistently improves as set variation decreases}, suggesting that clustering is more accurate when records are evenly distributed among entities within a set. This trend holds across all datasets and evaluation metrics, regardless of the set size. 
Moreover, the LLM {\em maintains relatively stable and robust performance} across balanced, relatively balanced, and unbalanced settings as the set size increases from 4 to 9. 
However, a marked decline in clustering accuracy is observed once the set size {\em exceeds 9 or 10}. This degradation can be attributed to the increased cognitive load placed on the model: larger sets require the LLM to process and compare more records simultaneously, which dilutes semantic focus and heightens the risk of clustering errors. 
Additionally, managing coherent contextual relationships across more records becomes increasingly challenging, further reducing accuracy.
Across different set variation settings, the trends in FP-measure {\em remain largely consistent}, as this metric reflects the alignment between predicted clusters and ground truth labels. Taken together, these results suggest that {\em the optimal configuration for in-context clustering occurs at a set size of 9}, and that {\em minimizing variation within each set} is key to maximizing LLM performance.

\begin{figure}[tb!]
\centering
\vspace{-1.5mm}
\includegraphics[width=0.20\textwidth]{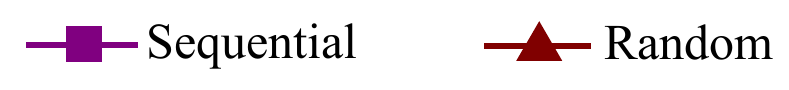}\\
  \vspace{-2mm}
  \hspace{-7mm}
  \subfigcapskip=-2mm
  \subfigure[Music 20K]{
   \includegraphics[width=1.16in]{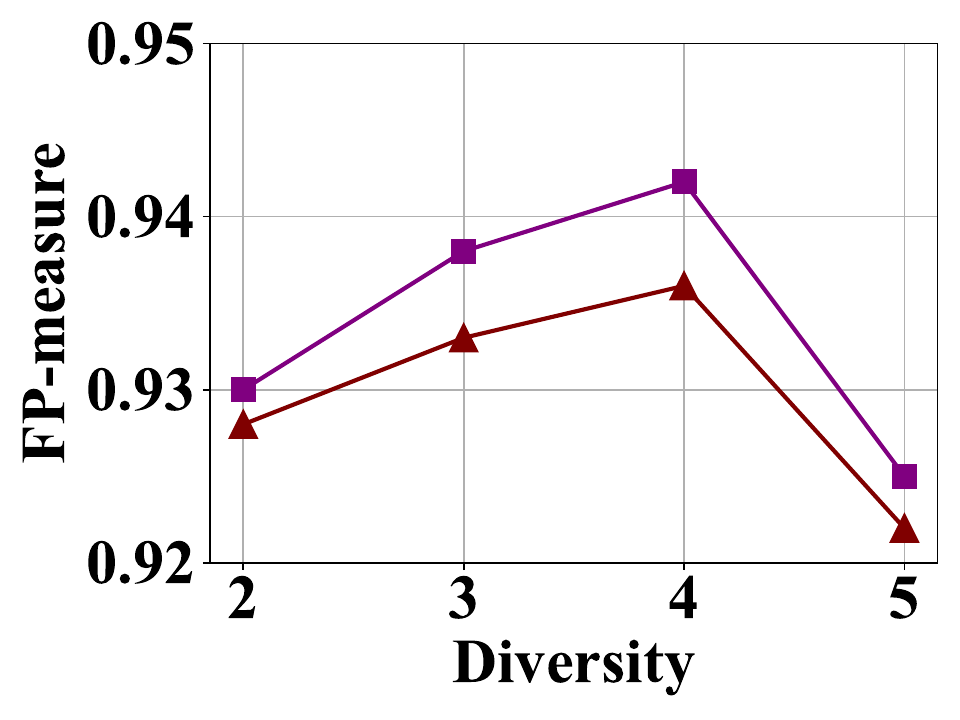}
   \label{fig:fp-music-order}
  }
 \hspace{ -3mm}
  \subfigure[Cora]{
   \includegraphics[width=1.16in]{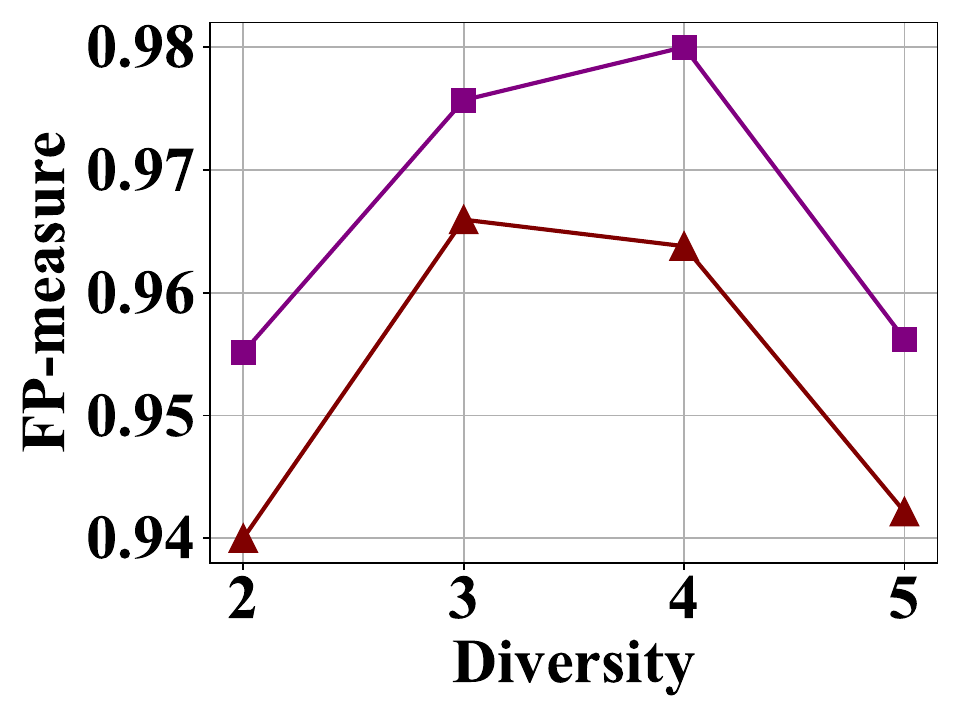} 
   \label{fig:fp-cora-order}
  }
  \hspace{-3mm}
  \subfigure[Alaska]{
   \centering
   \includegraphics[width=1.16in]{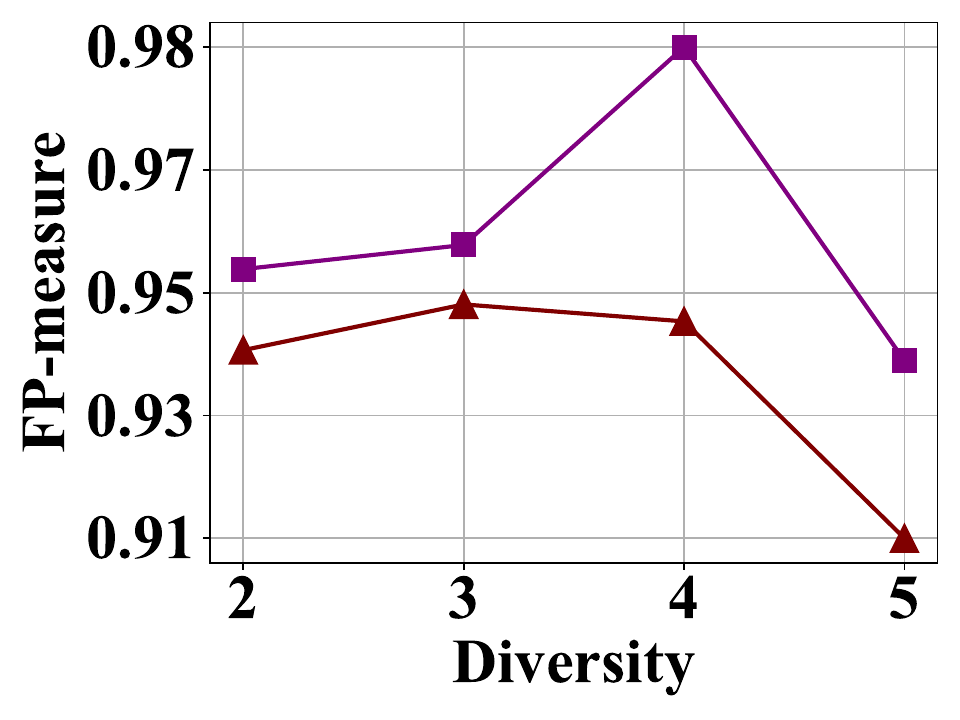}
   \label{fig:fp-sigmod-order}
  }
\vspace{-6.2mm}
\caption{ { \centering {Clustering performance vs. set diversity, ordering} } }
\label{fig:cluster-diversity-order}
\vspace{-5.8mm}
\end{figure}

Fixing the optimal values of $S_s$ and $S_v$ based on previous experiments, we further investigate the impact of $S_d$ and the ordering of records within record sets on the performance of LLM-based in-context clustering (Figure \ref{fig:cluster-diversity-order}). 
We notice that when $S_d$ is set to 4, the in-context clustering performance of the LLM is better across all datasets in terms of FP-measure compared to other diversity settings. Thus, we consider {\em $S_d = 4$ as our optimal diversity setting}.
With 4 entities, the clustering task provides enough granularity to differentiate entities while avoiding excessive sparsity or overlap in record distribution. 
In contrast, fewer entities (e.g., 2 or 3) might result in overly homogeneous clusters that fail to capture subtle differences, while a larger number of entities (e.g., 5) might introduce higher complexity and noise, negatively impacting the LLM’s ability to identify distinct clusters accurately. 
Meanwhile, it is observed that the LLM achieves better in-context clustering performance when the {\em sequential record order} is used. 
This improvement can be attributed to the sequential arrangement of similar records, which likely enhances the coherence of contextual information presented to the LLM. By maintaining entity-related records in close proximity, this ordering reduces potential context-switching overhead and provides a more structured input sequence~\cite{rag24}, thereby enabling the model to better discern subtle patterns and relationships essential for accurate clustering.

In conclusion, for the LLM's in-context clustering performance, 
our optimal configuration includes a set size of 9, a balanced distribution ($S_v$ close to 0), a set diversity of 4, and the sequential record order. Larger set sizes should be avoided, as the LLM’s performance under extremely unbalanced ratios deteriorates rapidly beyond this point. Conversely, smaller set sizes should also be avoided, as maximizing the set size is crucial for reducing API call frequency, improving efficiency and minimizing token usage and cost. 
More analyses about the choice of key factors can be found in \S~\ref{sec:dataset_key_factor}.

\begin{figure*}[tb!]
\centering
\vspace{-5mm}
\includegraphics[width=0.8\textwidth]{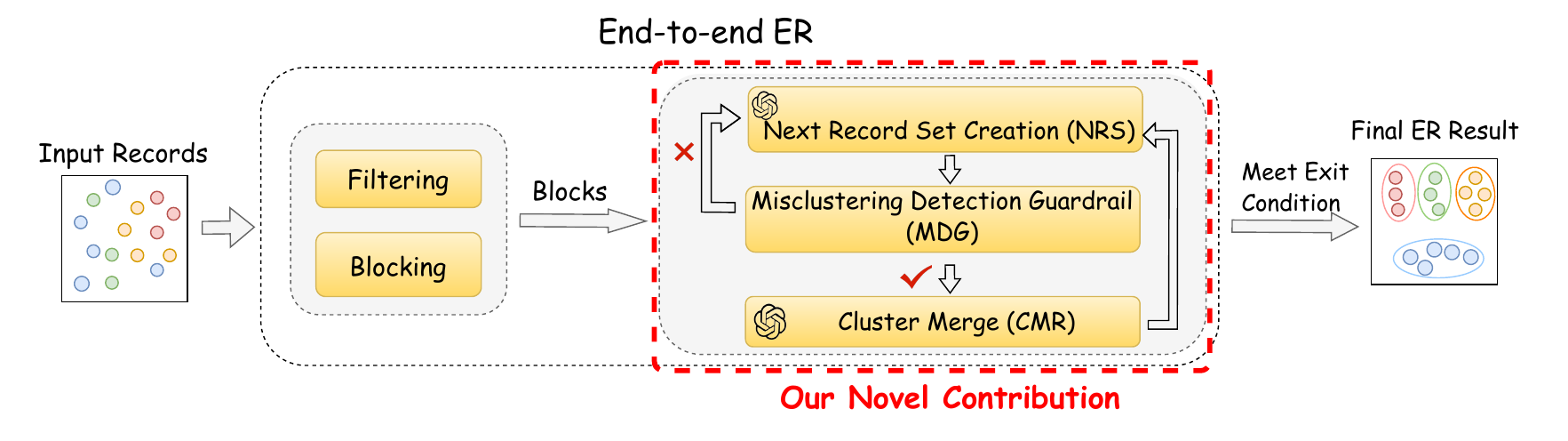}
\vspace{-6mm}
\caption{overview of our end-to-end entity resolution solution {\sf LLM-CER}}
\label{fig:overview}
\vspace{-5mm}
\end{figure*}

\subsection{Problem Statements for End-to-end ER}
\label{sec:problem for er}
\vspace{-1mm}

We are now ready to define our main problems.
\begin{myProblem}
    \vspace{-1mm}
    \label{def-NRSS}
    {\rm \textbf{({\scshape Next Record Set Selection}).}} This problem is relevant for an LLM's in-context clustering during the entity matching 
    
    \noindent phase. Given a set of records $\mathcal{R}$, the current clustering result $\mathbb{C}$, and predefined constraints on the set size ($S_s)$, diversity ($S_d$), and variation ($S_v$), the {\scshape{Next Record Set Selection}} problem identifies a subset of records $\mathcal{R}^* \subseteq \mathcal{R}$ satisfying the constraints $\langle S_s, S_d, S_v \rangle$ with the objective of minimizing the number of LLM API calls
    by leveraging the identified transitivity and anti-transitivity relationships.
\end{myProblem}

\begin{myProblem}
\vspace{-1mm}
\label{def-ClusterCombination}
{\rm \textbf{({\scshape Cluster Combination}).}}
Given the current clustering result $\mathbb{C}$ and the LLM's in-context clustering output $\mathbb{C}^{*}$ 
derived from a record set $\mathcal{R}^*$, the {\scshape{Cluster Combination}} problem updates $\mathbb{C}$ by combining the clusters in $\mathbb{C}$ based on the merge decisions of the LLM over $\mathbb{C}^{*}$, leveraging the identified transitivity and anti-transitivity relationships.
This problem is relevant in combining phase to gradually refine the LLM's clustering outputs and obtain the final ER result.
\vspace{-2mm}
\end{myProblem}
The number of record sets generated by our algorithm determines the number of LLM API calls. Since each LLM API call roughly consists of the same number of tokens (e.g., see Figure \ref{fig:cluster ER}), and an LLM's monetary cost and inference time are determined by the total number of input and output tokens, minimizing the number of LLM API calls also reduces the number of LLM tokens, monetary cost, as well as the end-to-end ER time. We discuss our solutions in the following section. In particular, the next record set selection is considered in \S\ref{sec:in_context_cluster}, while cluster combination is detailed in \S\ref{sec:cluster_merge}. Notice that the existing pairwise ER scheme is a {\em special case} of our in-context clustering-based ER paradigm when the record set size $S_s=2$. Therefore, our problems and solutions generalize the state-of-the-art ER approaches. 

\vspace{-1mm}
\section{Solution: End-to-End ER}
\label{sec:methodology}
\vspace{-1mm}

We present an overview of our solution, {\sf LLM-powered Clustering-based ER} ({\sf LLM-CER}), before delving into details. Our key technical contributions are: 
(1) an effective \textsf{Next Record Set Creation (NRS)} algorithm for constructing optimal record sets for in-context clustering from initial blocks generated by standard blocking techniques.;  
(2) a guardrail technique, \textsf{Misclustering Detection Guardrail (MDG)} for assessing and subsequently improving the accuracy of the LLM's in-context clustering outputs; 
and (3) an efficient heuristic algorithm, \textsf{Cluster Merge (CMR)} for merging the in-context clustering outputs. 
Putting them together, we systematically explore the design space of clustering-based entity resolution powered by LLMs.

As illustrated in Figure~\ref{fig:overview}, {\sf LLM-CER} begins with blocking to generate initial candidate blocks (Section~\ref{sec:Blocking}), which are then processed by the \textsf{NRS} algorithm (Algorithm~\ref{alg:block classification}) to construct record sets for in-context clustering by an LLM (Section~\ref{sec:in_context_cluster}).
The clusterings are assessed and refined by the \textsf{MDG} algorithm (Algorithm~\ref{alg:Guardrail}), which identifies and corrects likely misclusterings.
Improved clusters are then hierarchically merged using the \textsf{CMR} algorithm (Algorithm~\ref{alg:in-context cluster merge}), which repeatedly invokes LLM-based clustering over newly formed record sets while leveraging transitivity and anti-transitivity (Section~\ref{sec:cluster_merge}).
This process continues until a termination condition is met, producing the final entity resolution result (Section~\ref{sec:end2end}).

\vspace{-2mm}
\subsection{Filtering and Blocking}
\label{sec:Blocking}
\vspace{-1mm}

We begin by introducing the filtering and blocking strategies used as a pre-processing step in the ER pipeline. Given a collection of records $\mathcal{R}$, a naïve approach compares all pairs, incurring a computational complexity of O($|\mathcal{R}|^2$). To mitigate this, filtering and blocking techniques are employed~\cite{papadakis2016comparative,hassanzadeh2009framework,meduri2020comprehensive}. 
Blocking groups similar records into blocks, while filtering typically applies similarity joins to identify candidate matches for each record.  
Although not the focus of our technical contributions, this work is the first to explore their trade-offs in the context of (1) LLM-based ER, which leverages large language models for ER~\cite{li2024booster,fan2024cost}, and (2) clustering-based ER, where resolution is achieved by iteratively clustering subsets of records rather than pairwise comparisons.
We consider the following filtering and blocking strategies. 

\vspace{-1mm}
\spara{Filtering-based Block Creation.} 
Similarity joins identify records with similarity scores above a threshold using metrics like Jaccard similarity ~\cite{broder1997resemblance,BohmK01}. 
While effective for detecting near-duplicates, pairwise comparisons incur O$(|\mathcal{R}|^2)$ complexity, making filtering computationally expensive. 
To address this, positional filtering~\cite{xiao2011efficient} prunes candidate records by leveraging token ordering, reducing complexity to O$(|\mathcal{R}|\cdot k)$ when $k$ candidates are allowed per record.
The similarity threshold $b_t$ is set empirically by maximizing the F1-score of clustering on a validation dataset~\cite{PAPADAKIS2020101565}, iterating thresholds from 0.05 to 0.95 in 0.05 increments. 
When ground truth is unavailable, we exploit LLM-based clustering results over a certain number of record subsets as ground truth.

\vspace{-1mm}
\spara{Locality Sensitive Hashing-based Blocking.} Distributed representations effectively capture semantic similarities (e.g., ``p53'' and ``cancer'') in ER tasks ~\cite{gionis1999similarity}.
While cosine similarity can measure embedding proximity, its O$(|\mathcal{R}|^2)$ complexity becomes impractical for large datasets. 
Locality-sensitive hashing (LSH)~\cite{mests2018distributed} hashes similar embeddings into identical buckets with high probability, reducing complexity to O$(|\mathcal{R}| \cdot k)$, where $k$ is the number of buckets. 
However, the stochastic hash functions may co-locate dissimilar records, introducing false positives. To mitigate this issue, we retain only pairs with similarity exceeding a threshold $b_t$ to purify the initial blocks, where $b_t$ is determined via the aforementioned approach.

\vspace{-1mm}
\spara{Canopy Blocking~\cite{mccallum2000efficient}.} Two thresholds $b_s$ and $m_s$, with $b_s \geq m_s$, are utilized to balance efficiency and accuracy. 
A computationally inexpensive metric (e.g., inverted index-based edit distance on a single attribute) assigns record pairs with similarity $>b_s$ to the same block. Pairs with similarity $>m_s$ are grouped into overlapping canopies. Within these canopies, a refined metric (e.g., Jaccard similarity across all attributes) computes pairwise similarities. 
Matching pairs trigger transitive block merging: If records from two blocks match, the blocks are combined iteratively until convergence.
While initial canopy formation retains quadratic complexity, its reliance on lightweight metrics makes it orders of magnitude faster than refined similarity computations~\cite{mccallum2000efficient}. 
We tune $b_s$ and $m_s$ empirically on validation datasets, or when ground truth is unavailable, derive labels via LLM-based clustering (as described earlier).

\vspace{-0.8mm}
\spara{Remark.} We employ {\sf LSH} as the default blocking method based on empirical results (\S\ref{sec:exp_block_filter}). We adopt the above filtering and blocking techniques for their simplicity and compatibility with our framework. As our primary technical contributions lie in the in-context clustering for matching (\S\ref{sec:in_context_cluster}) and the merging strategy for combining clusters (\S\ref{sec:cluster_merge}), we leave the integration of advanced blocking methods—such as progressive blocking~\cite{galhotra2021efficient}—to future studies.

\vspace{-2.5mm}
\subsection{In-context Clustering of a Record Set}
\label{sec:in_context_cluster}
\vspace{-1.5mm}

We develop novel in-context clustering for the matching phase. 

\vspace{-1.2mm}
\spara{Next Record Set Creation Algorithm.} 
Consider the records $\mathcal{B}_{remain}$ in a block that are not used in forming a record set yet. We design an algorithm \textsf{NRS} (Algorithm \ref{alg:block classification})
to create the \underline{n}ext \underline{r}ecord \underline{s}et with $\mathcal{B}_{remain}$. 
If $|\mathcal{B}_{remain}|$ is smaller than a predefined set size ($S_s$), the next record set is created by grouping similar records sequentially, which improves an LLM's in-context clustering performance by facilitating better context differentiation, as shown in our experiment (\S\ref{sec:experiment on key factors}) (Lines 2-6). Otherwise, the next record set generation is optimized based on the empirical findings from \S\ref{sec:experiment on key factors}. Our results show that an LLM’s in-context clustering accuracy depends on the record set design. In particular, we select an optimal record set size $S_s$, diversity $S_d$, and variation $S_v$ based on experimental results. 
We then apply the elbow method and $k$-means~\cite{cui2020introduction} to perform a preliminary clustering of records in $\mathcal{B}_{remain}$, while also assessing its diversity $k$. Next, we construct the next record set while ensuring the set size constraint $S_s$. For the record set, we minimize $S_v$ as much as possible and aim to meet the $S_d$ requirement (Lines 8-17).

Experimental results indicate that the optimal values are $S_s = 9$ and $S_d = 4$, while LLM-based in-context clustering achieves better performance when $S_v$ is minimized, ideally approaching zero.

\vspace{-0.8mm}
\spara{Mitigating In-context Clustering Errors of LLMs.} In practice, LLMs are prone to hallucination~\cite{gu2023hallucinations}, leading to seemingly plausible, yet factually incorrect contents. As a result, LLM-based in-context clustering outcomes for record sets may not be entirely reliable. To ensure high accuracy of the output, we further devise a \underline{m}isclustering \underline{d}etection \underline{g}uardrail ({\sf MDG}) algorithm to verify the clustering results of the LLM, as detailed in Algorithm \ref{alg:Guardrail}. We first define some key terms.

\begin{myDef}
    \vspace{-1.5mm}
    \label{def-cluster-simi}
    {\rm \textbf{(Inter- and Intra-cluster Similarity)}}
    Given a similarity function \( F(\cdot) \) to measure the similarity between record pairs and a known record \( r \), the \textit{intra-cluster similarity} of \( r \) is defined as the minimum similarity between \( r \) and other records within the same cluster as $r$. The \textit{inter-cluster similarity} of \( r \) is defined as the maximum similarity between \( r \) and records from other clusters.
\end{myDef}

Specifically, for any cluster from the in-context clustering output, if the intra-cluster similarity of a record inside it is lower than its inter-cluster similarity, Algorithm \ref{alg:Guardrail} indicates the record to be {\em misclustered}; and hence, the result is not acceptable. When computing similarity between records or clusters, our choice of similarity function is closely linked to the block creation method. For example, when using Filtering-based Block Creation, we employ Jaccard similarity to measure record similarity. In contrast, with LSH-based Block Creation, we assess similarity through cosine similarity applied to record embeddings. By default, we use cosine similarity for record embeddings unless otherwise specified.
The overall time complexity of the {\sf MDG} algorithm is O$(S_s^2)$, where $S_s$ is typically 9. Given that $S_s$ is small and the computation is highly parallelizable, the actual runtime overhead remains low in practice.

\setlength{\textfloatsep}{0pt}
\begin{algorithm}[tb!]
\small
\caption{Next record set creation ({\sf NRS}) 
}
\label{alg:block classification}
    \begin{flushleft}
    \hspace*{0.02in} {\bf Input:}
	remaining records in a block $\mathcal{B}_{remain}$, $S_s$, $S_d$, $S_v$\\
	\hspace*{0.02in} {\bf Output:}
	  record set $\mathcal{R}_{set}$\\
    \end{flushleft}
	\begin{algorithmic}[1]
        \STATE $\mathcal{R}_{set}\leftarrow \emptyset$
        \IF{$|\mathcal{B}_{remain}|\leq S_s$}
            \STATE $r_{pre}\leftarrow$ first element in $\mathcal{B}_{remain}$
            \WHILE{$|\mathcal{B}_{remain}|>0$}
                \STATE add $r_{pre}$ to $R_{set}$ and remove it from $\mathcal{B}_{remain}$
                \STATE $r_{pre}\leftarrow$ update $r_{pre}$ w/ its most similar record from $\mathcal{B}_{remain}$
            \ENDWHILE
        \ELSE
            \STATE $k \leftarrow$ compute diversity of $\mathcal{B}_{remain}$ using the elbow method
            \STATE $\mathbb{B}_{remain,\,k} \leftarrow$  perform $k$-means on $\mathcal{B}_{remain}$
            \STATE $target_{size} \leftarrow \lfloor S_s / S_d \rfloor$
            \FOR{$\mathcal{B}_{remain,\,i}$ \textbf{in} $\mathbb{B}_{remain,\,k}$}
                \IF{$|\mathcal{R}_{set}|<S_s$ \textbf{and} $|\mathcal{B}_{remain,\,i}|\geq target_{size}$}
                \STATE select $target_{size}$ records from $\mathcal{B}_{remain,\,i}$ and add to $\mathcal{R}_{set}$
                \STATE delete those records from $\mathcal{B}_{remain}$
                \ENDIF
            \ENDFOR
            \WHILE{$|\mathcal{R}_{set}|<S_s$}
                \STATE $r_{set} \leftarrow$ find record in $\mathcal{B}_{remain}$ to least increase in $S_v$ \textcolor{deepgreen}{(via Eq. \ref{set deviation})}
                \STATE add $r_{set}$ to $\mathcal{R}_{set}$ and delete from $\mathcal{B}_{remain}$
            \ENDWHILE
            \STATE order similar records together in $\mathcal{R}_{set}$ (similar to Lines 3-6)
        \ENDIF
        \RETURN $\mathcal{R}_{set}$
        \end{algorithmic}
\end{algorithm}
\setlength{\textfloatsep}{12pt plus 2pt minus 2pt}

\setlength{\textfloatsep}{0pt}
\begin{algorithm}[tb!]
\small
\caption{Misclustering Detection Guardrail ({\sf MDG})}
\label{alg:Guardrail}
    \begin{flushleft}
    \hspace*{0.02in} {\bf Input:}
	In-context clustering result $\mathbb{C}_{set}$ of a record set\\
	\hspace*{0.02in} {\bf Output:}
	  whether in-context clustering result is acceptable $(True/False)$\\
    \end{flushleft}
	\begin{algorithmic}[1]
        \FOR{cluster $\mathcal{C}_i$ \textbf{in} $\mathbb{C}_{set}$}
            \FOR{record $r_j$ \textbf{in} $\mathcal{C}_i$}
                \IF{intra-cluster‌ sim. of $r_j$ $<$ inter-cluster‌ sim. of $r_j$}
                    \RETURN $False$
                \ENDIF 
            \ENDFOR
        \ENDFOR
        \RETURN $True$
        \end{algorithmic}
\end{algorithm}

\vspace{-1mm}
\spara{Record Set Regeneration.} 
For each misclustered record $r$, our end-to-end algorithm {\sf LLM-CER} identifies the cluster with the highest inter-cluster similarity to $r$, and relocates it immediately after that cluster within the record set. This targeted adjustment leaves all other records unchanged and aims to bring semantically similar records closer together inside the record set, resulting in an overall time complexity of O$(S_s)$. As shown in \S\ref{sec:experiment on key factors}, LLM's clustering accuracy improves when records from the same entity appear consecutively. Next, we conduct in-context clustering with this more sequentially-ordered record set, enhancing the LLM's performance significantly, which is evident in our empirical results (\S~\ref{sec:exp_block_filter}).


\vspace{-1mm}
\subsection{Hierarchical Cluster Merging}
\label{sec:cluster_merge}
\vspace{-1mm}

Leveraging the optimal set size $S_s$ = 9 as stated earlier, each block is divided into several record sets (e.g., a block of size 27 is split into 3 record sets). Each record set is then in-context clustered independently by the LLM. 
For example, if record set $A$ is clustered into subsets $\{A_1, A_2, A_3\}$, record set $B$ into $\{B_1, B_2\}$, and record set $C$ into $\{C_1, C_2, C_3, C_4, C_5\}$, it becomes necessary to merge subsets across record sets to obtain the final partition.
%
\begin{figure}[tb!]
\centering
\vspace{-8mm}
\includegraphics[width=0.5\textwidth]{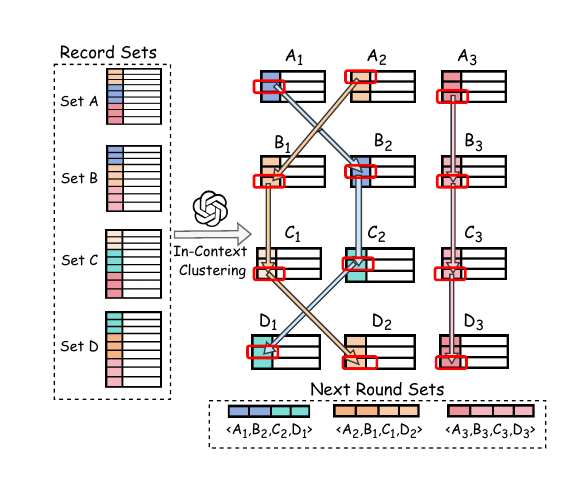}
\vspace{-13mm}
\caption{An example of the next round of record sets creation. For simplicity, consider that each of the record sets $A, B, C, D$ generates three in-context clusters: $A=\{A_1, A_2, A_3\}$, $B=\{B_1, B_2, B_3\}$, etc. Assume that $S_s\geq 4$ and $S_d=1$ for the sake of brevity. Then, the next round of record sets should pack the most similar clusters from $A, B, C,$ and $D$, forming three new sets for the next round as follows: $\{A_1, B_2, C_2, D_1\}, \{A_2, B_1, C_1, D_2\}$, and $\{A_3, B_3, C_3, D_3\}$. Each cluster appears exactly once in the next round of record sets to minimize the overall LLM API calls. Clusters
from the same record set (e.g., both $A_1$ and $A_2$) are not packed together in the same record set of the next round owing to anti-transitivity. 
} 
\vspace{1mm}
\label{fig:next round record generate}
\end{figure}

To achieve this, we note that clusters within the same record set (e.g., $A_1$ and $A_2$) are already disjoint (i.e., belong to different entity groups) due to the LLM’s in-context clustering and hence exhibit non-transitive relationships. However, clusters from different record sets (e.g., $A_1$ and $B_1$) may belong to the same entity group. Merging clusters across different record sets can be formulated as a $K$-dimensional maximum matching problem (analogous to a 3-dimensional maximum matching problem, which is proven to be \textbf{NP}-Hard~\cite{crescenzi2000maximum}), where $K$ is the number of record sets. 
\begin{myProblem}[Cluster Merging]
\vspace{-1mm}
Given $K$ record sets and their in-context clustering outputs, the \textsf{Cluster Merging} problem treats each output cluster as a ``new'' individual record for the next round and packs them optimally to construct record sets for the LLM's next round of in-context clustering.  
The problem is subject to the following conditions: 
(1) Each cluster (i.e., the corresponding ``new'' individual record) is selected exactly once in the next round of record sets to minimize the total number of LLM API calls. 
(2) Clusters from the same record set are never packed together in the same record set of the next round owing to anti-transitivity. (3) The next round of record sets should also ensure the set size constraint $S_s$, set diversity $S_d$, and minimize $S_v$ as much as possible.
\end{myProblem}

The complexity of evaluating all combinations in the cluster merging problem is O$(K \cdot n!)$, where $n$ is the average number of clusters per record set. This results in an exponential time cost, making it impractical for real-world applications with large blocks and multiple record sets. 

To address this issue, we propose a heuristic \textsf{Cluster Merge (CMR)} algorithm (Algorithm \ref{alg:in-context cluster merge}) that avoids exhaustive enumeration. Specifically, for each cluster of a given record set, we identify its most similar cluster from the next record set based on a defined similarity measure. For instance, consider two record sets $A$ and $B$ with their in-context clustering results $A=\{A_1, A_2, A_3\}$, $B=\{B_1, B_2\}$. If $A_1$ is most similar to $B_2$, these two clusters are packed together. This process is repeated iteratively across all record sets (e.g., finding a cluster of the next record set $C$ that is most similar to the group $A_1B_2$), gradually forming a new record set for the next round of in-context clustering, while satisfying $S_s$, $S_d$, and $S_v$ requirements. Figure \ref{fig:next round record generate} shows an example of our cluster merging procedure that constructs record sets for the next round of clustering. 

For simplicity, assume $K \leq S_s$, where $K$ denotes the number of record sets in the current round, and $S_s$ is the optimal record set size. For ease of understanding, we also assume that each of these record sets generates $n$ in-context clusters by the LLM. Algorithm \ref{alg:in-context cluster merge} outlines the process of constructing record sets for the next round of in-context clustering. Given $K$ record sets, the algorithm first generates clusters for each record set using the LLM, then replaces each cluster with a representative ``new'' record. 
This replacement is possible because the records within a cluster represent the same entity and satisfy transitivity. The representative record can either be uniformly chosen at random or selected as the one with the smallest distance from the average embedding of the records in that cluster (Lines 1-3). Then, it partitions the $K$ record sets into groups of size $\lceil K/S_d \rceil$, iterating over these groups to construct a record set $\mathcal{R}_{next}$ for the next round (Lines 5-6). For each group, it selects an unselected cluster $\mathcal{C}$ from the first record set of the group and uses its representative record as the initial record (Lines 7-8). Subsequently, for each remaining record set in the group, it identifies the most similar unselected cluster $\mathcal{C}'$ to the previously selected cluster $\mathcal{C}$ and adds its representative record to $\mathcal{R}{next}$ (Lines 9-11). Finally, the resulting record set is returned for the next round of in-context clustering (Line 12). This process also ensures that (1) similar elements of $\mathcal{R}{next}$ are sequentially ordered; and (2) we meet the $S_d$ requirement, while minimizing $S_v$ as much as possible.

Considering an average of $n$ clusters from a record set, the complexity of comparing clusters between two record sets is O($n^2$). Since there are $K$ record sets in the current round, the total complexity of {\sf CMR} algorithm is O($K \cdot n^2$) in the current round. This represents a significant reduction from the original exponential complexity O($K \cdot n!$).
 The maximum value of $n$ can be up to the record set size, which is typically 9 in our experiments. This is significantly smaller than the total number of records $|\mathcal{R}|$, leading to a reduced complexity in real-world usage. 
Moreover, the total hierarchy layers remain limited as observed in our experiments and summarized in Table~\ref{tab:hierarchy} (\S \ref{sec:exp_end2end}). 

\setlength{\textfloatsep}{0pt}
\begin{algorithm}[tb!]
\small
\linespread{0.7}\selectfont
	\caption{{\sf Cluster Merge (CMR)}: Depicting the creation of one record set for the next round}
	\label{alg:in-context cluster merge}
	\begin{flushleft}
    \hspace*{0.02in} {\bf Input:}
	$K$ ($\leq S_s$) record sets 
    $\{\mathcal{R}_{1},\dots,\mathcal{R}_{K}\}$ each with $n$  clusters; $S_s$, $S_d$, $S_v$\\
	\hspace*{0.02in} {\bf Output:}
	one record set $\mathcal{R}_{next}$ for the next round with size $K$ \\
    \end{flushleft}
	\begin{algorithmic}[1] 
            \FOR{$i\leftarrow 1$ \textbf{to} $K$}
                \STATE \(\mathbb{C}_{i} \leftarrow n\) output clusters of record set \(\mathcal{R}_{i}\)

                \STATE replace each cluster  $\mathcal{C} \in \mathbb{C}_{i}$ by a representative record $r_\mathcal{C} \in \mathcal{C}$          
            \ENDFOR
           \STATE $\mathcal{R}_{next} \leftarrow \phi$
            \FOR{$j\leftarrow 1$ \textbf{to} $S_d$}
            \STATE $w\leftarrow (j-1)\cdot \lceil K/S_d \rceil + 1 $ 
            \STATE select a previously unselected cluster $\mathcal{C} \in \mathbb{C}_w$ 
            \STATE insert $\mathcal{C}$ (i.e., its representative record $r_{\mathcal{C}}$) into $\mathcal{R}_{next}$               
            \FOR{$i\leftarrow (j-1)\cdot \lceil K/S_d \rceil + 2$ \textbf{to} $\min\{j\cdot \lceil K/S_d \rceil, \, K\}$}     
            \STATE find cluster $\mathcal{C}' \in \mathbb{C}_i$, previously unselected \& most similar to $\mathcal{C}$ 
            \STATE Insert $\mathcal{C}'$ (i.e., its representative record $r_{\mathcal{C}'}$) into $\mathcal{R}_{next}$
            \ENDFOR
            \ENDFOR
            \RETURN $\mathcal{R}_{next}$
        \end{algorithmic}
\end{algorithm}
\setlength{\textfloatsep}{12pt plus 2pt minus 2pt}

\vspace{-1mm}
\subsection{End-to-End ER Solution}
\label{sec:end2end}
\vspace{-1mm}

Algorithm \ref{alg:end2end} presents our end-to-end entity resolution framework outlining the key steps, given the input set of records  $\mathcal{R}$. 
The algorithm begins by applying blocking functions to $\mathcal{R}$, generating initial set of blocks $\mathbb{B}$ (\S\ref{sec:Blocking}), after which we build a disconnected graph based on it to keep track of similar records. For each block in $\mathbb{B}$, record sets are generated using the \textsf{NRS} algorithm (Algorithm~\ref{alg:block classification}), they are in-context clustered leveraging an LLM, with acceptable outputs verified by the \textsf{MDG} algorithm (Algorithm~\ref{alg:Guardrail}) given in \S\ref{sec:in_context_cluster}. Subsequently, the clustering outputs of these record sets are merged hierarchically using the \textsf{CMR} algorithm (Algorithm~\ref{alg:in-context cluster merge}) and next rounds of in-context clustering (\S\ref{sec:cluster_merge}).

\spara{Exit Condition.}
We keep generating hierarchical record sets for
in-context clustering and subsequent cluster merging.
The process continues until no more clusters can be merged owing to anti-transitivity. Here, anti-transitivity is identified by in-context clustering of the record sets. In particular, consider a current round when the in-context clustering outputs only singleton clusters, i.e., each cluster having only one element from the current round. The exit condition is thus satisfied. A naive method for the ``final check'' is to conduct pairwise comparisons of all singleton clusters to ensure that, indeed, no more merging is feasible, which costs quadratic time in the number of singleton clusters. 
During this final check, we, however, adopt a more efficient method by packaging multiple singleton clusters in a record set to reduce the number of LLM API calls, which is discussed in the following theoretical analysis.

Finally, we use each singleton cluster to reconstruct the full partition $\mathbb{C} = \{\mathcal{C}_1, \dots, \mathcal{C}_n\}$ and complete the ER process. 

\setlength{\textfloatsep}{0pt}
\begin{algorithm}[tb!]
\small
\linespread{0.7}\selectfont
	\caption{{\sf End-to-end ER} }
	\label{alg:end2end}
	\begin{flushleft}
    \hspace*{0.02in} {\bf Input:}
	A set of records $\mathcal{R}$ \\
	\hspace*{0.02in} {\bf Output:}
	 final partition $\mathbb{C}$ of $\mathcal{R}$ \\
    \end{flushleft}
	\begin{algorithmic}[1] 
            \STATE $\mathbb{B} \leftarrow $ apply blocking functions on $\mathcal{R}$ and get initial blocks
            \STATE generate record sets for each block $\mathcal{B} \in \mathbb{B}$ using \textsf{NRS} algorithm 
            \STATE in-context clustering of record sets with LLM
            \STATE check correctness of in-context clustering by \textsf{MDG} algorithm
            \STATE record sets regeneration to improve in-context clustering if needed
            \WHILE{True}
                \STATE generate record sets for next round using \textsf{Cluster Merge} algorithm
                \STATE in-context clustering of record sets with LLM
                \STATE check correctness of in-context clustering by \textsf{MDG} algorithm
                \STATE record sets regeneration to improve in-context clustering
                \IF{\textsf{Exit condition}}
                \STATE \textbf{Break}
                \ENDIF
            \ENDWHILE
            \STATE Use each singleton cluster to reconstruct the final partition $\mathbb{C}$
            \RETURN $\mathbb{C}$
        \end{algorithmic}
\end{algorithm}
\setlength{\textfloatsep}{12pt plus 2pt minus 2pt}

\spara{Theoretical Analysis.}
We analyze the number of LLM API calls required by our {\sf LLM-CER} method, focusing on a single block of $m$ records. To highlight the efficiency benefits, we consider two extreme cases that capture the boundaries of practical scenarios. 
{\bf (1)} {\em All records belong to same entity.} For each hierarchy level, the number of records is reduced by a factor of the record set size ($S_s$). 
Hence, the total number of LLM API calls equals $(\lceil m/S_s \rceil+ \lceil\lceil m/S_s \rceil/S_s\rceil+\ldots+1)$ $\approx \lceil m/(S_s-1)\rceil$. 
For larger $S_s$, the series converges quickly, and the total number of LLM API calls reduces significantly, justifying the superiority of our in-context clustering approach over pairwise comparisons (whose $S_s=2$) in the end-to-end ER. Specifically in this case, the API calls reduce from $O(m)$ in pairwise with transitivity to only $O(\frac{m}{S_s})$ in {\sf LLM-CER}.
{\bf (2)} {\em All records belong to different entities.} For hierarchy level 0, there are $p=\lceil m/S_s \rceil$ record sets, resulting in $p$ LLM API calls. 
For simplicity, assume $m=p\cdot S_s$, i.e., each record set has exactly $S_s$ records. Since all records are unique, each record forms a singleton cluster, satisfying the clustering exit condition immediately. 
However, a final disambiguation step is still needed to verify cross-set matches. This involves comparing each record to all others outside its record set, resulting in $\frac{m(m-S_s)}{2}$ comparisons. 
Using our in-context framework, we batch these comparisons into new record sets. Each set of size $S_s$ enables $\frac{S_s(S_s-1)}{2}$ pairwise comparisons, meaning we require $\frac{m(m-S_s)}{S_s(S_s-1)}$ LLM API calls for the final check. 
Including level 0, the total number of calls is $\frac{m}{S_s} + \frac{m(m-S_s)}{S_s(S_s-1)}$ = $\frac{m(m-1)}{S_s(S_s-1)}$, which again is much smaller than the $\frac{m(m-1)}{2}$ pairwise calls needed in conventional methods. 
This yields a non-trivial reduction by a factor of O($S_s^2$). 
Since all practical scenarios are between these two extreme cases, our theoretical analysis demonstrates that our design significantly reduces the LLM API calls compared to pairwise ER.
\section{Experimental Results}
\label{sec:exp}
In this section, we empirically evaluate our proposed algorithms as well as the competing approaches. All methods are implemented in Python and executed on a Linux server with 2.20 GHz CPU and 128GB of RAM.
\begin{table}[tb!]
      \vspace{-0.5mm}
      \caption{Dataset statistics: Rec. stands for records, ent. represents entities, attr. denotes attributes, $E_d$ indicates entity dispersion\cite{chen2005exploiting} (=\#Rec./\#Ent.). The `T' in Attr. Types denotes `Textual', `N' represents `Numeric', `C' indicates `Categorical', the number after the attribute category indicates the count of attributes of the corresponding type.} 
      \label{tab:datasets}
      \vspace{-3.2mm}
      \centering
      \small
      \begin{tabular}{m{0.95cm}<{\centering}|m{1.1cm}<{\centering}|m{0.7cm}<{\centering}|m{0.6cm}<{\centering}|m{0.6cm}<{\centering}|m{0.85cm}<{\centering}|m{1.2cm}<{\centering}}
        \hline 
        \multirow{2}{*}{\textbf{Datasets}} & \multirow{2}{*}{\textbf{Domain}} & \multirow{2}{*}{$\textbf{\#Rec.}$} & \multirow{2}{*}{$\textbf{\#Ent.}$} & \multirow{2}{*}{\textbf{$E_d$}} & $\textbf{\#Attr.}$ & $\textbf{Attr.}$ \\
        $\textbf{}$ & $\textbf{}$ & $\textbf{}$ & $\textbf{}$ & \textbf{} & $\textbf{per Rec.}$ & $\textbf{Types}$ \\
        \hline \hline
        \em{Alaska} & Product & 12k & 1.48k &  $\approx8$ & 9 & T(9) \\
        \hline
        \em{AS} & Geo & 2.26k & 0.33k & $\approx7$  & 1 & T(1) \\
        \hline
        \em{Song} & Music & 4.85k & 1.19k & $\approx3$  & 7 & T(4), N(3)\\
        \hline
        \shortstack{\em{Music-}\\\em{20K}} & Music & 19.3k & 10k & $\approx2$  & 6 & T(4), N(1), C(1) \\
        \hline
        \em{DBLP-Google} & Citation & 7.63k & 2.35k &  $\approx3$ & 4 &  T(3), N(1) \\
        \hline
        \em{Cora} & Citation & 1.29k & 0.11k &  $\approx12$ & 12 &  T(12) \\
        \hline
        \shortstack{\em{Cite-}\\\em{seer}} & Citation & 9.13k & 2.49k & $\approx4$  & 6 &  T(4), N(1), C(1) \\ \hline
        \shortstack{\em{Amazon-}\\\em{Google}} & Software & 2.16k & 0.99k & $\approx2$  & 3 & T(2), N(1) \\ \hline
        \shortstack{\em{Walmart-}\\\em{Amazon}} & Electronics & 1.81k & 0.85k & $\approx2$  & 5 & T(3), N(1), C(1) \\
        \hline
      \end{tabular}
      \vspace{-2mm}
\end{table}
\vspace{-2mm}
\subsection{Experimental Setup}
\label{sec:setup}
\vspace{-1mm}
\spara{Datasets.} 
We utilize nine real-world entity resolution (ER) datasets spanning various domains, including e-commerce products, academic citations, geography, and music. 
Details of these datasets are provided in Table \ref{tab:datasets}. 
Specifically, CORA is a text-based dataset comprising duplicate references to scientific publications. For this study, we use the structured CSV version provided by~\cite{nikoletos2022pyjedai}, with the raw data accessible online\footnote{\url{https://www.gabormelli.com/RKB/CORA_Citation_Benchmark_Task}.}. 
{\em{Alaska}} \cite{crescenzi2021alaska} is an e-commerce product dataset collected from multiple websites and previously featured in the SIGMOD 2020 programming contest \cite{sigmod20}. 
{\em{DBLP-Google~\cite{18sigmodEM}}} and {\em{Citeseer-DBLP}} belong to the citation domain, containing bibliographic duplicates sourced from DBLP, Google Scholar, and Citeseer. The {\em{Song}} dataset represents the music domain and includes music information from various platforms, with variations in attributes such as duration and release year. Both {\em{Song}} and {\em{Citeseer-DBLP}} can be found here\footnote{\url{https://pages.cs.wisc.edu/~anhai/data}.}. 
The {\em{AS}} and {\em{Music 20K}} datasets are widely recognized for Dirty ER benchmarks ~\cite{saeedi2017comparative}. 
Lastly, we consider {\em{Amazon-Google}}~\cite{18sigmodEM} and {\em{Walmart-Amazon}}~\cite{18sigmodEM} datasets, originating from the Software and Electronics domains, respectively. These domains are more specialized and require deeper domain knowledge, making the corresponding ER tasks more complex than others.

\spara{Evaluation Metrics.} We select several widely-used metrics to evaluate the in-context clustering results, as well as the final ER clustering results. 
These metrics are FP-measure (a variant of the F-measure), Accuracy (ACC), Normalized Mutual Information (NMI), and Adjusted Rand Score (ARI)~\cite{chen2007adaptive,ahmed2012mcl,wang2024integrated,yeung2001details}. 
All experiments are repeated three times to ensure robustness, and we report the average values over three runs. Due to space limitations, we present results for ACC and FP-measure in the main paper. 

The definition of ACC is as follows: Let $\mathbb{Y} = \{\mathcal{Y}_1, \mathcal{Y}_2, \dots, \mathcal{Y}_m\}$ represent the ground truth clusters and $\mathbb{X} = \{\mathcal{X}_1, \mathcal{X}_2, \dots, \mathcal{X}_n\}$ denote the predicted clusters. The total number of records in both sets is $|\mathcal{R}| = \bigcup_{i} |\mathcal{X}_i| = \bigcup_j |\mathcal{Y}_j|$.
The accuracy (\(\text{ACC}\)) is defined as:
\begin{equation}
    \label{ACC}
    \text{ACC} = \dfrac{\text{CorrectCount}}{\left|\mathcal{R}\right|}
\end{equation}
\begin{equation}
    \label{part-ACC}
    \text{CorrectCount} = \sum_{\mathcal{X}_j \in \mathbb{X}} \sum_{x \in \mathcal{X}_j} \mathbb{I}(\exists y \in \mathcal{Y}_j^{'}, y = x)
\end{equation}
where $\mathbb{Y}^{'} = \{\mathcal{Y}_1^{'}, \dots, \mathcal{Y}_m^{'}\}$ is obtained by reordering $\mathbb{Y}$ based on their intersection sizes with the predicted clusters, and $\mathbb{I}(\cdot)$ is the indicator function that equals to 1 if the condition satisfies.

FP-measure evaluates clustering results from the perspective of homogeneity and stability. It is the harmonic mean of purity and inverse-purity.
\begin{equation}
    \label{purity}
        \textit{purity\,($\mathbb{X}$,$\mathbb{Y}$)} = \sum_{\mathcal{X}_i \in \mathbb{X}} \frac{|\mathcal{X}_i|}{|\mathcal{R}|}\max_{\mathcal{Y}_i \in \mathbb{Y}} \textit{Overlap}(\mathcal{X}_i,\mathcal{Y}_i)
\end{equation}
\begin{equation}
    \label{purity}
        \textit{inverse-purity\,($\mathbb{X}$,$\mathbb{Y}$)} = \sum_{\mathcal{Y}_i \in \mathbb{Y}} \frac{|\mathcal{Y}_i|}{|\mathcal{R}|} \max_{\mathcal{X}_i \in \mathbb{X}} \textit{Overlap}(\mathcal{Y}_i,\mathcal{X}_i)
\end{equation}
\begin{equation}
    \label{overlap}
        \textit{Overlap}(\mathcal{X}_i,\mathcal{Y}_i) := \frac{|\mathcal{X}_i\cap \mathcal{Y}_i|}{|\mathcal{X}_i|} 
\end{equation}

Finally, we derive the FP-measure:
\begin{equation}
    \label{fp-measure}
        \textit{FP-measure}(\mathbb{X},\mathbb{Y}) = \dfrac{2}{1 / {\textit{purity($\mathbb{X},\mathbb{Y}$)}} + 1 /{\textit{inverse-purity($\mathbb{X},\mathbb{Y}$)}}}
\end{equation}

\subsection{Performance on End-to-End ER}
\label{sec:exp_end2end}
We use \texttt{gpt-4o-mini} for end-to-end tests and set the temperature parameter to zero. For embedding generation in blocking, we adopt widely-used \texttt{all-MiniLM-L6-v2} model\footnote{\url{https://www.sbert.net/}}. 
\subsubsection{In-context clustering vs. pairwise questioning}
We compare the performance of our in-context clustering-based ER ($S_s=9$) with the pairwise matching method ($S_s=2$) to demonstrate the effectiveness and efficiency of in-context clustering that can process multiple records simultaneously. For fair comparison, we also adopt our guardrail strategy to mitigate LLM-induced errors for pairwise matching 
as stated in \S~\ref{sec:in_context_cluster}. Table \ref{tab:ablation} shows that when $S_s$ is set to 9, the final ER results achieved by our algorithm are nearly identical or even better than those of the pairwise matching approach. On the other hand, our in-context clustering significantly reduces
the number of LLM API calls by 12-108$\times$, token consumption by 3-28$\times$, monetary cost by 3-22$\times$, and end-to-end ER time by 6-55$\times$, compared to pairwise matching.

Table \ref{tab:hierarchy} illustrates the distribution of the number of record sets across hierarchy levels, providing insights into our hierarchical clustering approach. For all datasets, the number of record sets decreases significantly from Level 0 to higher levels, and are nearly completed before Level 5, demonstrating the effectiveness of the hierarchical structure in progressively merging similar entities. Larger datasets generally require more hierarchy levels for processing, as evidenced by {\em{Alaska}} (the largest dataset) extending to Level 5, while {\em{Cora}} (the smallest dataset) completes by Level 3.

\begin{table}[tb!]
\vspace{-2mm}
\caption{Comparison with pairwise matching-based ER}
\label{tab:ablation}
\vspace{-3.5mm}
    \centering
    \resizebox{8.5cm}{!}{
    \renewcommand{\arraystretch}{1.2}
    \begin{tabular}{c|c|c|c|c|c|c}
    \hline
    \multirow{2}{*}{\textbf{Metrics}} & \multicolumn{2}{c|}{\em{Cora}} & \multicolumn{2}{c|}{\em{Alaska}} & \multicolumn{2}{c}{\em{AS}} \\
    \cline{2-7}
     & $\boldsymbol{S_s=2}$ & $\boldsymbol{S_s=9}$ & $\boldsymbol{S_s=2}$ & $\boldsymbol{S_s=9}$ & $\boldsymbol{S_s=2}$ & $\boldsymbol{S_s=9}$ \\
     & ${(pairwise)}$ & ${(clustering)}$ & ${(pairwise)}$ & ${(clustering)}$ & ${(pairwise)}$ & ${(clustering)}$ \\
    \hline
    \textbf{ACC} & 0.88 & {\bf 0.90} & 0.81 & {\bf 0.82} & {\bf 0.70} & {\bf 0.70} \\
    \hline
    \textbf{FP-measure} & 0.67 & {\bf 0.71} & 0.78 & {\bf 0.79} & 0.60 & {\bf 0.63} \\
    \hline
    \textbf{Cost (USD)} & 0.67 & {\bf 0.03} & 0.43 & {\bf 0.15} & 0.08 & {\bf 0.02} \\
    \hline
    \textbf{Tokens (M)} & 3.45 & {\bf 0.12} & 2.29 & {\bf 0.73} & 0.35 & {\bf 0.07} \\
    \hline
    \textbf{Time (min)} & 297.27 & {\bf 5.42} & 241.31 & {\bf 39.57} & 77.2 & {\bf 8.01} \\
    \hline
    \textbf{\# API Calls (K)} & 30.23 & {\bf 0.28} & 24.54 & {\bf 2.04} & 7.85 & {\bf 0.41} \\
    \hline
    \end{tabular}
}
\vspace{-4mm}
\end{table}

\begin{table}[tb!]
      \caption{Number of records set in each hierarchy level}
      \vspace{-3.5mm}
      \label{tab:hierarchy}
      \small
      \begin{tabular}{m{1.0cm}<{\centering}|m{0.8cm}<{\centering}|m{0.8cm}<{\centering}|m{0.8cm}<{\centering}<{\centering}|m{0.8cm}<{\centering}|m{0.8cm}<{\centering}|m{0.8cm}<{\centering}}
        \hline 
        $\textbf{Dataset}$ & $\textbf{Level 0}$ & $\textbf{Level 1}$ & $\textbf{Level 2}$ & $\textbf{Level 3}$ & $\textbf{Level 4}$ & $\textbf{Level 5}$ \\
        \hline \hline
        \em{Cora}  & 183 & 76 & 15 & 5 & / & / \\
        \hline
        \em{Alaska}  & 1312 & 604 & 101 & 20 & 5 & 1 \\
        \hline
        \em{AS}  & 251 & 107 & 48 & 6 & 1 & / \\
        \hline
      \end{tabular}
      \vspace{-1mm}
\end{table}

\subsubsection{Comparison with state-of-the-art.} 
\label{sec:compare sota}
We compare our {\sf LLM-CER} framework with several recent LLM-based ER methods to highlight its effectiveness and efficiency.

\begin{table}[tb!]
\vspace{-2mm}
\caption{End-to-end comparison of our in-context clustering-based ER with state-of-the-art LLM-based ER}
\vspace{-3mm}
\label{tab:performance of alg}
\footnotesize
\centering
\begin{tabular}{m{0.9cm}<{\centering}|c|c|c|c|m{1.15cm}<{\centering}}
\hline
\multirow{2}{*}{\textbf{Dataset}} & \multirow{2}{*}{\textbf{Metrics}} & \multirow{2}{*}{\textbf{LLM-CER}} & \multirow{2}{*}{\textbf{Booster}} & \multirow{2}{*}{\textbf{BQ}} & \textbf{CrowdER} \\
        $\textbf{}$ & $\textbf{}$ & $\textbf{}$ & $\textbf{}$ & $\textbf{}$ & $\textbf{+LLM}$
        \\ \hline \hline
\multirow{6}{*}{ \em{Alaska} } & ACC & \textbf{0.82} & 0.71 & 0.33 & 0.68 \\ \cline{2-6}
                          & FP & \textbf{0.79} & 0.55 & 0.49 & 0.62 \\ \cline{2-6}
                          & Cost (\$) & 0.15 & \textbf{0.02} & 1.55 &  0.42\\ \cline{2-6}
                          & Tokens (M) & 0.73 & \textbf{0.19} & 5.59 & 2.04\\ \cline{2-6}
                          & Time (s) & \textbf{2374.2} & 2450.1 & 8798.9 & 6547.2 \\ \cline{2-6}
                          & \# API Calls & \textbf{2043} & 2606 & 8035 & 5845\\
                          \hline \hline
\multirow{6}{*}{ \em{AS} } & ACC & \textbf{0.70} & 0.62 & 0.54 & 0.52 \\ \cline{2-6}
                          & FP & \textbf{0.63} & 0.62 & 0.51 & 0.50 \\ \cline{2-6}
                          & Cost (\$) & 0.02 & \textbf{0.01} & 0.29 & 0.11\\ \cline{2-6}
                          & Tokens (M) & 0.07 & \textbf{0.03} & 0.34 & 0.37 \\ \cline{2-6}
                          & Time (s) & \textbf{480.6} & 622.9 & 925.5 & 2356.2 \\ 
                          \cline{2-6}
                          & \# API Calls & \textbf{413} & 723 & 842 & 2084 \\
                          \hline \hline
\multirow{6}{*}{ \em{Song} } & ACC & \textbf{0.72} & 0.52 & 0.59 & 0.52\\ \cline{2-6}
                          & FP & \textbf{0.78} & 0.68 & 0.67 & 0.64\\ \cline{2-6}
                          & Cost (\$) & 0.06 & \textbf{0.02} & 0.77 & 0.12 \\ \cline{2-6}
                          & Tokens (M) & 0.22 & \textbf{0.11} & 1.98 & 0.43\\ \cline{2-6}
                          & Time (s) & \textbf{933.2} & 903.3 & 2581.5 & 1856.3\\ 
                          \cline{2-6}
                          & \# API Calls & \textbf{668} & 921 & 2338 & 1247\\
                          \hline \hline
\multirow{6}{*}{ \em{Music} } & ACC & \textbf{0.71} & 0.59 & 0.60 & 0.62 \\ \cline{2-6}
                          & FP & \textbf{0.61} & 0.60 & 0.54 & 0.55\\ \cline{2-6}
                          & Cost (\$) & 0.19 & \textbf{0.02} & 2.18 & 0.39\\ \cline{2-6}
                          & Tokens (M) & 0.90 & \textbf{0.15} & 8.96 & 1.82\\ \cline{2-6}
                          & Time (s) & \textbf{2388.4} & 2585.1 & 17515.8 & 4562.3\\
                          \cline{2-6}
                          & \# API Calls & \textbf{3859} & 3915 & 17365 & 7782\\
                          \hline \hline
\multirow{6}{*}{ \em{DG} } & ACC & \textbf{0.81} & 0.56 & 0.62 & 0.72 \\ \cline{2-6}
                          & FP & \textbf{0.70} & 0.68 & 0.63 & 0.65 \\ \cline{2-6}
                          & Cost (\$) & 0.07 & \textbf{0.02} & 1.12 &  0.34 \\ \cline{2-6}
                          & Tokens (M) & 0.37 & \textbf{0.18} & 3.92 & 1.79 \\ \cline{2-6}
                          & Time (s) & \textbf{1552.4} & 2552.2 & 6052.2 & 7456.3 \\  \cline{2-6}
                          & \# API Calls & \textbf{1285} & 3085 & 6456 & 6504 \\
                          \hline \hline
\multirow{6}{*}{ \em{Cora} } & ACC & \textbf{0.90} & 0.75 & 0.62 & 0.51 \\ \cline{2-6}
                          & FP & \textbf{0.71} & 0.60 & 0.56 & 0.61 \\ \cline{2-6}
                          & Cost (\$) & 0.03 & \textbf{0.01} & 1.45 & 0.07 \\ \cline{2-6}
                          & Tokens (M) & 0.12 & \textbf{0.06} & 4.23 & 0.29 \\ \cline{2-6}
                          & Time (s) & \textbf{325.5} & 605.4 & 4085.3 & 598.5 \\ 
                          \cline{2-6}
                          & \# API Calls & \textbf{279} & 698 & 4882 & 483 \\
                          \hline \hline
\multirow{6}{*}{ \shortstack{\em{Cite-}\\\em{seer}} } & ACC & \textbf{0.88} & 0.72 & 0.64 & 0.60
 \\ \cline{2-6}
                          & FP & \textbf{0.95} & 0.78 & 0.79 & 0.69 \\ \cline{2-6}
                          & Cost (\$) & 0.03 & \textbf{0.01} & 0.63 &  0.08 \\ \cline{2-6}
                          & Tokens (M) & 0.13 & \textbf{0.05} & 1.64 & 0.37 \\ \cline{2-6}
                          & Time (s) & \textbf{1360.8} & 1585.2 & 6228.9 & 3895.6 \\ 
                          \cline{2-6}
                          & \# API Calls & \textbf{1302} & 2169 & 6420 & 3858 \\
                          \hline \hline
\multirow{6}{*}{ \shortstack{\em{Amazon-}\\\em{Google} } }  & ACC & \textbf{0.71} & 0.58 & 0.53 & 0.50
 \\ \cline{2-6}
                          & FP & \textbf{0.64} & 0.55 & 0.50 & 0.48 \\ \cline{2-6}
                          & Cost (\$) & 0.02 & \textbf{0.01} & 0.62 &  0.09 \\ \cline{2-6}
                          & Tokens (M) & 0.07 & \textbf{0.03} & 0.86 & 0.42 \\ \cline{2-6}
                          & Time (s) & \textbf{465.6} & 785.2 & 1658.2 & 1985.2 \\ 
                          \cline{2-6}
                          & \# API Calls & \textbf{452} & 998 & 1895 & 2025 \\
                          \hline \hline
\multirow{6}{*}{ \shortstack{\em{Walmart-}\\\em{Amazon}} }  & ACC & \textbf{0.61} & 0.50 & 0.42 & 0.51
 \\ \cline{2-6}
                          & FP & \textbf{0.56} & 0.48 & 0.41 & 0.50 \\ \cline{2-6}
                          & Cost (\$) & 0.02 & \textbf{0.01} & 0.59 &  0.08 \\ \cline{2-6}
                          & Tokens (M) & 0.06 & \textbf{0.03} & 0.68 & 0.39 \\ \cline{2-6}
                          & Time (s) & \textbf{375.8} & 475.2 & 1498.5 & 3895.6 \\ 
                          \cline{2-6}
                          & \# API Calls & \textbf{398} & 825 & 1585 & 1958 \\
                          \hline 
\end{tabular}
\vspace{-4mm}
\end{table}

\textbf{Booster~\cite{li2024booster}}: \textsf{Booster} employs traditional blocking to generate multiple candidate partitions and iteratively queries the LLM using a {\em next-question selection}: identifying record pairs that are most informative in distinguishing between different partitioning. The final output is one of the partitionings with the highest probability. 
Unlike \textsf{Booster}, which is limited to selecting from predefined partitions, our method integrates blocking with hierarchical clustering and misclustering correction to produce refined clusters beyond the initial partitions—resulting in higher flexibility and accuracy.

\textbf{BQ~\cite{fan2024cost}}: \textsf{BQ} reduces LLM usage by batching multiple pairwise match questions into a single prompt, allowing contextual information to improve prediction quality across questions. It also uses an enhanced demonstration selection strategy for better few-shot generalization. Each batch of $k$ records forms $\frac{k}{2}$ pairwise comparisons. In contrast, our method with set size $k$ performs clustering over the entire set, avoiding up to $\frac{k(k-1)}{2}$ comparisons, yielding significantly fewer API calls and lower monetary cost.

\textbf{CrowdER+LLM~\cite{wang2012crowder}:} \textsf{CrowdER+LLM} follows a clustering-based design that filters irrelevant pairs and generates record sets to resolve ambiguous cases, merging clusters via transitive closure. 
Our approach improves on this by incrementally constructing record sets and leveraging both transitivity and anti-transitivity to prune unnecessary sets. 
For a fair comparison, we adopt their record grouping strategy but replace crowdsourcing with LLM-based clustering, using the same blocking and record set size.

\vspace{-1mm}
\spara{Implementation.} For \textsf{BQ}, we follow~\cite{fan2024cost} by including 8 demonstrations per prompt and applying the same blocking strategy as ours. Each prompt contains 5 pairwise questions (i.e., 10 records), approximately matching our 9-record clustering prompts for a fair workload comparison. As \textsf{BQ} requires labeled pairs for demonstrations, we include their annotation cost following the same accounting method. For \textsf{Booster}, we apply our blocking method with tuned parameters. We evaluate all methods using ACC and FP-measure.

\vspace{-1mm}
\spara{End-to-end Performance.} As shown in Table \ref{tab:performance of alg}, the following observations can be made: 
(1) Our {\sf LLM-CER} method consistently outperforms all baselines across datasets in both clustering quality (ACC and FP-measure) and operational efficiency (API calls and runtime), while maintaining competitive monetary cost. 
(2) \textsf{Booster} selects from multiple blocking-based partitionings using an LLM-driven scoring mechanism but does not refine these partitions. While it may choose a relatively strong candidate, its inability to correct or merge partitions limits its clustering quality: its ACC and FP scores remain suboptimal.  In terms of monetary cost, however, \textsf{Booster} benefits from minimal token consumption, as it only selects the best partitioning without further modifying it. 
(3) \textsf{BQ} performs exhaustive pairwise matching within blocks using few-shot prompts, incurring substantial API and token costs—5–35$\times$ higher than ours—even after applying transitivity and anti-transitivity. 
Although demonstrations slightly improve pairwise accuracy, the lack of result verification often leads to incorrect merges, resulting in the lowest ACC and FP scores. 
Furthermore, \textsf{BQ} requires labeled examples for few-shot prompting, adding additional annotation cost, whereas our method operates entirely without supervision.
(4) \textsf{CrowdER+LLM} uses clustering and achieves the second-best accuracy. However, its HIT design permits overlapping records and lacks alignment with LLM-optimized record set formats (\S~\ref{sec:related works}), resulting in greater API calls and monetary cost. Moreover, like \textsf{BQ}, it lacks verification for LLM outputs, limiting its clustering quality compared to our {\sf LLM-CER} approach.

\vspace{-1mm}
\spara{Remark.} Some of the competing methods are also {\em complementary to ours}. For example, {\sf BQ}’s batch processing can be used in our case to batch clustering of multiple record sets simultaneously. Similarly, \textsf{Booster}’s best partitioning selection can serve as an effective pre-processing step, providing high-quality initial blocks that our algorithm can refine for enhanced accuracy. These complementarities underscore the potential for integrating strengths from different methods to advance ER performance.

\vspace{-1mm}
\subsection{Impact of Dataset Characteristics} 
\label{sec:dataset_key_factor}
\vspace{-0.5mm}

\noindent
In this section, we analyze how dataset characteristics—namely the number of attributes, attribute types, and overall dataset difficulty—affect the optimal configuration of key parameters and the performance of our end-to-end ER framework. To ensure fair evaluation, we retain critical attributes (e.g., title) across all settings.

\begin{table}[!tb]
      \vspace{-1mm}
      \caption{Optimal values vs. attribute count ($A_n$) and attribute types. The `T' denotes `Textual', `N' represents `Numeric', `C' indicates `Categorical'. Inc. denotes `Included'}
      \label{tab:key values an and at}
      \vspace{-3mm}
      \centering
        \begin{tabular}{c|c|c|c}
        \hline
        \textbf{Dataset} & \textbf{$A_n$} & $S_s$ & $S_d$
                \\ \hline \hline
        \multirow{3}{*}{ {\em {Cora}} } & $4$ & 9 & 3 \\ \cline{2-4}
                                  & $8$ & 9 & 4 \\ \cline{2-4}
                                  & $12$ & 9 & 4 \\ \cline{2-4}
                                 
                                  \hline 
        \multirow{3}{*}{ {\em {Alaska}} } & $3$ & 9 & 4 \\ \cline{2-4}
                                  & $6$ & 9 & 4 \\ \cline{2-4}
                                  & $9$ & 9 & 4  \\ \cline{2-4}
                                  \hline 
        \end{tabular}
        \quad
       \begin{tabular}{c|c|c|c}
        \hline
        \textbf{Dataset} & \textbf{Inc. type} & $S_s$ & $S_d$
                \\ \hline \hline
        \multirow{4}{*}{ {\em WA} } & T, N, C & 7 & 3 \\ \cline{2-4}
                                  & N, C & 12 & 4 \\ \cline{2-4}
                                  & T, C & 8 & 3 \\ \cline{2-4}
                                  & T, N & 8 & 4 \\ \cline{2-4}
                                  \hline 
        \multirow{4}{*}{ \shortstack{\em{Cite-}\\\em{seer}} } & T, N, C & 9 & 4 \\ \cline{2-4}
                                  & N, C & 8 & 4 \\ \cline{2-4}
                                  & T, C & 9 & 4 \\ \cline{2-4}
                                  & T, N & 9 & 4  \\ \cline{2-4}
                                  \hline 
        \end{tabular}
        \vspace{-3mm}
\end{table}

\begin{table}[t]
\caption{ End-to-end ER performance vs. attribute count}
\label{tab:e2e with varying an}
\centering
\small
\vspace{-3mm}
\begin{tabular}{m{0.98cm}<{\centering}|c|c|c|c}
\hline
\multirow{2}{*}{\textbf{Dataset}} & \multirow{2}{*}{\textbf{Metrics}} & \multirow{2}{*}{$A_n=4$} & \multirow{2}{*}{\textbf{$A_n=8$}} & \multirow{2}{*}{\textbf{$A_n=12$}}\\
        $\textbf{}$ & $\textbf{}$ & $\textbf{}$ & $\textbf{}$ & $\textbf{}$
        \\ \hline \hline
\multirow{6}{*}{ \em{Cora} } & ACC & 0.82 & 0.85 &  \textbf{0.90} \\ \cline{2-5}
                          & FP & 0.66 & 0.67 & \textbf{0.71} \\ \cline{2-5}
                          & Cost (\$) & \textbf{0.02} & 0.03 & 0.03  \\ \cline{2-5}
                          & Tokens (M) & \textbf{0.05} & 0.09 & 0.12  \\ \cline{2-5}
                          & Time (min) & \textbf{5.04} & 5.21 & 5.43  \\ 
                          \cline{2-5}
                          & API Calls & 288 & 283 & \textbf{279}  \\
                          \hline 
\multirow{6}{*}{ \em{Alaska} } & ACC & 0.74 & 0.77 &  \textbf{0.82} \\ \cline{2-5}
                          & FP & 0.74 & 0.75 & \textbf{0.79} \\ \cline{2-5}
                          & Cost (\$) & \textbf{0.06} & 0.11 & 0.15  \\ \cline{2-5}
                          & Tokens (M) & \textbf{0.26} & 0.51 & 0.73  \\ \cline{2-5}
                          & Time (min) & \textbf{37.54} & 38.24 & 39.57  \\ 
                          \cline{2-5}
                          & API Calls & 2064 & 2055 & \textbf{2043}  \\
                          \hline 
\end{tabular}
\vspace{-2mm}
\end{table}

\spara{Impact of Attribute Count.} 
To isolate the effect of increasing attribute count, we use the \textit{Cora} and \textit{Alaska} datasets, which contain many single-type (textual) attributes.
As shown in Table~\ref{tab:key values an and at}, the optimal values for $S_s$ and $S_d$ remain largely stable. For \textit{Cora}, $S_s$ stays at 9 across all attribute counts, while $S_d$ increases slightly from 3 to 4. For \textit{Alaska}, both values remain fixed at 9 and 4, respectively. 
This suggests that, for single-type textual datasets, the parameter configuration is robust to changes in attribute count.

Building on these configurations, we conduct end-to-end ER experiments (Table~\ref{tab:e2e with varying an}). Results show that increasing attribute count consistently improves clustering performance: for \textit{Cora}, ACC improves from 0.84 to 0.90 and FP-measure from 0.66 to 0.71 when moving from 4 to 12 attributes. Similar trends are observed for \textit{Alaska}. 
These improvements persist with {\sf MDG}, indicating that richer attribute sets enhance record distinguishability. 
While token usage increases moderately (0.05M to 0.12M for \textit{Cora}), the accuracy gains justify retaining all attributes in single-type datasets.

\spara{Impact of Attribute Types and Dataset Difficulty.} 
To assess the role of attribute types, we use \textit{Walmart-Amazon} and \textit{Citeseer}, both containing textual, numerical, and categorical fields. Controlled ablations omit one attribute type at a time to evaluate its contribution.

As shown in Table~\ref{tab:key values an and at}, the simpler \textit{Citeseer} dataset exhibits stable parameter values, with only a minor drop when textual attributes are removed—likely due to increased reliance on less informative modalities.
In contrast, the more complex \textit{Walmart-Amazon} dataset shows notable deviations. When all attributes are retained, optimal values are $S_s = 7$ and $S_d = 3$, lower than the typical $9$ and $4$, indicating higher clustering difficulty. Removing textual attributes increases $S_s$ and $S_d$ to 12 and 4, suggesting that pruning noisy fields simplifies the task and enables larger input sets. Excluding numerical or categorical attributes also shifts $S_s$ to 8, with corresponding adjustments in $S_d$. 
These results highlight that attribute type plays a critical role in complex domains, where noisy or redundant fields may obscure true entity boundaries.

\begin{table}[t]
\vspace{-1mm}
\caption{ End-to-end ER performance vs. attribute types } 
\label{tab:e2e with varying type}
\centering
\footnotesize
\vspace{-4mm}
\begin{tabular}{m{1.05cm}<{\centering}|c|c|c|c|c}
\hline
\multirow{2}{*}{\textbf{Dataset}} & \multirow{2}{*}{\textbf{Metrics}} & \multirow{2}{*}{\textbf{Original}} & {\textbf{w/o}} & {w/o} & {w/o}\\
        $\textbf{}$ & $\textbf{}$ & $\textbf{}$ & $\textbf{Textual}$ & $\textbf{Numeric}$ & $\textbf{Categorical}$
        \\ \hline \hline
\multirow{6}{*}{ \shortstack{\em{Walmart-}\\\em{Amazon}} } & ACC & 0.61 & \textbf{0.72} & 0.66 & 0.60 \\ \cline{2-6}
                          & FP & 0.56 & \textbf{0.66} &  0.58 & 0.54 \\ \cline{2-6}
                          & Cost (\$) & 0.02 & \textbf{0.01} & 0.02 & 0.02 \\ \cline{2-6}
                          & Tokens (M) & 0.06 & \textbf{0.04} & 0.05 & 0.06 \\ \cline{2-6}
                          & Time (min) & 6.25 & \textbf{5.89} & 6.02 & 6.36 \\ 
                          \cline{2-6}
                          & API Calls & 398 & \textbf{374} & 393 & 409 \\
                          \hline 
\multirow{6}{*}{ \shortstack{\em{Cite-}\\\em{seer}} } & ACC & \textbf{0.88} & 0.82 & 0.86 & 0.86 \\ \cline{2-6}
                          & FP & \textbf{0.95} & 0.90 & 0.92  & 0.93 \\ \cline{2-6}
                          & Cost (\$) & 0.03 & \textbf{0.02} & 0.03 & 0.03 \\ \cline{2-6}
                          & Tokens (M) & 0.13 & \textbf{0.11} & 0.12 & 0.12 \\ \cline{2-6}
                          & Time (min) & 22.68 & \textbf{20.98} & 21.88 & 22.03 \\ 
                          \cline{2-6}
                          & API Calls & \textbf{1302} & 1331 & 1312 & 1314 \\
                          \hline 
\end{tabular}
\vspace{-2.5mm}
\end{table}

End-to-end results (Table~\ref{tab:e2e with varying an}) further confirm this. 
For \textit{Walmart-Amazon}, removing noisy textual attributes—while preserving key fields like title—boosts ACC from 0.61 to 0.72 and FP-measure from 0.56 to 0.66, with token usage dropping from 0.06M to 0.04M. This is due to common extraction errors (e.g., `brand' values in `name', or `leather red' misclassified as `color') that introduce significant noise. 
Eliminating these sources of noise allows the model to focus on cleaner signals, enhancing clustering quality. Similar observations are reported in~\cite{18sigmodEM}. 
Excluding numerical attributes yields a modest ACC gain (to 0.66), while removing categorical attributes slightly degrades performance, suggesting categorical fields offer informative structure. 
In contrast, \textit{Citeseer} suffers across all ablations, reflecting the complementary value of all attribute types in well-structured datasets. 
These findings underscore the importance of selective attribute pruning for noisy, domain-specific datasets, and comprehensive attribute inclusion for simpler, structured ones.

\vspace{-1mm}
\subsection{Impact of LLM Guardrails}
\label{sec:exp_block_filter}
\vspace{-0.5mm}

\begin{table}[t]
\caption{Effect of MDG algorithm and record set regeneration on end-to-end performance}
\label{tab:effect_MDG}
\vspace{-3.5mm}
    \centering
    \resizebox{8.5cm}{!}{
    \renewcommand{\arraystretch}{1.2}
    \begin{tabular}{c|c|c|c|c|c|c}
    \hline
    \multirow{2}{*}{\textbf{Metrics}} & \multicolumn{2}{c|}{\em{Cora}} & \multicolumn{2}{c|}{\em{Alaska}} & \multicolumn{2}{c}{\em{AS}} \\
    \cline{2-7}
     & {\em {w/o MDG}} & {\em {w/ MDG}} & {\em {w/o MDG}} & {\em {w/ MDG}} & {\em {w/o MDG}} & {\em {w/ MDG}} \\
     
    \hline
    \textbf{ACC} & 0.60 & {\bf 0.90} & 0.35 & {\bf 0.82} & 0.52 & {\bf 0.70} \\
    \hline
    \textbf{FP-measure} & 0.58 & {\bf 0.71} & 0.47 & {\bf 0.79} & 0.52 & {\bf 0.63} \\
    \hline
    \textbf{Cost (USD)} & {\bf 0.03} & {\bf 0.03} & {\bf 0.14} & 0.15 & {\bf 0.02} & {\bf 0.02} \\
    \hline
    \textbf{Tokens (M)} & {\bf 0.11} & 0.12 & {\bf 0.66} & 0.73 & {\bf 0.06} &  0.07 \\
    \hline
    \textbf{Time (min)} & {\bf 5.04} & 5.42 & {\bf 36.88} & 39.57 & {\bf 7.35} & 8.01 \\
    \hline
    \textbf{\# API Calls (K)} & {\bf 0.26} & 0.28 & {\bf 1.85} &  2.04 & {\bf 0.37} & 0.41 \\
    \hline
    \end{tabular}
}
\vspace{-2mm}
\end{table}

The results in Table \ref{tab:effect_MDG} show that the incorporation of the {\sf Misclustering Detection Guardrail (MDG)} algorithm leads to improvements in ACC and FP-measure.
For instance, ACC increases by up to 50\% and FP-measure by 22\% on {\em Cora}. This supports the claim that {\sf MDG} reduces the occurrence of misclassifications in the ER process.

Moreover, these performance gains are achieved with minimal additional computational cost. The total cost, token usage, processing time, and number of API calls remain almost unchanged or only slightly increase with the addition of {\sf MDG}. This suggests that {\sf MDG} enhances ER performance without incurring a significant resource overhead, making it an efficient approach for improving entity resolution in practical applications.

\vspace{-1mm}
\subsection{Scalability Test}
\label{sec:scalability}
\vspace{-1mm}

To assess the scalability of our end-to-end ER algorithm, we conduct experiments on progressively larger subsets of the publicly available {\textit{Music}} dataset~\cite{saeedi2017comparative}, containing 10K, 20K, 30K, and 50K records. These subsets preserve consistent entity dispersion, enabling a controlled evaluation of how performance and resource usage evolve with scale. 
We also compare our method with {\sf BQ} \cite{fan2024cost} (introduced in \S\ref{sec:compare sota}), a state-of-the-art baseline designed for efficient LLM-based ER. To ensure fairness, all experiments are subject to a maximum execution time limit of 24 hours.

\begin{figure}[t]
\centering
\vspace{-2mm}
\includegraphics[width=0.16\textwidth]{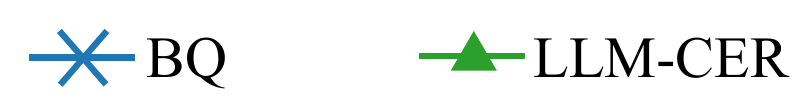}\\
  \vspace{-2mm}
  \hspace{-7.5mm}
  \subfigcapskip=-2mm
  \subfigure[FP-measure]{
   \includegraphics[width=1.6in]{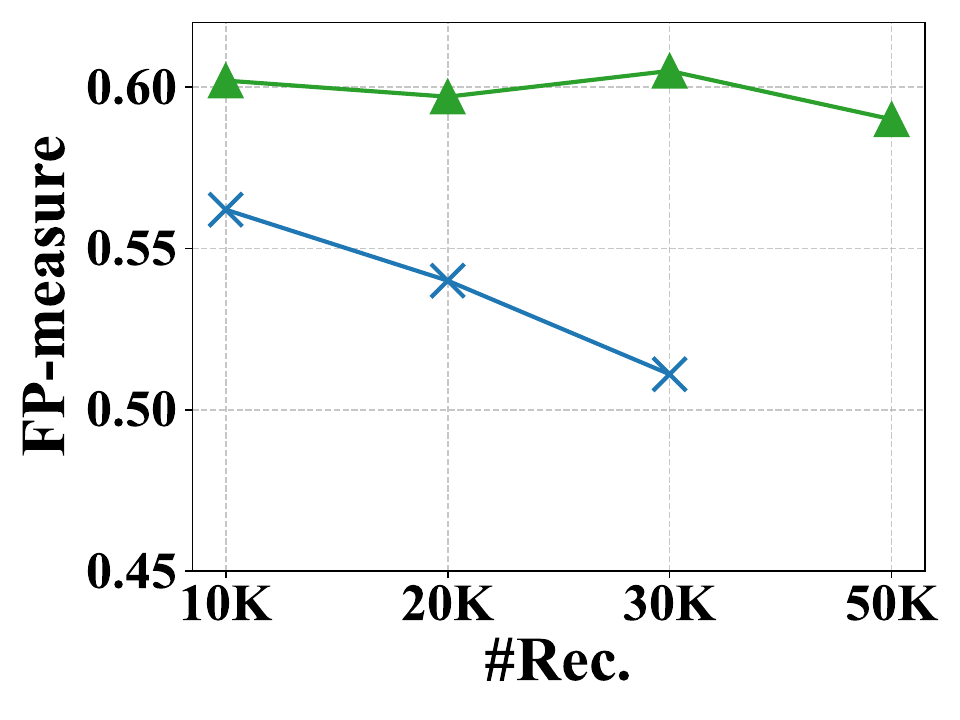}
   \label{fig:fp-scala}
  }
  \hspace{-2.7mm}
  \subfigure[Cost(\$)]{
   \centering
   \includegraphics[width=1.6in]{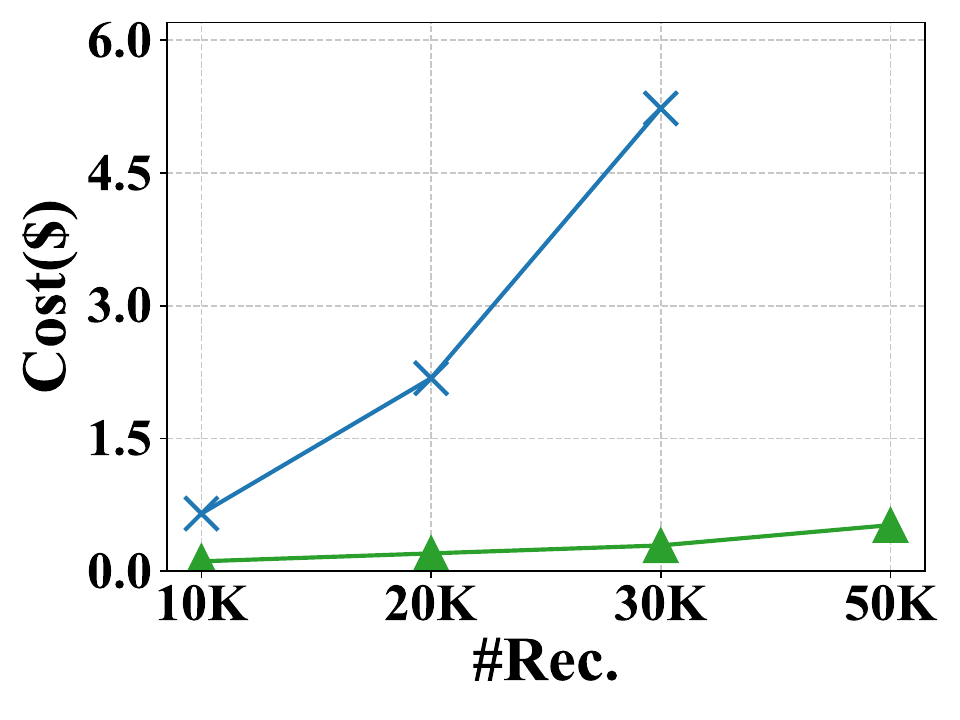}
   \label{fig:cost-scala}
  }\\ \vspace{-3mm}
  \hspace{-7.5mm}
  \subfigcapskip=-2mm
  \subfigure[Times (min)]{
   \includegraphics[width=1.6in]{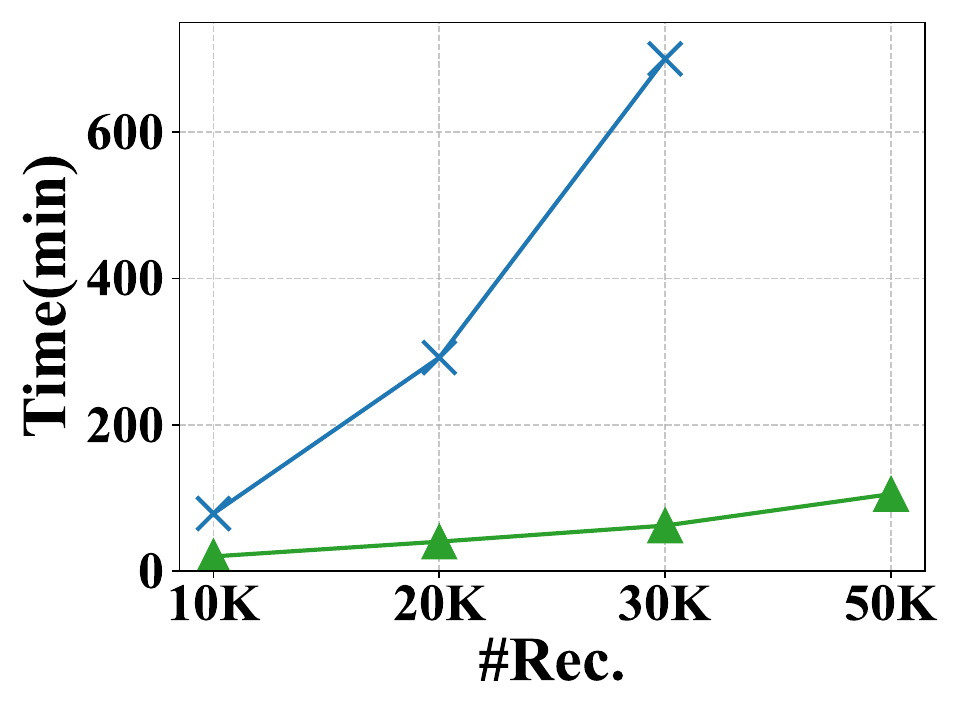}
   \label{fig:time-scala}
  }
  \hspace{-2.7mm}
  \subfigure[API Calls (K)]{
   \centering
   \includegraphics[width=1.6in]{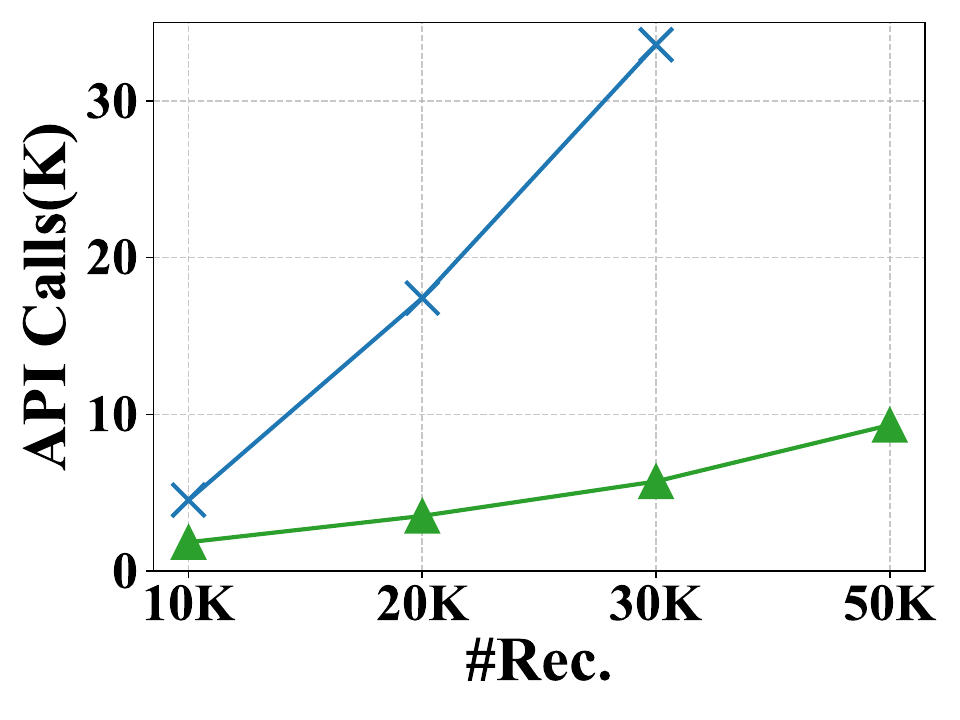}
   \label{fig:api-scala}
  }
\vspace{-5.5mm}
\caption{ \centering { End-to-end performance vs. \#Records} } 
\vspace{-4mm}
\label{fig:scala-perform}
\end{figure}

Figure~\ref{fig:scala-perform} shows that our {\sf LLM-CER} consistently maintains high FP-measure as the dataset size increases, demonstrating strong robustness to scale. 
In contrast, {\sf BQ} experiences a notable accuracy loss.
Regarding efficiency, our method exhibits a predictable and modest increase in both cost and the number of API calls as data size grows, whereas {\sf BQ} incurs substantially higher overheads. 
Execution time for our method scales linearly from approximately 20 minutes on {\textit{Music 10K}} to just over 100 minutes on {\textit{Music 50K}}. 
{\sf BQ} becomes prohibitively slow on larger datasets, particularly failing to complete on the {\textit{Music 50K}} dataset within the time limit.

\vspace{-1mm}
\section{Conclusion}
\label{sec:conclusions}
In this work, we proposed an end-to-end in-context clustering-based entity resolution solution utilizing the Large Language Models (LLMs). 
We thoroughly explored the design space for in-context clustering, investigating how key factors -- including set size, set diversity, set variation, and record order --influence clustering performance. 
Building on these insights, we introduced a robust end-to-end framework that incorporates an innovative record set creation strategy, a hierarchical clustering merge algorithm, and a tailored misclustering detection mechanism.
Comprehensive evaluations on nine real-world datasets demonstrated that our approach, {\sf LLM-CER} not only achieves substantial improvements in result quality but also significantly optimizes resource efficiency, including token usage, running time, and API call volume. 
In the future, ideas from other LLM-based ER methods could complement our approach, such as batch clustering of multiple record sets simultaneously and advanced prompt engineering for in-context clustering.


\section*{Acknowledgments}

Xiangyu Ke, and Yunjun Gao are supported in part by the NSFC under Grants No. (62025206, U23A20296), Zhejiang Province's ``Lingya'' R\&D Project under Grant No. 2024C01259, Ningbo Yongjiang Talent Introduction Programme (2022A-237-G). Arijit Khan acknowledge support from the Novo Nordisk Foundation grant NNF22OC0072415. Sharad Mehrotra is supported in part by NSF Grants No. 1952247, 2245372, 2420846, 2133391, 2008993, 2008993, 1952247, 2032525. Xiangyu Ke is the corresponding author.


\clearpage
\bibliographystyle{ACM-Reference-Format}
\bibliography{ref}

\clearpage
\section{Appendix}
\label{sec-appendix}

This section presents the additional experimental results omitted from \S~\ref{sec:exp} due to space constraints.

\spara{More Evaluation Metrics.} We select several widely-used metrics to evaluate the in-context clustering results, as well as the final ER clustering results, including FP-measure (a variant of the F-measure), Accuracy (ACC), Normalized Mutual Information (NMI), and Adjusted Rand Score (ARI)~\cite{chen2007adaptive,ahmed2012mcl,wang2024integrated,yeung2001details}. 
All experiments are reported as the average of 3 repeats. 
Due to space limitations, we present results only for ACC and FP-measure in the main paper. We define the other two measures below.

Let $\mathbb{Y} = \{\mathcal{Y}_1, \mathcal{Y}_2, \dots, \mathcal{Y}_m\}$ represents the ground truth clusters and $\mathbb{X} = \{\mathcal{X}_1, \mathcal{X}_2, \dots, \mathcal{X}_n\}$ denotes the predicted clusters. The total number of records in both sets is $|\mathcal{R}| = \bigcup_{i} |\mathcal{X}_i| = \bigcup_j |\mathcal{Y}_j|$.

NMI measures the similarity between clustering results and ground truth, from the perspective of clustering entropy. 
\begin{equation}
    \label{NMI}
    \textit{NMI}(\mathbb{X},\mathbb{Y}) = \dfrac{2I(\mathbb{X},\mathbb{Y})}{H(\mathbb{X})+H(\mathbb{Y})}    
    \vspace{-1mm}
\end{equation}
where the function $I(\cdot)$ measures the degree of information sharing between the two clusters and the function $H(\cdot)$ is used to measure the uncertainty of elements in clusters.
\begin{equation}
    \label{mutual information}
    \textit{I}(\mathbb{X},\mathbb{Y}) = \sum_{i=1}^{|\mathbb{X}|} \sum_{j=1}^{|\mathbb{Y}|} p(i,j)\log\left(\dfrac{p(i,j)}{p(i)p(j)}\right)   
    \vspace{-1mm}
\end{equation}
\begin{equation}
    \label{entropy}
    \textit{H}(\mathbb{X}) = -\sum_{i=1}^{|\mathbb{X}|}p(i)\log p(i)
    \vspace{-1mm}
\end{equation}
where \( p(i) \) is the probability of element \( i \) being assigned to cluster \( \mathcal{X}_i \), i.e., \( p(i) = \frac{|\mathcal{X}_i|}{|\mathcal{R}|} \), and \( p(i,j) \) is the joint probability of element \( i \) correctly being assigned to both the predicted cluster \( \mathcal{X}_i \) and the ground truth cluster \( \mathcal{Y}_j \), i.e., \( p(i,j) = \frac{|\mathcal{X}_i \cap \mathcal{Y}_j|}{|\mathcal{R}|} \).

Different from the above, ARI considers both the similarity and dissimilarity of all pairs of data records. The definition of ARI in our work is the same as~\cite{wu2019deep}. 
Consider a contingency table [$t_{ij}$] of overlaps between $\mathcal{X}_i$ and $\mathcal{Y}_j$. Each element in [$t_{ij}$] shows the number of overlapping objects between $\mathcal{X}_i$ and $\mathcal{Y}_j$. We denote by $a_i$ the number of elements in $\mathcal{X}_i$, i.e., $a_i=|\mathcal{X}_i|$,  where \( b_j \) also represents the cardinality of \( \mathcal{Y}_j \), i.e., \( b_j = |\mathcal{Y}_j| \).

\begin{equation}
    \label{map-func}
        \textit{ARI}=\dfrac{\displaystyle \sum_{i,j} \binom{t_{i,j}}{2} - \left[\sum_i \binom{a_i}{2}\sum_j \binom{b_j}{2}\right]/ \binom{n}{2}}{\frac{1}{2} \displaystyle \left[\sum_i \binom{a_i}{2}+\sum_j \binom{b_j}{2}\right]-\left[\sum_i \binom{a_i}{2}\sum_j \binom{b_j}{2}\right]/ \binom{n}{2}}
\end{equation}

\begin{table}[h]
      \caption{Optimal key factor values with different LLMs}
      \vspace{-3mm}
      \label{tab:values of llms}
      \small
      \begin{tabular}{m{1.5cm}<{\centering}|m{2.1cm}<{\centering}|m{1.9cm}<{\centering}}
        \hline 
        $\textbf{Factors}$ & $\textbf{GPT 4o-mini-8B}$ & $\textbf{Llama 3.2-1B}$  \\
        \hline \hline
        $S_s$  & 9 & 6\\
        \hline
        $S_d$  & 4 & 3 \\
        \hline
        $S_v$  & $\sim$ 0 & $\sim$ 0 \\
        \hline
        $\textit{Orders}$  & Sequential & Sequential \\
        \hline
      \end{tabular}
      \vspace{-1mm}
\end{table}

\begin{table}[t]
\vspace{-1mm}
\caption{Comparisons of {\sf LLM-CER} using GPT and Llama}
\vspace{-3mm}
\label{tab:performance of llms}
\centering
\small
\begin{tabular}{m{1.5cm}<{\centering}|m{1.9cm}<{\centering}|m{1.3cm}<{\centering}|m{1.3cm}<{\centering}}
\hline
{\textbf{Dataset}} & {\textbf{Metrics}} & {\textbf{GPT}} & {\textbf{Llama}} 
        \\ \hline \hline
\multirow{5}{*}{ \em{Alaska} } & ACC & \textbf{0.82} & 0.64  \\ \cline{2-4}
                          & FP & \textbf{0.79} & 0.46 \\ \cline{2-4}
                          & NMI & \textbf{0.79} & 0.48  \\ \cline{2-4}
                          & ARI & \textbf{0.65} & 0.41 \\ \cline{2-4}
                          & \# API Calls & \textbf{2043} & 3215 \\
                          \hline \hline
\multirow{5}{*}{ \em{AS} } & ACC & \textbf{0.70} & 0.42  \\ \cline{2-4}
                          & FP & \textbf{0.63} & 0.52 \\ \cline{2-4}
                          & NMI & \textbf{0.73} & 0.46  \\ \cline{2-4}
                          & ARI & \textbf{0.62} & 0.49 \\ \cline{2-4}
                          & \# API Calls & \textbf{413} & 685  \\
                          \hline \hline
\multirow{5}{*}{ \em{Song} } & ACC & \textbf{0.72} & 0.45 \\ \cline{2-4}
                          & FP & \textbf{0.78} & 0.52 \\ \cline{2-4}
                          & NMI & \textbf{0.74} & 0.53  \\ \cline{2-4}
                          & ARI & \textbf{0.66} & 0.48 \\ \cline{2-4}
                          & \# API Calls & \textbf{668} & 1025\\
                          \hline \hline
\multirow{5}{*}{ \em{Music} } & ACC & \textbf{0.71} & 0.52  \\ \cline{2-4}
                          & FP & \textbf{0.61} & 0.57 \\ \cline{2-4}
                          & NMI & \textbf{0.74} & 0.53  \\ \cline{2-4}
                          & ARI & \textbf{0.62} & 0.45 \\ \cline{2-4}
                          & \# API Calls & \textbf{3859} & 5745 \\
                          \hline \hline
\multirow{5}{*}{ \em{DG} } & ACC & \textbf{0.81} & 0.49 \\ \cline{2-4}
                          & FP & \textbf{0.70} & 0.57 \\ \cline{2-4}
                          & NMI & \textbf{0.84} & 0.51  \\ \cline{2-4}
                          & ARI & \textbf{0.68} & 0.49 \\ \cline{2-4}
                          & \# API Calls & \textbf{1285} & 1865 \\
                          \hline \hline
\multirow{5}{*}{ \em{Cora} } & ACC & \textbf{0.90} & 0.63  \\ \cline{2-4}
                          & FP & \textbf{0.71} & 0.48 \\ \cline{2-4}
                          & NMI & \textbf{0.82} & 0.52  \\ \cline{2-4}
                          & ARI & \textbf{0.69} & 0.43 \\ \cline{2-4}
                          & \# API Calls & \textbf{279} & 412 \\
                          \hline \hline
\multirow{5}{*}{ \shortstack{\em{Cite-}\\\em{seer}} } & ACC & \textbf{0.88} & 0.61 
 \\ \cline{2-4}
                          & FP & \textbf{0.95} & 0.58  \\ \cline{2-4}
                          & NMI & \textbf{0.85} & 0.59  \\ \cline{2-4}
                          & ARI & \textbf{0.74} & 0.61 \\ \cline{2-4}
                          & \# API Calls & \textbf{1302} & 2005  \\
                          \hline 
\end{tabular}
\vspace{-3mm}
\end{table}

\stitle{A.1 Impact of Different LLMs on Optimal Key Factor Values and End-to-end ER Performance}

\vspace{-0.8mm}
\spara{Optimal Key Factor Values with Different LLMs.} Table \ref{tab:values of llms} shows optimal key factor values for GPT-4o mini-8B and Llama 3.2-1B, reflecting their distinct characteristics. For $S_s$ (record set size), GPT-4o mini’s optimal value is 9, indicating its ability to handle larger contexts due to its 8B parameters and better long-range dependency processing. Llama 3.2-1B, with 1B parameters, has an optimal $S_s$ of 6, performing better with smaller record sets due to its limited capacity. For $S_d$ (entities per record set), GPT-4o mini’s optimal value is 4, managing more entities effectively, while Llama’s is 3, reflecting its reduced ability to handle complex prompts. Both models prefer smaller $S_v$ ($\sim$0) and sequential record ordering, benefiting from coherence in ER tasks regardless of model size.

\vspace{-1mm}
\spara{End-to-end ER Performance Comparison.} 
Table~\ref{tab:performance of llms} compares in-context clustering-based ER performance of GPT and Llama across seven datasets, focusing on ACC, FP-measure, NMI, ARI, and API calls. Token usage, API costs, and time-based metrics are excluded, as Llama is open-source (no costs) and lacks network latency.
GPT consistently outperforms Llama in all quality metrics across all datasets, likely due to its larger model size and stronger context understanding. Llama requires more API calls, indicating greater computational resource use despite its cost-free nature. Larger models like GPT enhance clustering performance, but Llama remains a viable, cost-effective option when accuracy is less critical than cost.

\stitle{A.2 Impact of Entity Dispersion on Optimal Key Factor Values and End-to-end ER Performance}

\noindent
In this section, we investigate the impact of entity dispersion ($E_d$), defined as the average number of records per entity, on optimal in-context clustering parameters ($S_s$, $S_d$, $S_v$, and sequential/random ordering) and the performance of our end-to-end ER approach using the semantically diverse \textit{Cora} dataset.
In the first experiment, we fix 100 entities and vary duplicate records per entity to 4, 8, or 12, creating datasets with $E_d = 4, 8, 12$ (sizes 400, 800, 1200 records, respectively). This tests the effect of increasing record density per entity with constant entity count. In the second experiment, we fix total records at $\approx$600 and adjust entity counts to achieve $E_d = 12$ (50 entities, 12 records each), $E_d = 16$ (38 entities, 16 records each), and $E_d = 20$ (30 entities, 20 records each), examining how redistributing records across varying entity counts affects clustering.

\begin{table}[tb!]
      \caption{Optimal key factor values with entity dispersion} 
      \vspace{-3mm}
      \label{tab:values of diversity}
      \small
      \begin{tabular}{m{1.5cm}<{\centering}|m{1.2cm}<{\centering}|m{1.2cm}<{\centering}|m{1.2cm}<{\centering}}
        \hline 
        $\textbf{Factors}$ & $E_d=4$ & $E_d=8$ & $E_d=12$ \\
        \hline \hline
        $S_s$  & 8 & 9 & 9\\
        \hline
        $S_d$  & 4 & 4 & 4 \\
        \hline
        $S_v$  & $\sim$ 0 & $\sim$ 0 & $\sim$ 0 \\
        \hline
        $\textit{Orders}$  & Sequential & Sequential & Sequential \\
        \hline
      \end{tabular}
      \vspace{-1mm}
\end{table}

\begin{table}[tb!]
\caption{ End-to-end ER performance with varying $E_d$ (with \#Entity fixed to 100)} 
\label{tab:impact of dd on e2e}
\centering
\small
\vspace{-3mm}
\begin{tabular}{m{0.98cm}<{\centering}|c|c|c|c}
\hline
\multirow{2}{*}{\textbf{Dataset}} & \multirow{2}{*}{\textbf{Metrics}} & \multirow{2}{*}{$E_d=4$} & \multirow{2}{*}{\textbf{$E_d=8$}} & \multirow{2}{*}{\textbf{$E_d=12$}}\\
        $\textbf{}$ & $\textbf{}$ & $\textbf{}$ & $\textbf{}$ & $\textbf{}$
        \\ \hline \hline
\multirow{6}{*}{ \em{Cora} } & ACC & 0.87 & 0.84 &  0.89 \\ \cline{2-5}
                          & FP & 0.71 & 0.68 & 0.72 \\ \cline{2-5}
                          & Cost (\$) & 0.02 & 0.03 & 0.03  \\ \cline{2-5}
                          & Tokens (M) & 0.06 & 0.10 & 0.12  \\ \cline{2-5}
                          & Time (s) & 209.4 & 285.3 & 321.5  \\ 
                          \cline{2-5}
                          & API Calls & 192 & 251 & 277  \\
                          \hline 
\end{tabular}
\vspace{-2mm}
\end{table}

\begin{table}[tb!]
\caption{ End-to-end ER performance with varying $E_d$ (with \#Record fixed to 600)} 
\label{tab:e2e with fixed rec}
\centering
\small
\vspace{-3mm}
\begin{tabular}{m{0.98cm}<{\centering}|c|c|c|c}
\hline
\multirow{2}{*}{\textbf{Dataset}} & \multirow{2}{*}{\textbf{Metrics}} & \multirow{2}{*}{$E_d=12$} & \multirow{2}{*}{\textbf{$E_d=16$}} & \multirow{2}{*}{\textbf{$E_d=20$}}\\
        $\textbf{}$ & $\textbf{}$ & $\textbf{}$ & $\textbf{}$ & $\textbf{}$
        \\ \hline \hline
\multirow{6}{*}{ \em{Cora} } & ACC & 0.87 & 0.85 &  0.88 \\ \cline{2-5}
                          & FP & 0.73 & 0.73 & 0.72 \\ \cline{2-5}
                          & Cost (\$) & 0.02 & 0.02 & 0.01  \\ \cline{2-5}
                          & Tokens (M) & 0.06 & 0.05 & 0.04  \\ \cline{2-5}
                          & Time (s) & 163.2 & 134.5 & 114.6  \\ 
                          \cline{2-5}
                          & API Calls & 139 & 112 & 98  \\
                          \hline 
\end{tabular}
\vspace{-2mm}
\end{table}

\spara{Impact on Optimal Values of Key Factors.} Table \ref{tab:values of diversity} shows that entity dispersion ($E_d$) has little effect on optimal record set parameters across \textit{Cora} dataset with $E_d = 4, 8, 12$. Optimal $S_d$, $S_v$, and "Sequential" ordering remain consistent, with $S_s$ slightly dropping from 9 to 8 at $E_d = 4$. This stability reflects the LLM's strong semantic understanding, effectively distinguishing entities despite varying record counts. The minor $S_s$ adjustment at lower $E_d$ suggests that fewer records suffice for entity differentiation. These results underscore the method's adaptability to diverse datasets without extensive parameter tuning for LLM like GPT-4o-mini.

\spara{Impact on End-to-end ER Performance.} Table~\ref{tab:impact of dd on e2e} shows that higher entity dispersion ($E_d$) improves clustering performance with modestly increased resource usage. ACC slightly decreases from 0.87 at $E_d=4$ to 0.84 at $E_d=8$ but rises to 0.89 at $E_d=12$, with FP-measure showing a similar trend. Lower $E_d$ may lack sufficient duplicates for robust clustering, while higher $E_d$ leverages redundancy to resolve ambiguities and enhance stability. Despite larger datasets at higher $E_d$ (e.g., $E_d=12$ has $3\times$ records of $E_d=4$), resource use (costs, tokens, processing times, API calls) grows less than threefold, indicating efficiency from consolidating duplicates, as analyzed in \S \ref{sec:end2end}.

In the second scenario (Table~\ref{tab:e2e with fixed rec}), increasing $E_d$ from 12 to 20 minimally affects ACC and FP but significantly reduces resource consumption. API calls drop from 139 to 98, token usage from 0.06M to 0.04M, and costs from \$0.02 to \$0.01. Higher $E_d$ maintains clustering quality with lower computational overhead, improving efficiency, consistent with \S \ref{sec:end2end}.

\vspace{-2.5mm}
\begin{table}[h]
\caption{Impact of blocking and filtering}
\label{tab:impact of blocking}
\centering
\small
\vspace{-3mm}
\begin{tabular}{m{0.98cm}<{\centering}|c|c|c|c|c}
\hline
\multirow{2}{*}{\textbf{Dataset}} & \multirow{2}{*}{\textbf{Metrics}} & {\textbf{w/o}} & \multirow{2}{*}{\textbf{Filter}} & \multirow{2}{*}{\textbf{Canopy}} & \multirow{2}{*}{\textbf{LSH}} \\
        $\textbf{}$ & $\textbf{}$ & $\textbf{blocking}$ & $\textbf{}$ & $\textbf{}$ & $\textbf{}$
        \\ \hline \hline
\multirow{6}{*}{ \em{Cora} } & ACC & 0.62 & 0.81 & 0.67 & \textbf{0.90} \\ \cline{2-6}
                          & FP & 0.58 & 0.78 & 0.60 & \textbf{0.71} \\ \cline{2-6}
                          & Cost (\$) & 0.33 & 0.03 & 0.06 & \textbf{0.03} \\ \cline{2-6}
                          & Tokens (M) & 1.33 & 0.13 & 0.22 & \textbf{0.12} \\ \cline{2-6}
                          & Time (s) & 3708.8 & 326.1 & 529.6 & \textbf{325.5} \\ 
                          \cline{2-6}
                          & API Calls & 1996 & 301 & 440 & \textbf{279} \\
                          \hline 
\multirow{6}{*}{ { \em{AS} } } & ACC & 0.61 & 0.68 & 0.66 & \textbf{0.70} \\ \cline{2-6}
                          & FP & 0.58 & \textbf{0.64} & 0.60 & 0.63\\ \cline{2-6}
                          & Cost (\$) & 0.11 & \textbf{0.02} & 0.03 & \textbf{0.02} \\ \cline{2-6}
                          & Tokens (M) & 0.34 & \textbf{0.07} & 0.09 & \textbf{0.07} \\ \cline{2-6}
                          & Time (s) & 1542.3 & \textbf{305.2} & 406.5 & 325.5 \\ 
                          \cline{2-6}
                          & API Calls & 2156 & \textbf{402} & 526 & 413 \\
                          \hline 
\multirow{6}{*}{ { \em{Alaska} } } & ACC & 0.70 & 0.77 & 0.74 & \textbf{0.82} \\ \cline{2-6}
                          & FP & 0.69 & 0.74 & 0.72 & \textbf{0.79} \\ \cline{2-6}
                          & Cost (\$) & 0.82 & 0.17 & 0.18 & \textbf{0.15} \\ \cline{2-6}
                          & Tokens (M) & 3.85 & 0.80 & 0.84 & \textbf{0.73} \\ \cline{2-6}
                          & Time (s) & 12452.5 & 2612.5 & 2745.2 & \textbf{2374.2} \\ 
                          \cline{2-6}
                          & API Calls & 11542 & 2252 & 2354 & \textbf{2043} \\
                          \hline 
\end{tabular}
\vspace{-4mm}
\end{table}

\stitle{A.3 Impact of Blocking and Filtering Techniques} 

\noindent
In this section, we evaluate the impact of different blocking methods on both the performance (e.g., ACC, FP-measure) and resource consumption (e.g., tokens, cost, time, and number of API calls) of our algorithm on three representative datasets: \textit{Cora}, \textit{AS}, and \textit{Alaska}. 
All three blocking methods are unsupervised and are applied in an alternating manner throughout our experiments.

 Table~\ref{tab:impact of blocking} shows that blocking significantly reduces resource consumption compared to the no-blocking baseline, with varying performance across methods. For \textit{Cora}, no blocking yields low ACC and FP-measure ($\approx0.6$) with high resource use (>1M tokens, long runtime). \textsf{LSH} performs best, achieving ACC near 0.9 and FP-measure around 0.7 with minimal resources (<0.2M tokens, hundreds of seconds). \textsf{Filtering} follows with ACC near 0.8 and FP-measure close to 0.8, also resource-efficient. \textsf{Canopy} slightly improves over no blocking but uses more resources than \textsf{LSH} and \textsf{Filtering}. These trends hold for \textit{AS} and \textit{Alaska}.

The superior performance of \textsf{LSH} stems from forming high-quality clusters by capturing semantic similarities, while \textsf{Canopy}'s coarser clusters reduce accuracy. No blocking is costly and less effective due to noise and unstructured input. These findings highlight blocking’s role in enhancing efficiency and the need to choose suitable methods to optimize both performance and resource use in LLM-based ER pipelines.

\begin{table}[t]
\vspace{-1mm}
\caption{Effect of MDG algorithm and record set regeneration on end-to-end performance: NMI and ARI}
\label{tab:effect_MDG_nmi}
\vspace{-3mm}
    \centering
    \resizebox{8.5cm}{!}{
    \renewcommand{\arraystretch}{1.2}
    \begin{tabular}{c|c|c|c|c|c|c}
    \hline
    \multirow{2}{*}{\textbf{Metrics}} & \multicolumn{2}{c|}{\em{Cora}} & \multicolumn{2}{c|}{\em{Alaska}} & \multicolumn{2}{c}{\em{AS}} \\
    \cline{2-7}
     & {\em {w/o MDG}} & {\em {w/ MDG}} & {\em {w/o MDG}} & {\em {w/ MDG}} & {\em {w/o MDG}} & {\em {w/ MDG}} \\
     
    \hline
    \textbf{NMI} & 0.61 & {\bf 0.82} & 0.52 & {\bf 0.79} & 0.54 & {\bf 0.73} \\
    \hline
    \textbf{ARI} & 0.48 & {\bf 0.69} & 0.33 & {\bf 0.65} & 0.48 & {\bf 0.62} \\
    \hline
    \end{tabular}
}
\vspace{-4mm}
\end{table}

\vspace{-1mm}
\stitle{A.4 LLM Guardrails on Misclustering Detection and Record
Set Regeneration: MRI and ARI}

The results in Table \ref{tab:effect_MDG_nmi} show that the incorporation of the {\sf MDG} algorithm also leads to improvements in NMI and ARI. 
In addition to ACC and FP-measure, NMI and ARI also show substantial improvements, indicating that {\sf MDG} not only enhances classification accuracy but also produces more consistent and accurate clustering results. 
These improvements can be attributed to the ability of {\sf MDG} to more effectively prevent misclustering, ensuring that entities which should remain separate are not incorrectly merged. This is reflected in higher values of the clustering metrics (NMI and ARI), indicating better alignment between predictions and truths.

\vspace{-0.8mm}
\stitle{A.5 Comparisons with PLM-based ER Methods}
\vspace{-0.8mm}

 \spara{Baselines.} We compare our algorithm with two widely-used PLM (pre-trained language model)-based ER methods, {\sf Ditto}~\cite{0001LSDT20} and {\sf DeepMatcher}~\cite{18sigmodEM}, both of which follow a pairwise matching paradigm. We use transitivity and anti-transitivity to obtain their final clustering results for comparison.

\begin{itemize}[left=4pt]
\vspace{-0.7mm}
\item \textbf{Ditto~\cite{0001LSDT20}}:  \textsf{Ditto}~\cite{0001LSDT20} is a PLM-based entity resolution method that fine-tunes a pre-trained language model using labeled entity pairs. By leveraging task-specific supervision, it enhances entity matching performance. 
To ensure a fair comparison, we use the original implementation and default hyperparameters from the paper. 
\item \textbf{DeepMatcher~\cite{18sigmodEM}}: \textsf{DeepMatcher}~\cite{18sigmodEM} is a deep learning framework for entity and text matching, offering pre-built neural networks and utilities for training advanced models. It excels in resolving complex entity alignment and text correspondence tasks with high precision. Given that the authors have released it as a Python library, we adopt it directly for evaluation. 
\vspace{-0.7mm}
\end{itemize}

\spara{Implementation Details and Cost Evaluation.}
 {\sf Ditto} \cite{0001LSDT20} and {\sf DeepMatcher} \cite{18sigmodEM} are PLM-based ER methods designed for general applicability across diverse datasets. To systematically evaluate their performance, we consider two alternative modes: 
(1) {\em Without fine-tuning}: The pre-trained models are used directly for ER without additional training, we report the ACC and FP-measure obtained from applying these models to the entire dataset (100\% treated as test data). Additionally, we report the monetary cost (discussed below) required for inference over the whole dataset; 
and (2) {\em With fine-tuning}: The models are fine-tuned using 20\%/80\% of the ground truth pairs for each dataset to adapt to domain-specific variations, the remaining 80\%/20\% of the data is used as test data for obtain ACC and FP-measure. We choose to use 80\% of the data for training, as performance improvements for PLM-based methods plateaued at this proportion across nearly all datasets.
This ensures a consistent evaluation framework across all approaches. Following~\cite{WLCHWZS25}, we estimate the monetary cost of PLM methods by incorporating both inference and additional fine-tuning costs, based on the required training time and the hourly price of cloud NVIDIA A40 GPUs\footnote{\url{https://www.runpod.io/pricing}}.  

For our proposed method, no fine-tuning is required and we report the results  over the entire dataset. Due to the network latency associated with LLM API calls, a direct comparison of our algorithm with PLM-based methods in terms of running time would be unfair, thus we omit time-based comparisons in this analysis.

\begin{table}[tb!]
\caption{ End-to-end comparison of our in-context clustering-based ER with PLM-based ER, "FT" denotes fine-tuning} 
\vspace{-2mm}
\label{tab:performance of plm}
\footnotesize
\centering
\begin{tabular}{m{0.68cm}<{\centering}|c|c|c|c|m{0.42cm}<{\centering}|c|c|m{0.42cm}<{\centering}}
\hline
\multirow{3}{*}{\textbf{Dataset}} & \multirow{3}{*}{\textbf{Metrics}} & \multirow{3}{*}{\textbf{Ours}} & \multicolumn{3}{c|}{\textbf{Ditto} \cite{0001LSDT20}} & \multicolumn{3}{c}{\textbf{DeepMatcher} \cite{18sigmodEM}}\\
    \cline{4-9}
     & {\em {}} & {\em {}} & {\em {w/ FT}} & {\em {w/ FT}} & {\em {w/o}} & {\em {w/ FT}} & {\em {w/ FT}} & {\em {w/o}} 
        \\ 
     & {\em {}} & {\em {}} & {\em {(20\%)}} & {\em {(80\%)}} & {\em {FT}} & {\em {(20\%)}} & {\em {(80\%)}} & {\em {FT}} \\    
    \hline \hline
\multirow{3}{*}{ \em{Alaska} } & ACC & \textbf{0.82} & 0.70 & 0.81 & 0.64 & 0.65 & 0.74 & 0.58 \\ \cline{2-9}
                          & FP & \textbf{0.79} & 0.62 & 0.77 & 0.55 & 0.51 & 0.70  & 0.43 \\ \cline{2-9}
                          & Cost (\$) & 0.15 & 65.67 & 260.21 & \textbf{0.12} & 65.81 & 260.93 & 0.14 \\ \cline{2-9}
                          \hline \hline
\multirow{3}{*}{ \em{AS} } & ACC & 0.70 & 0.60 & \textbf{0.72}  & 0.58 & 0.57 & 0.65  & 0.54 \\ \cline{2-9}
                          & FP & 0.63 & 0.53 & \textbf{0.66}  & 0.51 & 0.49 & 0.61  & 0.47 \\ \cline{2-9}
                          & Cost (\$) & \textbf{0.02} & 7.06 & 28.85  & 0.08 & 7.18 & 29.29  & 0.07 \\ \cline{2-9}
                          \hline \hline
\multirow{3}{*}{ \em{Song} } & ACC & \textbf{0.72} & 0.61 &  0.70  & 0.54 & 0.59 & 0.66  & 0.53 \\ \cline{2-9}
                          & FP & \textbf{0.78} & 0.63 & 0.76  & 0.58 & 0.52 & 0.72  & 0.51 \\ \cline{2-9}
                          & Cost (\$) & \textbf{0.06} & 13.74 & 41.34 & 0.09 & 12.82 & 38.65  & 0.15 \\ \cline{2-9}
                          \hline \hline
\multirow{3}{*}{ \em{Music} } & ACC & 0.71 & 0.63 & \textbf{0.74}  & 0.57 & 0.59 & 0.66  & 0.55 \\ \cline{2-9}
                          & FP & 0.61 & 0.53 & \textbf{0.65}  & 0.44 & 0.52 &  0.56  & 0.47 \\ \cline{2-9}
                          & Cost (\$) & 0.19 & 79.30 &  239.22  & 0.22 & 79.08 & 238.54   & \textbf{0.14} \\ \cline{2-9}
                          \hline \hline
\multirow{3}{*}{ \em{DG} } & ACC & \textbf{0.81} & 0.68 & 0.80  & 0.60 & 0.67 &  0.77  & 0.60 \\ \cline{2-9}
                          & FP & \textbf{0.70} & 0.60 & 0.69  & 0.52 & 0.53 & 0.66   & 0.50 \\ \cline{2-9}
                          & Cost (\$) & \textbf{0.07} & 12.93 &  42.04 & 0.13 & 13.01 &  42.22  & 0.16 \\ \cline{2-9}
                          \hline \hline
\multirow{3}{*}{ \em{Cora} } & ACC & \textbf{0.90} & 0.76 &  \textbf{0.90} & 0.67 & 0.71 &  0.88  & 0.66 \\ \cline{2-9}
                          & FP & 0.71 & 0.56 &  \textbf{0.72}  & 0.48 & 0.54 &  0.70  & 0.50 \\ \cline{2-9}
                          & Cost (\$) & \textbf{0.03} & 10.71 & 42.92  & 0.07 & 10.74 &  43.08 & 0.06 \\ \cline{2-9}
                          \hline \hline
\multirow{3}{*}{ \shortstack{\em{Cite-}\\\em{seer}} } & ACC &                                   0.88 & 0.76 & \textbf{0.89} & 0.62 & 0.60 &  0.73 & 0.59  \\ \cline{2-9}
                          & FP & \textbf{0.95} & 0.70 &  0.93 & 0.65 & 0.67 &  0.91 & 0.59 \\ \cline{2-9}
                          & Cost (\$) & \textbf{0.03} & 9.62 & 38.32  & 0.08 & 9.73 & 39.04  & 0.09 \\ \cline{2-9}
                          \hline \hline
\multirow{3}{*}{ \em{AG} } & ACC &                                  0.71  & 0.61 & \textbf{0.73} & 0.56 & 0.58 &  0.69 & 0.57  \\ \cline{2-9}
                          & FP & 0.64 & 0.57 &  \textbf{0.68} & 0.50 & 0.54 &  0.64 & 0.49 \\ \cline{2-9}
                          & Cost (\$) & \textbf{0.02} & 7.13 & 28.84  & 0.07 & 7.16 & 28.04  & 0.05 \\ \cline{2-9}
                          \hline  \hline
\multirow{3}{*}{ \em{WA} } & ACC &                                   0.61 & 0.56 & \textbf{0.65} & 0.48 & 0.50 &  0.60 & 0.44  \\ \cline{2-9}
                          & FP & 0.56 & 0.45 &  \textbf{0.60} & 0.40 & 0.44 &  0.57 & 0.39 \\ \cline{2-9}
                          & Cost (\$) & \textbf{0.02} & 6.98 & 28.03  & 0.06 & 6.89 & 27.96  & 0.06 \\ \cline{2-9}
                          \hline 
\end{tabular}
\vspace{-3mm}
\end{table}

\spara{Experiment Results.} 
 Table \ref{tab:performance of plm} shows that {\sf LLM-CER} outperforms PLM-based approaches like {\sf Ditto} and {\sf DeepMatcher} across datasets without fine-tuning (w/o FT) or with 20\% FT. Even with 80\% FT, our method achieves comparable performance at significantly lower costs. On the \textit{Alaska} dataset, our approach yields an ACC of 0.82 and FP-measure of 0.79, surpassing {\sf Ditto} with 20\% FT (ACC: 0.70, FP: 0.62) and {\sf DeepMatcher} with 20\% FT (ACC: 0.65, FP: 0.51). With 80\% FT, {\sf Ditto} (ACC: 0.81, FP: 0.77) and {\sf DeepMatcher} (ACC: 0.74, FP: 0.70) are close but still trail our method. Non-fine-tuned PLMs perform poorly, with {\sf Ditto} at 0.64 ACC and 0.55 FP, and {\sf DeepMatcher} at 0.58 ACC and 0.43 FP. This trend holds across datasets, likely due to our LLM’s contextual generalization, unlike PLMs’ reliance on fine-tuning. Monetarily, our method is far more efficient. On \textit{Alaska}, our cost is \$0.15, compared to \$260.21 for {\sf Ditto} (80\% FT) and \$260.93 for {\sf DeepMatcher} (80\% FT)—roughly 0.06\% of theirs. With 20\% FT, costs are \$65.67 for {\sf Ditto} and \$65.81 for {\sf DeepMatcher}, still significantly higher. Non-fine-tuned PLMs have closer costs (e.g., \$0.12 for {\sf Ditto} w/o FT) to ours, but poor performance.

 Overall, {\sf LLM-CER} balances high clustering quality and low cost, outperforming PLMs with 20\% FT or w/o FT, and matching 80\% FT PLMs at a fraction of the expense. While 80\% FT PLMs may slightly excel on some datasets, their marginal gains are cost-prohibitive, making our approach ideal for precision and cost-efficiency. 

\begin{figure}[h]
\centering
\vspace{-2mm}
\includegraphics[width=0.20\textwidth]{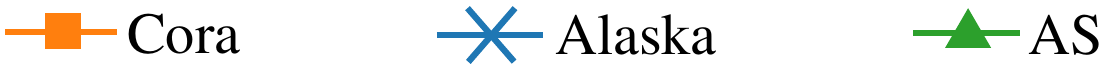}\\
  \vspace{-2mm}
  \hspace{-7mm}
  \subfigcapskip=-2mm
  \subfigure[Impact of Temp.]{
   \includegraphics[width=1.17in]{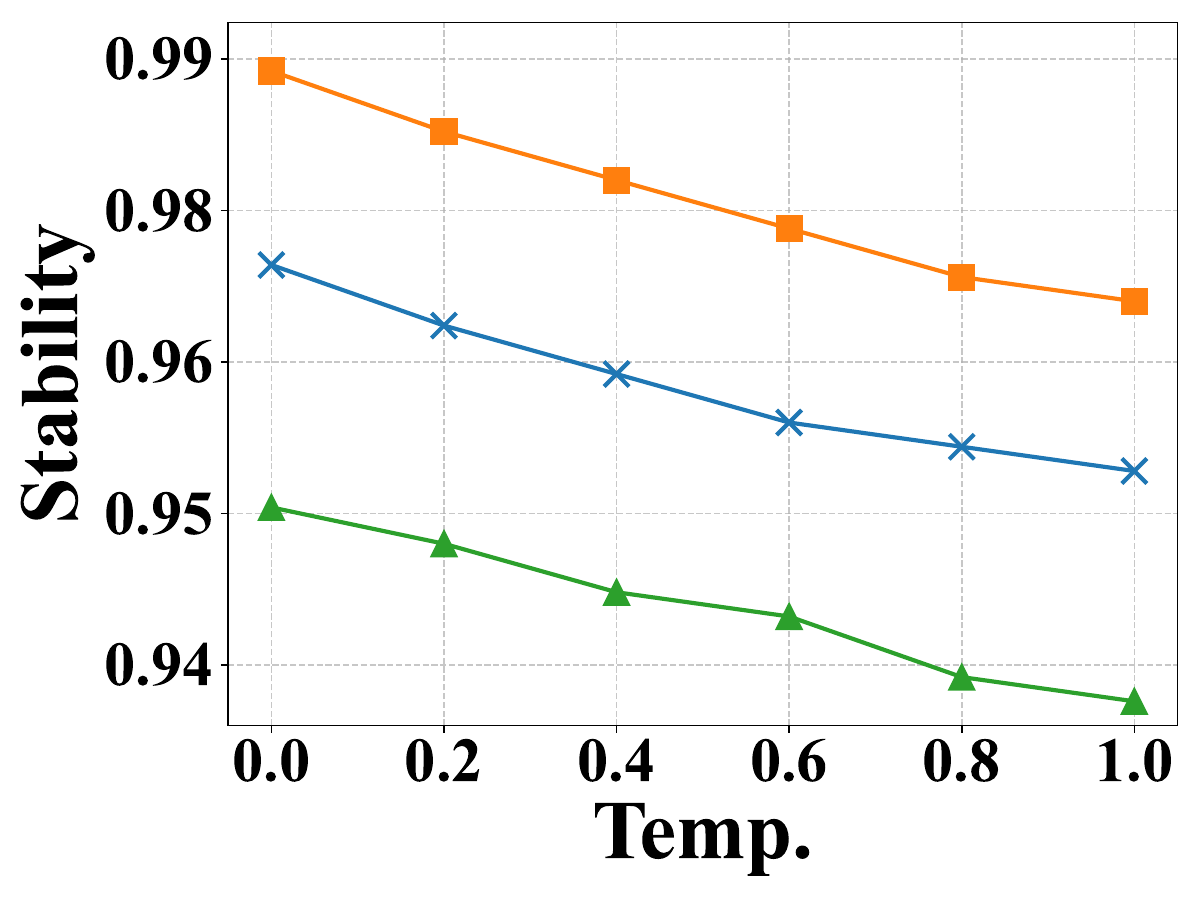}
   \label{fig:stable-temperature}
  }
 \hspace{ -3mm}
  \subfigure[Impact of context length]{
   \includegraphics[width=1.17in]{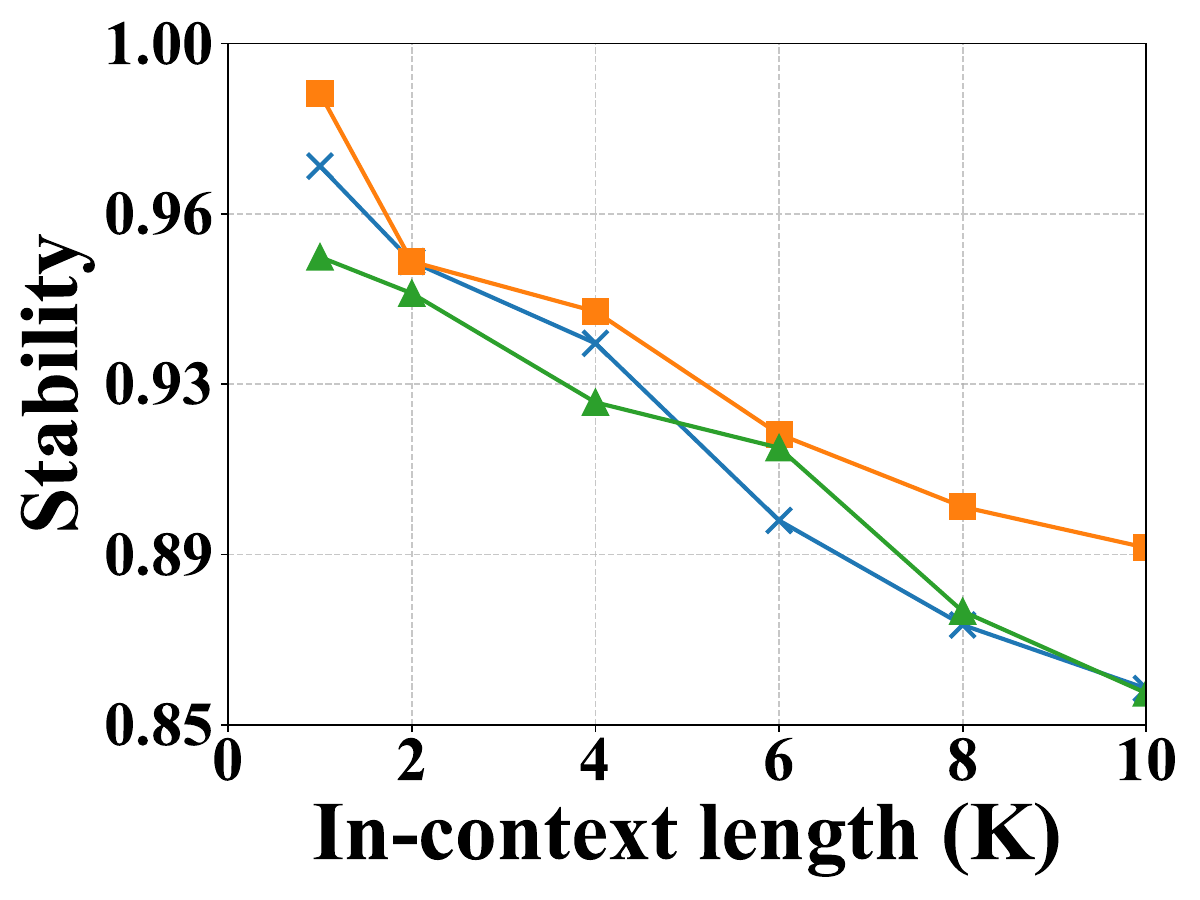} 
   \label{fig:stable-context}
  }
  \hspace{-3mm}
  \subfigure[Impact of Few-shot]{
   \centering
   \includegraphics[width=1.17in]{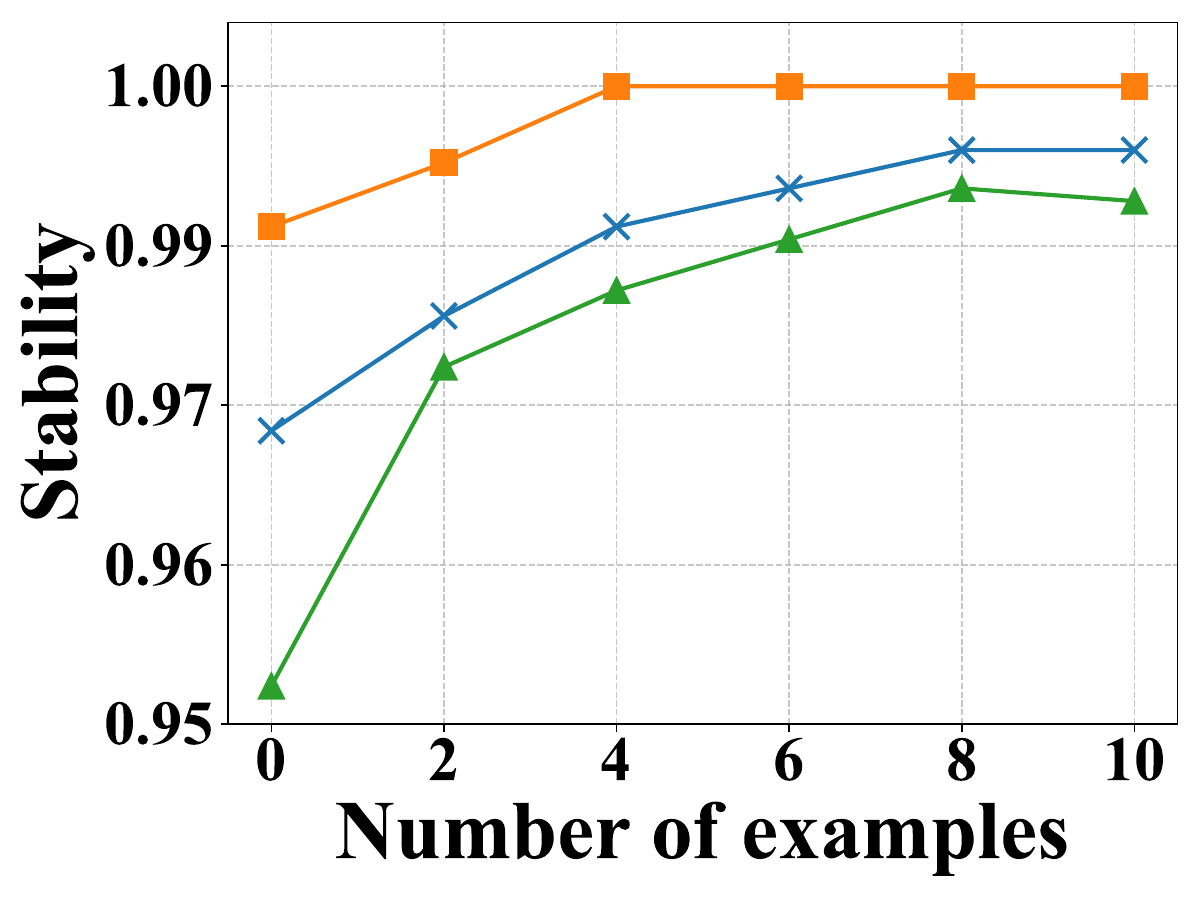}
   \label{fig:stable-few-shot}
  }
\vspace{-5mm}
\caption{ \centering {Stability of LLMs in in-context clustering under different settings} } 
\label{fig:impact of stability}
\vspace{-4.5mm}
\end{figure}

\stitle{A.6 Evaluating LLM In-context Clustering Stability in ER}

\noindent
To assess the stability of LLMs in ER tasks, we evaluate classification consistency across 50 repeated trials for a 40-query set with ground truth. We query the LLM using identical prompts and record sets, measuring stability via the Stability Score (identical responses divided by total queries, averaged across questions). We vary context length (record set size), LLM temperature, and few-shot example counts, with defaults of temperature 0, zero few-shot examples, and optimal record set size $S_s$ (9, $\approx1k$ tokens).

Figure \ref{fig:impact of stability} shows stability trends across \textit{Alaska}, \textit{Cora}, and \textit{AS} datasets. The first subplot indicates that Stability Scores slightly decrease as temperature rises from 0 to 1: \textit{Alaska} from 0.97 to 0.95, \textit{Cora} from 0.99 to 0.97, and \textit{AS} from 0.95 to 0.93, reflecting increased randomness. The second subplot shows Stability Scores declining with context length (1k to 10k tokens): \textit{Alaska} from 0.97 to 0.86, \textit{Cora} from 0.99 to 0.89, and \textit{AS} from 0.95 to 0.86, likely due to ambiguity from larger record sets, consistent with \S\ref{sec:experiment on key factors}. The third subplot reveals that more few-shot examples improve Stability Scores, enhancing generalization and decision consistency. 

In summary, lower temperatures and shorter context lengths enhance LLM stability in ER clustering, while few-shot examples boost consistency. Tuning prompt parameters is crucial for balancing randomness, cognitive load, and task guidance to optimize LLM reliability in ER tasks.

\begin{figure}[t]
\centering
\includegraphics[width=0.19\textwidth]{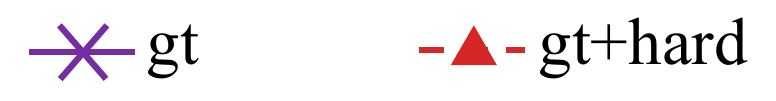}\\
  \vspace{-2mm}
  \hspace{-7.5mm}
  \subfigcapskip=-2mm
  \subfigure[Walmart-Amazon]{
   \includegraphics[width=1.75in]{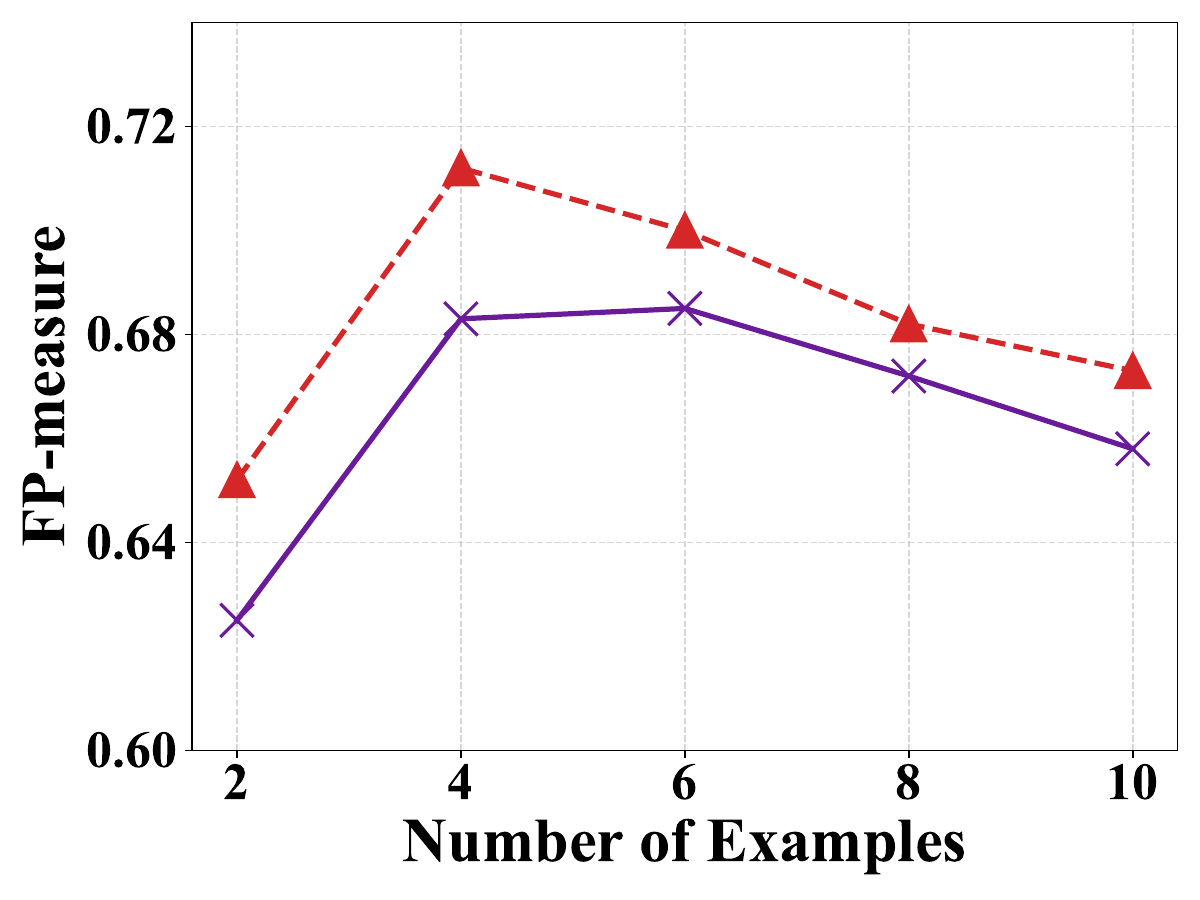}
   \label{fig:few-shot-wa}
  }
  \hspace{-2.7mm}
  \subfigure[Citeseer]{
   \centering
   \includegraphics[width=1.75in]{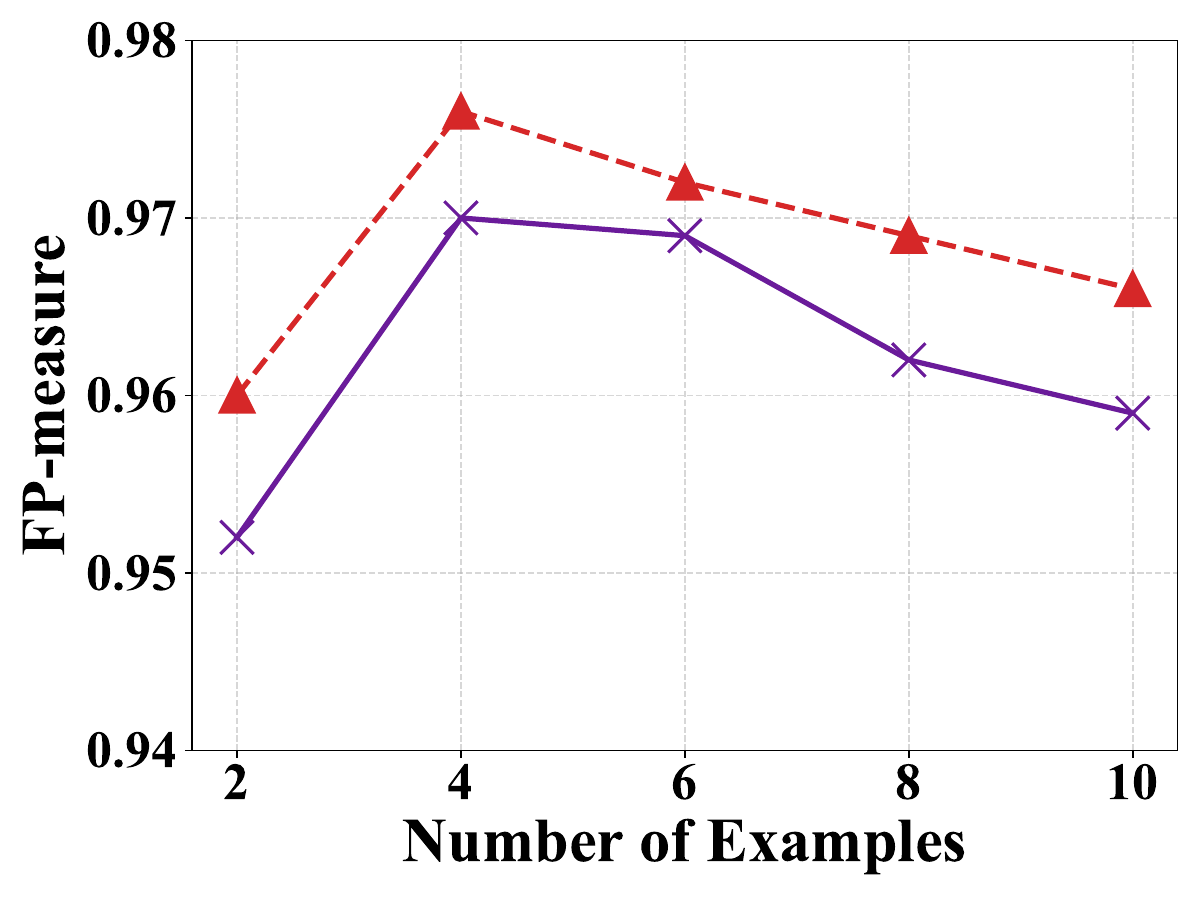}
   \label{fig:few-shot-citeseer}
  }
\vspace{-5mm}
\caption{ \centering {Impact of few-shot example counts} } 
\vspace{-5mm}
\label{fig:few-shot-num}
\end{figure}

\stitle{A.7 Few-shot Learning: Out-of-Domain Situation Exploration}

\noindent
In this section, we assess the role of few-shot learning in addressing out-of-domain generalization in end-to-end ER, using the challenging, heterogeneous \textit{Walmart-Amazon} dataset and the simpler, structured \textit{Citeseer} dataset. 

To adapt the LLM to domain-specific contexts, we create few-shot prompts with labeled examples from ground truth and hard examples (where the LLM often errs) to guide similarity judgments. We first examine how the number of few-shot examples affects ER performance under two sampling strategies: random sampling from ground truth (gt) and a balanced mix of ground truth and hard examples (gt + hard). Using the optimal example count, we analyze the contributions of few-shot learning and {\sf MDG} to performance, identifying conditions where each excels in end-to-end ER.

Our objective is to evaluate how few-shot learning improves {\sf LLM-CER}, focusing on its ability to address domain shift in complex, out-of-domain scenarios. We quantify its role in bridging the domain knowledge gap, compare its impact with {\sf MDG}, and identify conditions where few-shot learning is most effective. Additionally, we determine the optimal number of few-shot examples to maximize clustering performance while maintaining token efficiency and overall effectiveness.

\spara{Impact of the Number of Examples in Few-Shot Setting.} Figure \ref{fig:few-shot-num} shows that on the \textit{Citeseer} dataset, both few-shot modes slightly improve FP-measure from 0.95 to a peak of 0.97 with 4–6 examples, with little gain beyond. On the challenging \textit{Walmart-Amazon} dataset, FP-measure starts at 0.63–0.65, rising to 0.70–0.71 with 4–6 examples. The mixed mode (ground truth + hard examples) consistently outperforms the ground truth-only mode on \textit{Walmart-Amazon}, emphasizing hard examples’ value in domain-specific settings. However, performance slightly drops with 8–10 examples on both datasets, indicating that overly long contexts may impair the LLM’s use of few-shot prompts, reducing ER effectiveness.

\spara{Impact of Few-shot Setting.} In this experiment, we evaluate the impact of few-shot learning, with and without {\sf MDG}, compared to zero-shot learning, using a mixed example strategy (ground truth and hard examples) with four few-shot examples for optimal FP-measure and low token usage. 

As shown in Table~\ref{tab:impact of few shot}, on the domain-specific \textit{Walmart-Amazon} dataset, few-shot learning with {\sf MDG} significantly boosts ER performance, increasing ACC from 0.61 (zero-shot) to 0.77 and FP-measure from 0.56 to 0.71, though token usage rises from 0.06M to 0.27M. Without {\sf MDG}, performance declines (ACC: 0.58, FP: 0.52), underscoring {\sf MDG}'s critical role in effective clustering.

On the simpler \textit{Citeseer} dataset, few-shot learning with {\sf MDG} yields modest gains, with ACC improving from 0.88 to 0.90 and FP-measure from 0.95 to 0.97, while token usage increases from 0.13M to 0.60M. {\sf MDG} consistently enhances performance across both datasets, providing substantial benefits with minimal additional cost, particularly for \textit{Citeseer}. These findings highlight few-shot learning's value in complex domains and {\sf MDG}'s role as an efficient, effective enhancement.

\begin{table}[tb!]
\caption{Quantifying the impact of few-shot learning}
\label{tab:impact of few shot}
\centering
\small
\vspace{-3mm}
\begin{tabular}{m{1.05cm}<{\centering}|c|c|c|c}
\hline
\multirow{2}{*}{\textbf{Dataset}} & \multirow{2}{*}{\textbf{Metrics}} & \multirow{2}{*}{\textbf{Zero-Shot}} & {\textbf{Few-Shot}} & {\textbf{Few-Shot}}\\
        $\textbf{}$ & $\textbf{}$ & $\textbf{}$ & $\textbf{w/o MDG}$ & $\textbf{w/ MDG}$
        \\ \hline \hline
\multirow{6}{*}{ \shortstack{\em{Walmart-}\\\em{Amazon}} } & ACC & 0.61 & 0.58 & \textbf{0.77} \\ \cline{2-5}
                          & FP & 0.56 & 0.52 & \textbf{0.71} \\ \cline{2-5}
                          & Cost (\$) & \textbf{0.02} & 0.08 &  0.09 \\ \cline{2-5}
                          & Tokens (M) & \textbf{0.06} & 0.24 &  0.27 \\ \cline{2-5}
                          & Time (min) & 6.25 & \textbf{5.62} &  6.32 \\ 
                          \cline{2-5}
                          & API Calls & 398 & \textbf{332} &  365 \\
                          \hline 
\multirow{6}{*}{ \shortstack{\em{Cite-}\\\em{seer}} } & ACC & 0.88 & 0.74 & \textbf{0.90} \\ \cline{2-5}
                          & FP & 0.95 & 0.84 & \textbf{0.97} \\ \cline{2-5}
                          & Cost (\$) & \textbf{0.03} & 0.13 &  0.14 \\ \cline{2-5}
                          & Tokens (M) & \textbf{0.13} & 0.58 & 0.60 \\ \cline{2-5}
                          & Time (min) & 22.68 & \textbf{21.87} &  24.02 \\ 
                          \cline{2-5}
                          & API Calls & 1302 & \textbf{1174} &  1214 \\
                          \hline 
\end{tabular}
\vspace{-2mm}
\end{table}

\stitle{A.8 Advantages of Merging Similar Clusters}

\begin{table}[t]
\caption{ Comparison with random clusters merging} 
\label{tab:random merge ablation}
\vspace{-3.5mm}
    \centering
    \resizebox{8.5cm}{!}{
    \renewcommand{\arraystretch}{1.2}
    \begin{tabular}{c|c|c|c|c|c|c}
    \hline
    \multirow{2}{*}{\textbf{Metrics}} & \multicolumn{3}{c|}{\em{Cora}} & \multicolumn{3}{c}{\em{Alaska}}  \\
    \cline{2-7}
     & $\textbf{Sim.}$ & $\textbf{Ran.}$ & $\textbf{Ran. w/o MDG}$   & $\textbf{Sim.}$ & $\textbf{Ran.}$ & $\textbf{Ran. w/o MDG}$  \\
    \hline
    \textbf{ACC} & {\bf 0.90} & 0.87 ($\pm$0.03) &  0.61 ($\pm$0.08) & {\bf 0.82} & 0.79 ($\pm$0.02) & 0.39 ($\pm$0.11) \\
    \hline
    \textbf{FP-measure} & {\bf 0.71} & 0.69 ($\pm$0.02) &  0.57 ($\pm$0.07)  & {\bf 0.79} & 0.77 ($\pm$0.01) &  0.48 ($\pm$0.08) \\
    \hline
    \textbf{Cost (USD)} & {\bf 0.03} & 0.04 ($\pm$0.01) & 0.03 ($\pm$0.01)  & {\bf 0.15} & 0.16 ($\pm$0.02) & 0.15 ($\pm$0.02) \\
    \hline
    \textbf{Tokens (M)} & {\bf 0.12} & 0.15 ($\pm$0.04) & 0.11 ($\pm$0.02)  & {\bf 0.73} & 0.82 ($\pm$0.08) &  0.72 ($\pm$0.07) \\
    \hline
    \textbf{Time (min)} & {\bf 4.95} & 5.88 ($\pm$0.38) & 4.98 ($\pm$0.63) & {\bf 39.57} & 43.52 ($\pm$4.08) & 39.05 ($\pm$ 3.87) \\
    \hline
    \textbf{\# API Calls} & {\bf 279} & 334 ($\pm$61) & 254 ($\pm$32) & {\bf 2043} & 2308 ($\pm$225) &  1985 ($\pm$188) \\
    \hline
    \end{tabular}
}
\vspace{-3mm}
\end{table}

\noindent
In this section, we conduct an ablation study to assess our similarity-based hierarchical cluster merge strategy in the end-to-end ER pipeline, comparing it to random merging with and without {\sf MDG}. Random merging experiments are repeated five times, reporting average performance and standard deviations to evaluate stability.

Table \ref{tab:random merge ablation} shows that the similarity-based strategy outperforms random merging (Ran.) and random merging without {\sf MDG} (Ran. w/o MDG) in efficiency and effectiveness on \textit{Cora} and \textit{Alaska} datasets. On \textit{Cora}, it reduces costs (0.04 to 0.03 USD) and tokens (0.15M to 0.12M) compared to Ran. On \textit{Alaska}, it lowers costs (0.16 to 0.15 USD), tokens (0.82M to 0.73M), time (43.52 to 39.57 minutes), and API calls (2308 to 2043). Ran. w/o MDG matches Ran.’s average efficiency (e.g., 0.03 USD, 254 API calls on \textit{Cora}) but shows high variability (e.g., $\pm$0.01 USD, $\pm$32 API calls), indicating instability. The similarity-based approach’s efficiency stems from aligning with $S_d$ and $S_v$ constraints (\S\ref{sec:experiment on key factors}), creating cohesive record sets that reduce LLM prediction errors and {\sf MDG} regenerations, minimizing API usage, tokens, and costs with low variance. Random merging produces less coherent record sets, increasing errors and {\sf MDG} reliance, with higher variability (e.g., $\pm$61 API calls on \textit{Cora}). Ran. w/o MDG avoids regeneration but is unstable (e.g., $\pm$188 API calls on \textit{Alaska}) due to unstructured merging.

For clustering quality, the similarity-based approach achieves the highest ACC (0.90 on \textit{Cora}, 0.82 on \textit{Alaska}) and FP-measure (0.71 on \textit{Cora}, 0.79 on \textit{Alaska}) with minimal variability, leveraging cohesive record sets. Random merging maintains similar quality (e.g., ACC 0.87 on \textit{Cora}, $\pm$0.03) due to {\sf MDG} corrections, but Ran. w/o MDG shows lower quality and higher variability without error correction. This underscores {\sf MDG}’s role in robust clustering, while the similarity-based strategy excels in both efficiency and quality.

\stitle{A.9 Qualitative Analysis of our In-context Clustering-based Method with Other Baselines}

\begin{figure}[h]
\centering
\vspace{-2mm}
  \hspace{-7.5mm}
  \subfigcapskip=-2mm
  \subfigure[Booster]{
   \includegraphics[width=1.17in]{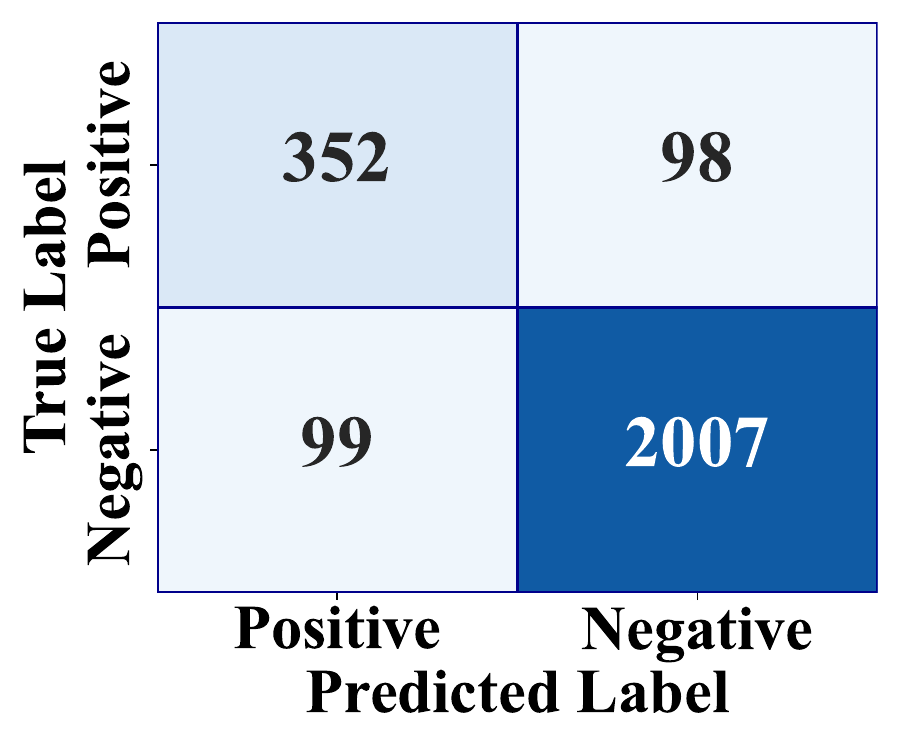}
   \label{fig:qualitative-booster}
  }
 \hspace{ -2.7mm}
  \subfigure[BQ]{
   \includegraphics[width=1.17in]{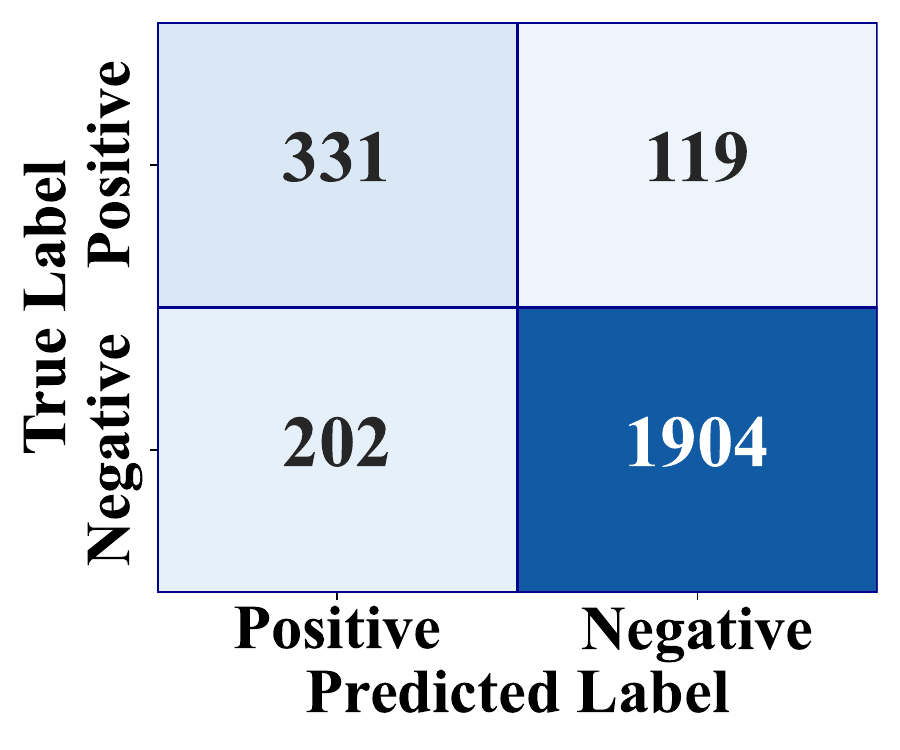} 
   \label{fig:qualitative-bq}
  }
  \hspace{-2.7mm}
  \subfigure[LLM-CER]{
   \centering
   \includegraphics[width=1.17in]{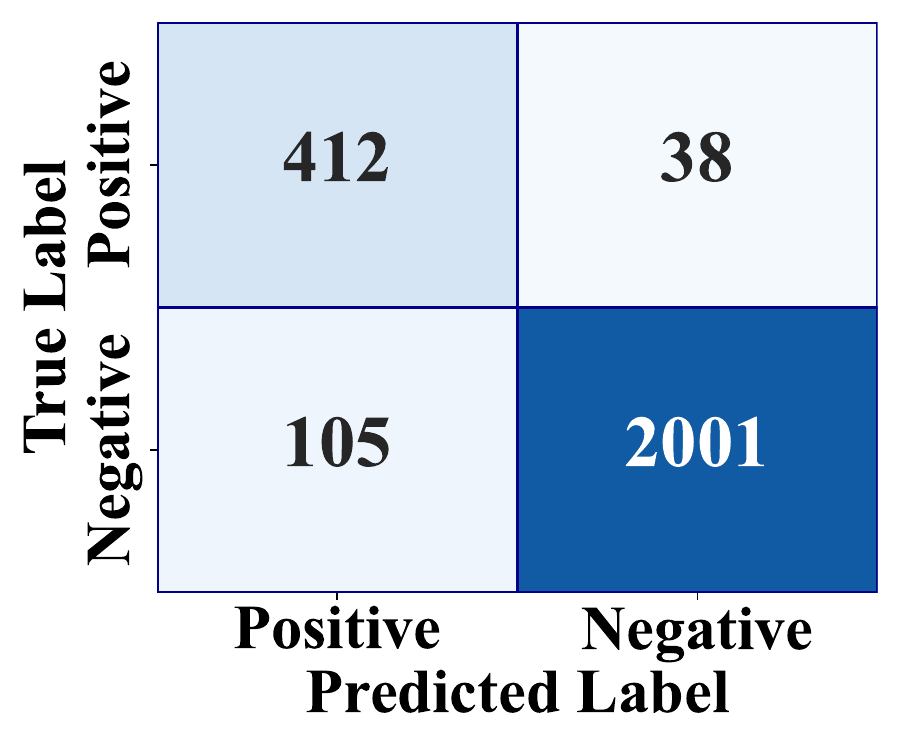}
   \label{fig:qualitative-ours}
  }
\vspace{-5mm}
\caption{ { \centering {Qualitative analysis of our experimental results  } } }
\vspace{-4mm}
\label{fig:qualitative analysis}
\end{figure}

\noindent
We present a qualitative analysis of our in-context clustering-based method compared to two baseline algorithms, {\sf BQ} \cite{fan2024cost} and {\sf Booster} \cite{li2024booster}, using the \textit{Cora} dataset. To evaluate performance, we employed confusion matrices, measuring True Positives (TP), False Positives (FP), False Negatives (FN), and True Negatives (TN) to highlight differences in clustering accuracy. Specifically, we first apply {\sf LSH} for preliminary blocking and select a block of 72 records, comprising 6 entities (4 entities with 15 duplicate records each and 2 entities with 6 duplicate records each). We then conduct experiments using our method, {\sf BQ}\,, and {\sf Booster} on this block and evaluate the ER results. Notably, with 72 records, there are total $72*71/2=2556$ record pairs, which is the sum of the four numbers indicating TP, TN, FP, FN obtained for each method shown in Figure \ref{fig:qualitative analysis}. 

\spara{Experiment Result Analysis.} As shown in Figure \ref{fig:qualitative analysis}, our algorithm achieves a TP count of 412, surpassing {\sf Booster} (352) and {\sf BQ} (331), indicating superior clustering accuracy in correctly grouping records of the same entity. Additionally, our method records an FN count of 38 and an FP count of 105, demonstrating a balanced performance with fewer errors compared to the baselines. 

{\sf BQ}\,, lacking {\sf MDG} detection, struggles with transitive errors, often merging records from different entities into the same cluster and splitting records of the same entity across clusters. This results in a high FP (202) and FN (119), significantly degrading its ER quality. In contrast, our {\sf MDG}-equipped approach effectively identifies and mitigates such errors, enhancing clustering precision. {\sf Booster}, while adept at distinguishing records from different entities (achieving the lowest FP of 99), tends to over-segment, leading to a higher FN (98) due to excessive differentiation. This over-separation reduces its FP-measure and overall effectiveness. Our algorithm, with an FP (105) close to {\sf Booster}’s and a substantially lower FN, strikes a better balance, yielding a superior overall ER performance.

\stitle{A.10 Integrating Batch Processing within Our Framework}

\noindent
In this section, we integrate batch processing from the baseline {\sf BQ} \cite{fan2024cost} into our framework to enhance end-to-end ER efficiency. Specifically, we incorporate multiple record sets within a single prompt, tasking the LLM with sequentially classifying them. We conduct two experiments: First, we investigate the impact of varying the number of record sets in the prompt on clustering performance. Based on these findings, we evaluate the end-to-end ER results under this setup and quantify the extent of improvement it brings. 

\begin{figure}[h]
\centering
\vspace{-3.5mm}
  \hspace{-7mm}
  \subfigcapskip=-2mm
  \subfigure[Citeseer]{
   \includegraphics[width=1.75in]{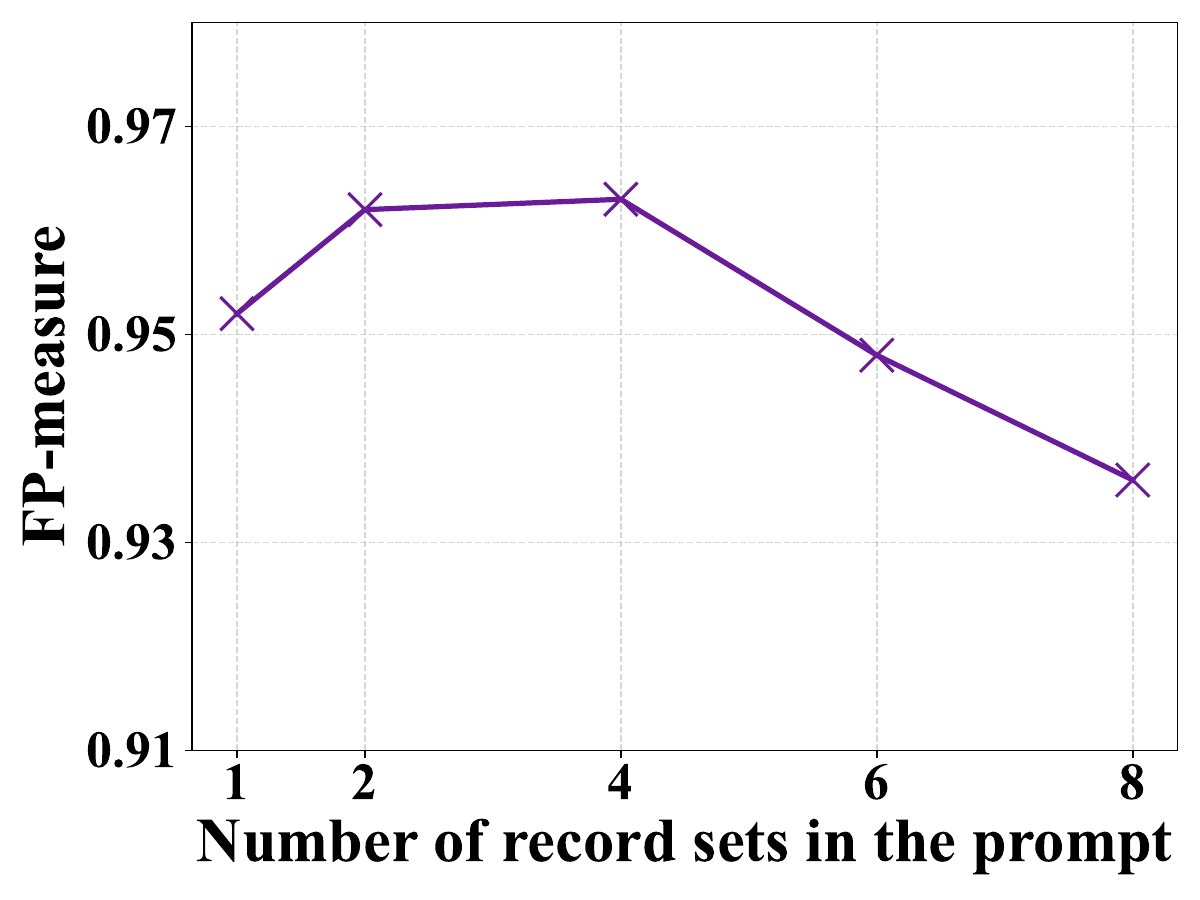}
   \label{fig:batch-citeseer}
  }
  \hspace{-2.7mm}
  \subfigure[Walmart-Amazon]{
   \centering
   \includegraphics[width=1.75in]{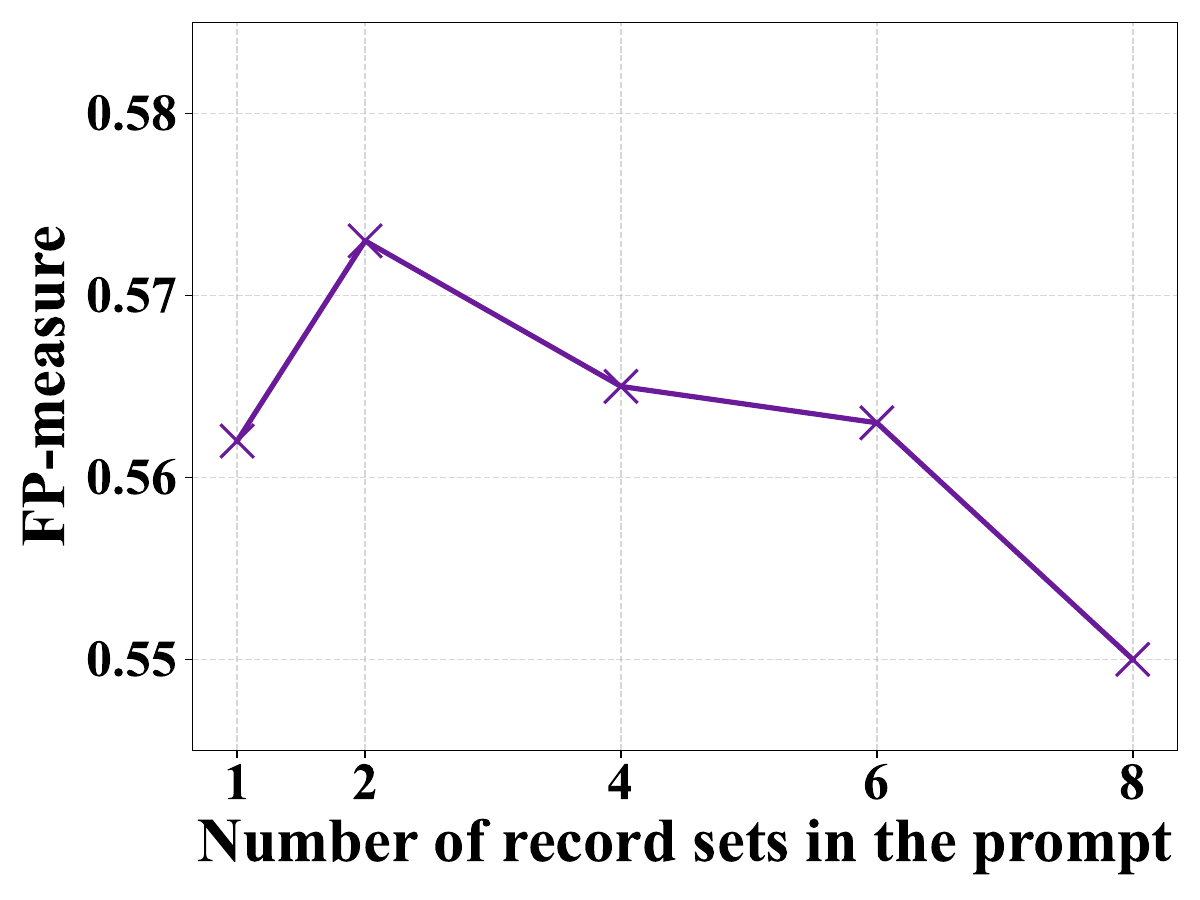}
   \label{fig:batch-wz}
  }
\vspace{-5mm}
\caption{ \centering {  FP-measure vs. batch count } } 
\vspace{-4mm}
\label{fig:batch-perform}
\end{figure}

\spara{Impact of Batch Size on In-context Clustering Performance.} As shown in Figure \ref{fig:batch-perform}, our experiments evaluate the impact of batch size (number of record sets in a prompt) on in-context clustering performance within our end-to-end ER framework integrated with batch processing. The results reveal that FP-measure initially increases with larger batch sizes, consistent with findings in \cite{fan2024cost}, as the LLM leverages prior classifications to improve subsequent ones. However, as the number of record sets grows further, excessive context length degrades LLM performance, leading to a decline in FP-measure.

\begin{table}[h]
\vspace{-2mm}
\caption{ {Impact of batch processing on end-to-end ER} }
\label{tab:batching e2e}
\vspace{-3mm}
    \centering
    \resizebox{8.5cm}{!}{
    \renewcommand{\arraystretch}{1.2}
    \begin{tabular}{c|c|c|c|c}
    \hline
    \multirow{2}{*}{\textbf{Metrics}} & \multicolumn{2}{c|}{\em{Citeseer}} & \multicolumn{2}{c}{\em{Walmart-Amazon}}  \\
    \cline{2-5}
     & $\textbf{w/ batching}$ & $\textbf{w/o batching}$  & $\textbf{w/ batching}$ & $\textbf{w/o batching}$ \\
    \hline
    \textbf{ACC} & {\bf 0.90} & 0.88  &  {\bf 0.64} & 0.61  \\
    \hline
    \textbf{FP-measure} & {\bf 0.96} & 0.95  & {\bf 0.57}  & 0.56  \\
    \hline
    \textbf{Cost (USD)} & {\bf 0.03} & 0.03 & {\bf 0.02}  & 0.02 \\
    \hline
    \textbf{Tokens (M)} & {\bf 0.12} & 0.13 & {\bf 0.05}  & 0.06  \\
    \hline
    \textbf{Time (min)} & {\bf 6.87} & 22.68 & {\bf 1.72} & 6.25 \\
    \hline
    \textbf{\# API Calls} & {\bf 318} & 1302 & {\bf 92} & 398 \\
    \hline
    \end{tabular}
}
\vspace{-3mm}
\end{table}

\spara{End-to-end ER Performance.}  Based on our prior experiments, we set the batch size to 4, as it optimizes in-context clustering performance while significantly accelerating the algorithm. We evaluate our end-to-end ER framework with batch processing on the \textit{Citeseer} and \textit{Walmart-Amazon} datasets, comparing performance with and without batching. As shown in Table \ref{tab:batching e2e}, batching improves ACC from 0.88 to 0.90 on \textit{Citeseer}, with FP-measure increasing from 0.95 to 0.96. Additionally, batching reduces computational overhead, lowering processing time more than $4\times$ on both \textit{Citeseer} and \textit{Walmart-Amazon}, while decreasing API calls from 1302 to 318 and 398 to 92, respectively. These improvements stem from the LLM’s ability to learn from prior classifications within a batch, enhancing clustering accuracy, and the reduced context overhead from fewer API calls. 

\begin{figure}[h]
\centering
\vspace{-1.5mm}
\includegraphics[width=0.35\textwidth]{figures/icon.pdf}\\
  \vspace{-1.5mm}
  \hspace{-7.5mm}
  \subfigcapskip=-2mm
  \subfigure[Music 20K]{
   \includegraphics[width=1.17in]{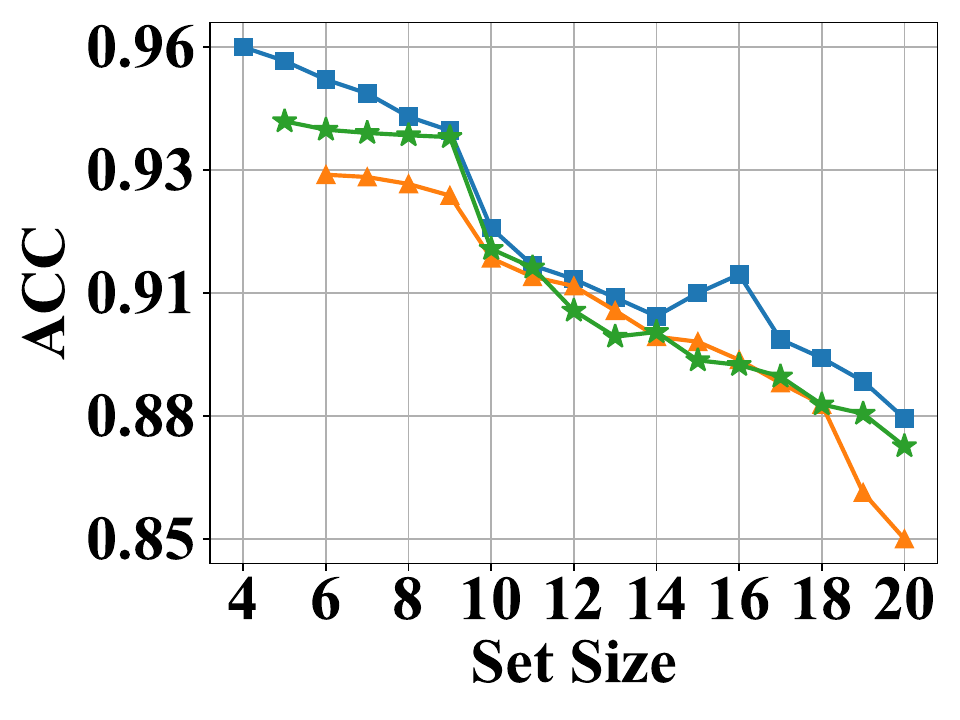}
   \label{fig:acc-music-diver4}
  }
 \hspace{ -2.7mm}
  \subfigure[Cora]{
   \includegraphics[width=1.17in]{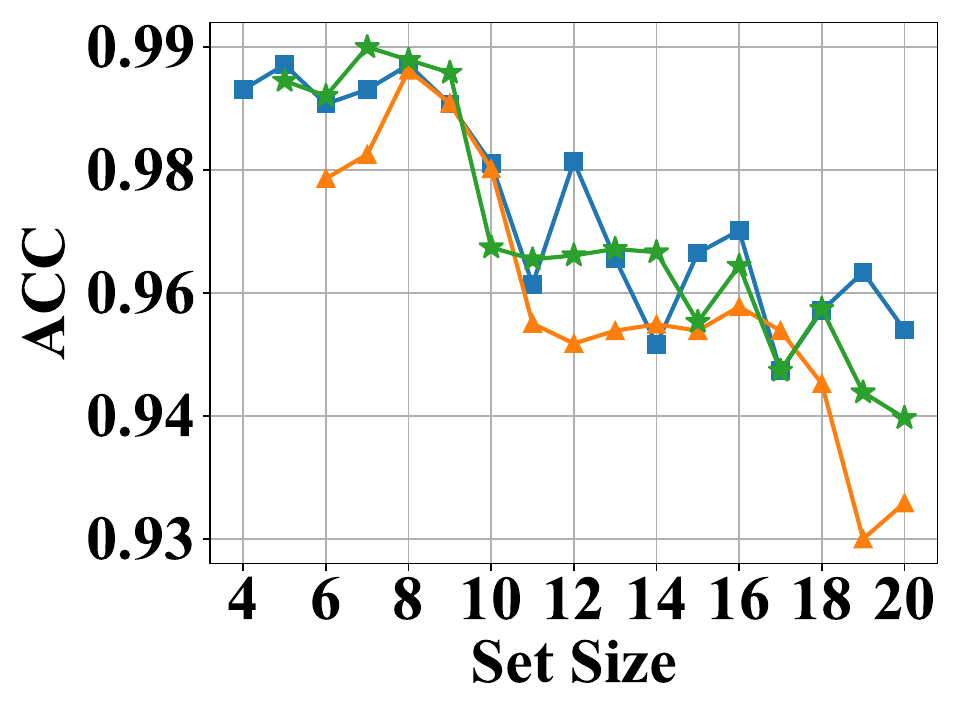} 
   \label{fig:acc-cora-diver4}
  }
  \hspace{-2.7mm}
  \subfigure[Alaska]{
   \centering
   \includegraphics[width=1.17in]{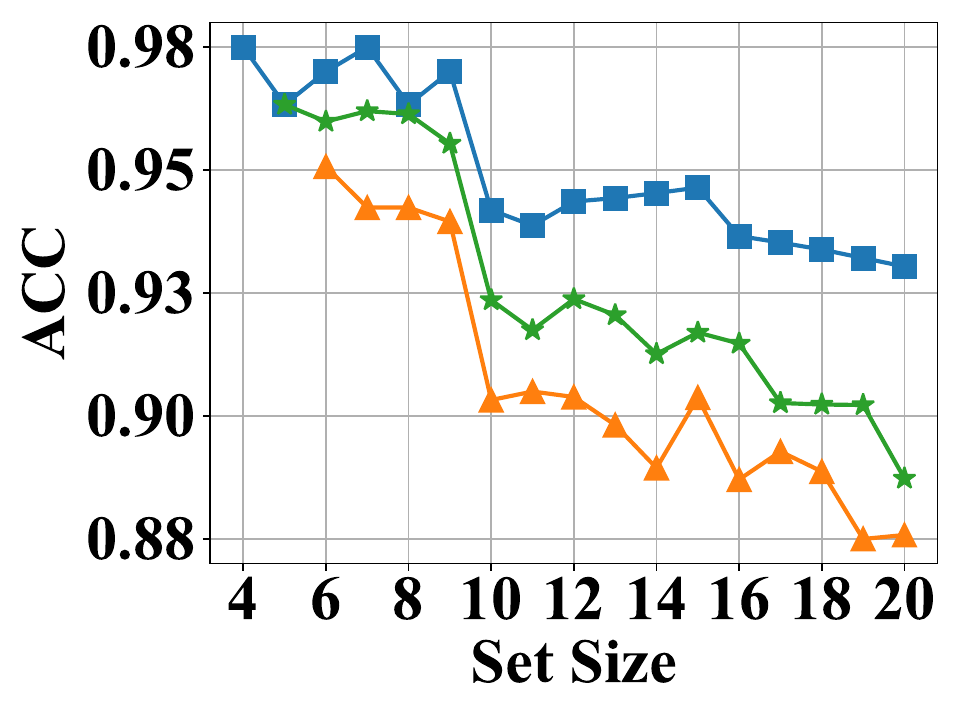}
   \label{fig:acc-sigmod-diver4}
  }
\vspace{-5mm}
\caption{ \centering {Clustering performance vs. $S_s$ and $S_v$ (ACC)} } 
\vspace{-4mm}
\label{fig:cluster-Ss-Sv-acc-diver4}
\end{figure}

\stitle{A.11 Key Factors Evaluations under ACC Metric}

\begin{figure}[h]
\centering
\vspace{-1.5mm}
\includegraphics[width=0.20\textwidth]{figures/icon_order.pdf}\\
  \vspace{-2mm}
  \hspace{-7mm}
  \subfigcapskip=-2mm
  \subfigure[Music 20K]{
   \includegraphics[width=1.17in]{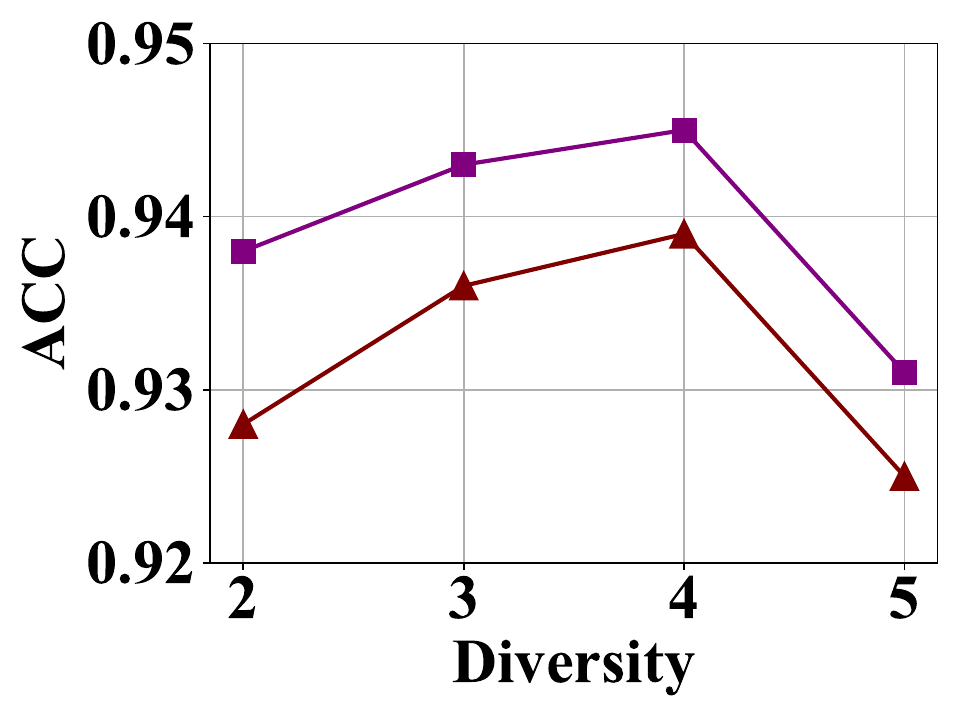}
   \label{fig:acc-music-order}
  }
 \hspace{ -3mm}
  \subfigure[Cora]{
   \includegraphics[width=1.17in]{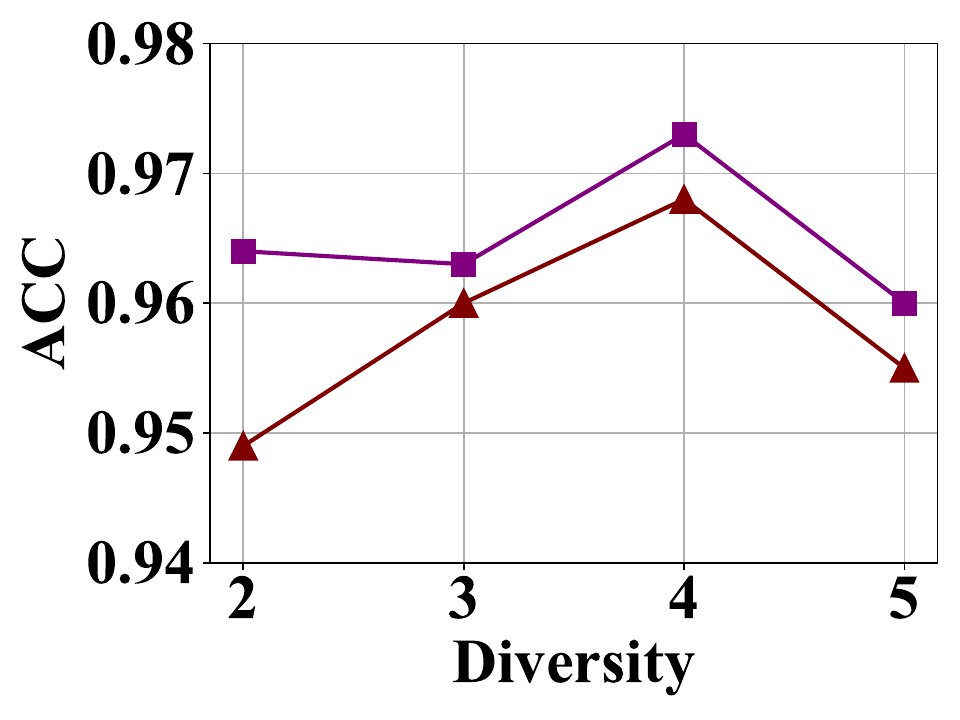} 
   \label{fig:acc-cora-order}
  }
  \hspace{-3mm}
  \subfigure[Alaska]{
   \centering
   \includegraphics[width=1.17in]{figures/performance_individual_batch/random_generate/sigmod/sigmod_sequential_vs_random.pdf}
   \label{fig:acc-sigmod-order}
  }
\vspace{-6mm}
\caption{ { \centering {Clustering performance vs. $S_d$, ordering (ACC)} } }
\label{fig:cluster-diversity-order-acc}
\vspace{-4mm}
\end{figure}

\noindent
In this section, we present the variation of key factors with respect to \( S_s \), \( S_d \), \( S_v \), and ordering under the ACC metric, as illustrated in the figures below. As shown in Figures~\ref{fig:cluster-Ss-Sv-acc-diver4} and~\ref{fig:cluster-diversity-order-acc}, the results are highly consistent with those under the FP-measure. For the \textit{Music 20K}, \textit{Alaska}, and \textit{Cora} datasets, the optimal \( S_s \) and \( S_d \) values remain 9 and 4, respectively, with \( S_v \) performing best when close to 0. Additionally, the records' sequential ordering enhances LLM's in-context clustering performance. These findings indicate that the optimal parameter settings are nearly identical across both clustering metrics.

\end{document}